\documentclass{article}
\usepackage{arxiv}

\usepackage[utf8]{inputenc} 
\usepackage[T1]{fontenc}    
\usepackage{hyperref}       
\usepackage{url}            
\usepackage{booktabs}       
\usepackage{amsfonts}       
\usepackage{amsmath}
\usepackage{nicefrac}       
\usepackage{microtype}      
\usepackage{lipsum}
\usepackage{graphicx}
\usepackage{float}
\usepackage{caption}
\usepackage{titlesec}
\usepackage{natbib}
\usepackage{setspace}
\titlespacing*{\section}{8pt}{12pt}{8pt}
\titlespacing*{\subsection}{8pt}{8pt}{8pt}
\titlespacing*{\subsubsection}{8pt}{8pt}{8pt}

\usepackage{lmodern}
\graphicspath{ {./images/} }
\usepackage{authblk}
\usepackage{float}
\usepackage{savesym}
\usepackage{moreverb}
\savesymbol{bigtimes}
\usepackage{mathabx}
\restoresymbol{TXF}{bigtimes}
\usepackage{makecell}
\usepackage{subcaption}
\usepackage{fancyvrb}
\usepackage{longtable}
\usepackage{multirow}
\usepackage{setspace}
\usepackage{rotating}
\usepackage{supertabular}

\def\bSig\mathbf{\Sigma}

\newcommand{\bbeta}{\mbox{\boldmath{$\beta$}}}

\title{Discrete-time Competing-Risks Regression with or without Penalization}

\author[1,*]{Tomer Meir}
\author[2]{Malka Gorfine}

\affil[1]{Faculty of Data and Decision Sciences, Technion - Israel Institute of Technology}
\affil[2]{Department of Statistics and Operations Research, Tel Aviv University}
\affil[*]{Corresponding Author: tomer1812@gmail.com}

\begin{document}
\maketitle
\begin{abstract}
Many studies employ the analysis of time-to-event data that incorporates competing risks and right censoring. Most methods and software packages are geared towards analyzing data that comes from a continuous failure time distribution. However, failure-time data may sometimes be discrete either because time is inherently discrete or due to imprecise measurement. 
This paper introduces a new estimation procedure for discrete-time survival analysis with competing events. The proposed approach offers a major key advantage over existing procedures and allows for straightforward integration and application of widely used regularized regression and screening-features methods. We illustrate the benefits of our proposed approach by  a comprehensive simulation study. Additionally, we showcase the utility of the proposed procedure by estimating a survival model for the length of stay of patients hospitalized in the intensive care unit, considering three competing events: discharge to home, transfer to another medical facility, and in-hospital death. A Python package, \texttt{PyDTS}, is available for applying the proposed method with additional features.
\end{abstract}



\section{Introduction}\label{sec1}
Most methods and software for survival analysis are tailored to data with continuous failure-time distributions. However, there are situations where failure times are discrete. This can be due to the nature of the time unit being discrete or because of inaccuracies in measurement. An example is time to pregnancy where the observation time is defined by the number of menstrual cycles. 
In some cases, events can happen at any point in time, but only the time interval in which each event occurred is recorded in available data. For instance, death from cancer recorded in months since diagnosis \citep{lee_analysis_2018}. 

Competing events occur when individuals are susceptible to several types of events but can only experience at most one event at a time. If multiple events can happen simultaneously, they can be treated as a separate event type \citep{kalbfleisch_statistical_2011}. For instance, competing risks in a study of hospital length of stay could be discharge and in-hospital death, where the occurrence of one of these events prevents observation of the other event for the same patient. Another classic example of competing risks is cause-specific mortality, such as death from heart disease, cancer, or other causes.  It is  acknowledged that using standard continuous-time models on discrete-time data with competing events  without proper adjustments  can lead to a {\it systematic} bias~\citep{lee_analysis_2018,wu2022analysis}. For example, \cite{lee_analysis_2018} noted the bias in the cumulative incidence function estimator, resulting in poor coverage rates, such as an empirical coverage of 0.66 versus a nominal level of 0.95.

The motivation for this project is to analyze data of length of stay (LOS) of patients in healthcare facilities. LOS typically refers to the number of days a patient stays in the hospital during a single admission \citep{lequertier2021hospital,awad_patient_2017}. Accurate prediction of LOS is crucial for hospital management and planning of bed capacity, as it affects healthcare delivery access, quality, and efficiency \citep{lequertier2021hospital}. In particular, hospitalizations in intensive care units (ICU) consume a significant amount of hospital resources per patient \citep{adhikari_critical_2010}. In this study, we use the publicly available Medical Information Mart for Intensive Care (MIMIC) - IV (version 2.0) data \citep{johnson_mimic-iv_2022, goldberger_physiobank_2000} to develop a model for predicting LOS in ICU based on patients' characteristics upon arrival in ICU. The study involves 25,170 ICU admissions from 2014 to 2020 with only 28 unique times, resulting in many tied events at each time point. The three competing events analyzed were: discharge to home (69.0\%), transfer to another medical facility (21.4\%), and in-hospital death (6.1\%). Patients who left the ICU against medical advice (1.0\%) were considered censored, and administrative censoring was imposed for patients hospitalized for more than 28 days (2.5\%).

Regression analysis of continuous-time survival data with competing risks can be performed using standard non-competing events tools because the likelihood function for the continuous-time setting can be factored into likelihoods for each cause-specific hazard function \citep{kalbfleisch_statistical_2011}. However, this is not the case for some regression models of discrete-time survival data with competing risks \citep{allison_discrete-time_1982}. 
The literature on discrete-time survival data with competing risks can be categorized into two primary groups. The first group involves cause-specific hazard functions that serve as a natural and direct analogy to those found in the continuous survival time context. In this case, the cause-specific hazard function is solely dependent on the parameters of the specific competing event \citep{allison_discrete-time_1982,lee_analysis_2018,wu2022analysis}. As this formulation results in a likelihood that cannot be decomposed into distinct components for each type of event, \cite{allison_discrete-time_1982} explored an alternative, more manageable formulation. In particular, he introduced a cause-specific hazard function that depends not only on the parameters associated with the specific competing event but also on the parameters related to all other event types. This cause-specific hazard formulation was later adopted and further developed by others \citep{tutz2016modeling,most2016variable,schmid2021competing}. As noted by \cite{allison_discrete-time_1982}, the advantage of the second approach is a significant simplification of the estimation procedure, albeit at the cost of interpretability. For additional technical details comparing these two approaches, please refer to Section \ref{sec:discussion} below.
In this work, we choose to align with the first suggestion of \cite{allison_discrete-time_1982} and the approach also adopted by \cite{lee_analysis_2018}, and focus on the natural and direct analogy to the cause-specific hazard function in the context of continuous survival time, as this formulation of cause-specific hazard models provides a clearer interpretation.

\cite{lee_analysis_2018} showed that if one naively treats competing events as censoring in the discrete-time likelihood, separate estimation of cause-specific hazard models for different event types may be accomplished using a collapsed likelihood which is equivalent to fitting a generalized linear model to repeated binary outcomes. Moreover, the maximum collapsed-likelihood estimators are consistent and asymptotically normal under standard regularity conditions, which gives rise to Wald confidence intervals and likelihood-ratio tests for the effects of covariates. \cite{wu2022analysis} focused on two competing events and used a different approach from that of \cite{lee_analysis_2018}. However, they noted that it leads to the same estimators. The contribution of \cite{wu2022analysis} is mainly by allowing an additional fixed effect of  medical center in the model. 

Consider a setting with $M$ competing events. Each cause-specific hazard model of \cite{lee_analysis_2018} includes $d+p$ parameters, where $d$ parameters can be viewed as the cause-specific baseline-hazard parameters and $p$ are the unknown cause-specific regression coefficients. As will be shown in Section \ref{sec:methods}, the standard maximum likelihood approach requires estimating $M(d+p)$ parameters simultaneously. \cite{lee_analysis_2018}  substantially simplified this by estimating $p+d$ parameters for each competing event separately. 

In this work we focus primarily on the popular logit-link function and introduce a new estimation technique that further simplifies the estimation procedure. Our new estimator separates the estimation procedure of the $d$ cause-specific baseline-hazard parameters and the $p$
cause-specific regression coefficients, within each event type. It will be demonstrated that this separation is highly useful 
for incorporating common penalized methods like lasso and elastic net among others \citep{hastie2009elements} and enables easy implementation of screening methods for high-dimensional data, such as sure independent screening \citep{fan2008sure,fan2010high,zhao2012principled,saldana2018sis}. Our Python software, \texttt{PyDTS} \citep{meir_pydts_2022}, implements both our method and the one from \cite{lee_analysis_2018} and other tools for discrete-time survival analysis.


It might seem that the advantages of using a proper discrete-time competing-events regression model over a continuous-time model would diminish when $d$ is large. However, Web Appendix A provides simulation results challenging this assumption. We compared our discrete analysis to a naive analysis using the standard partial-likelihood approach, like that in the R function \texttt{coxph}. With 2,000 observations, 9 time points, and 2 competing events, the naive approach's baseline hazard estimators—regardless of using Breslow, Efron, or Exact tie corrections—showed substantial biases. In contrast, our method produced almost unbiased results. This finding holds similarly with 5,000 observations and 50 time points. The bias in the naive approach arises from the inappropriate use of the Breslow estimator for baseline hazards, which theoretically is only justifiable when the likelihood can be decomposed into distinct components for each event type, a criterion our discrete-time model does not meet, as will be shown in the next section.


\section{Methods}
\label{sec:methods}
\subsection{Models and Likelihood Function}
\label{sec:definitions}
Consider $T$ as a discrete event time taking values ${1,2,\ldots,d}$, and $J$ as the type of event, where $J \in \{1,\ldots,M\}$. Let ${\bf Z}$ be a  $p \times 1$ vector of time-independent covariates. The setting of time-dependent covariates will be discussed later. The discrete cause-specific hazard function is defined as
$
\lambda_j(t|{\bf Z}) = \Pr(T=t,J=j|T\geq t, {\bf Z})  
$
for $t = 1,2,\ldots,d$ and $j=1,\ldots,M$. Following \cite{allison_discrete-time_1982} framework, the semi-parametric  hazard functions via 
 a regression transformation model are expressed as
$$
h(\lambda_{j}(t|{\bf Z}))  = \alpha_{jt} +{\bf Z}^T \bbeta_j  \,\,\,\,\,\,\,\, t = 1,2,\ldots, \,\,\,\,  \,\,\,\, j=1,\ldots,M,
$$
where  $h$ is a known function. The model's complexity is emphasized by its semi-parametric nature, handling $M(d+p)$ unknown parameters. The shared covariates ${\bf Z}$ among the $M$ models does not require that every model uses all the covariates. The  regression coefficient vectors, $\bbeta_j$, are specific to different event types, allowing for flexibility in model specification. By setting any coefficient to zero, its corresponding covariate can be excluded from that particular model. We adopt the popular logit transformation $h(a)=\log \{ a/(1-a) \}$, leading to the following  cause-specific hazard function
\begin{equation}\label{eq:logis}
\lambda_j(t|{\bf Z})=\frac{\exp(\alpha_{jt}+{\bf Z}^T\bbeta_j)}{1+\exp(\alpha_{jt}+{\bf Z}^T\bbeta_j)} \, .
\end{equation}
This approach, which leaves $\alpha_{jt}$ unspecified, parallels the method of an unspecified baseline hazard in the Cox model \citep{cox_regression_1972}, affirming the semi-parametric nature of our discrete-time model.

Define $S(t|{\bf Z}) = \Pr(T>t|{\bf Z})$ as the overall survival given ${\bf Z}$. Then, the probability that an event of type $j$ occurs at time $t$, $t=1,\ldots,d$, $j=1,\ldots,M$,  is given by
$$
\Pr(T=t,J=j|{\bf Z})=\lambda_j(t|{\bf Z})S(t-1|{\bf Z})=\lambda_j(t|{\bf Z}) \prod_{k=1}^{t-1} \left\{1-
\sum_{j'=1}^M\lambda_{j'}(k|{\bf Z}) \right\}  \, .
$$
The probability of event type $j$ by time $t$ given $Z$, also known as the cumulative incident function (CIF) is 
$
F_j(t|{\bf Z}) = 	 \sum_{k=1}^{t}\lambda_j(k|{\bf Z}) \prod_{l=1}^{k-1} \left\{1-\sum_{j'=1}^M\lambda_{j'}(l|{\bf Z}) \right\}
$, and the marginal probability of event type $j$  equals
$
\Pr(J=j|{\bf Z}) = \sum_{t=1}^{d} \lambda_j(t|{\bf Z}) \prod_{k=1}^{t-1} \left\{1-\sum_{j'=1}^M\lambda_{j'}(k|{\bf Z}) \right\} 
$.
Our goal is estimating the parameters 
$\Omega = (\alpha_{11},\ldots,\alpha_{1d},\bbeta_1^T,  \ldots,  \alpha_{M1},\ldots,\alpha_{Md},\bbeta_M^T)$.

For simplicity, we temporarily assume two competing events, i.e., $M=2$. The data consist of $n$ independent observations, each with $(X_i,\delta_i,J_i,{\bf Z}_i)$ where $X_i=\min(C_i,T_i)$, $C_i$ is a discrete right-censoring time, 
$\delta_i=I(T_i \leq C_i)$ is the event indicator and $J_i\in\{0,1,2\}$, where $J_i=0$ if and only if $\delta_i=0$, 
$i=1,\ldots,n$. It is assumed that given the covariates, the censoring and failure times are independent and non-informative in the sense  of Section 3.2 of \cite{kalbfleisch_statistical_2011}. In the case of grouped continuous-time data,  it is assumed that events always occur before censoring within the same interval. Then, the likelihood function is proportional to 
\begin{eqnarray*}
	L &=& \prod_{i=1}^n  \left\{\frac{\lambda_1(X_i|{\bf Z}_i)}{1-\lambda_1(X_i|{\bf Z}_i)-\lambda_2(X_i|{\bf Z}_i)}\right\}^{I(J_i=1)}
	 \left\{\frac{\lambda_2(X_i|{\bf Z}_i)}{1-\lambda_1(X_i|{\bf Z}_i)-\lambda_2(X_i|{\bf Z}_i)}\right\}^{I(J_{i}=2)} \\
  && \times  \prod_{t=1}^{X_i}\{1-\lambda_1(t|{\bf Z}_i)-\lambda_2(t|{\bf Z}_i)\} \, .
\end{eqnarray*}	
Equivalently,
\begin{eqnarray*}
L &=& \prod_{i=1}^n \left[ \prod_{j=1}^2 \prod_{t=1}^{X_i} \left\{  \frac{\lambda_j(t|{\bf Z}_i)}{1-\lambda_1(t|{\bf Z}_i)-\lambda_2(t|{\bf Z}_i)}\right\}^{\delta_{jit}}\right] \prod_{t=1}^{X_i}\{1-\lambda_1(t|{\bf Z}_i)-\lambda_2(t|{\bf Z}_i)\}
\end{eqnarray*}
and the log-likelihood (up to a constant) becomes
\begin{equation}
	\log L = \sum_{i=1}^n \sum_{t=1}^{X_i} \left[  \sum_{j=1}^2\delta_{jit} \log \lambda_j(t|{\bf Z}_i) 
  +  \{ 1-\delta_{1it}-\delta_{2it}\}\log\{ 1-\lambda_1(t|{\bf Z}_i)-\lambda_2(t|{\bf Z}_i)\}\right] 
\end{equation}
where $\delta_{jit}$ is set to one if subject $i$ experiences event of type $j$ at time $t$, and 0 otherwise. Evidently, in contrast to the continuous-time setting with competing events, $L$ cannot be decomposed into separate likelihoods for each cause-specific
hazard function $\lambda_j$. To estimate  $\Omega$, which encompasses $M(d+p)$ parameters, maximizing  $\log L$ becomes time-intensive. \cite{lee_analysis_2018} suggested estimating each set of $d+p$ parameters of each cause independently. We enhance this approach by separately estimating $(\alpha_{j1},\ldots,\alpha_{jd})$ and $\bbeta_j$ for each cause.

\subsection{The Collapsed Log-Likelihood Approach of Lee et al. (2018)}
\label{subsection:lee2018}
The estimation method of \cite{lee_analysis_2018} uses a collapsed log-likelihood approach, simplifying the analysis by expanding the dataset. Each subject $i$ is represented by multiple dummy observations up to time $X_i$. For each time $t \leq X_i$, indicators $\delta_{jit}=I(T_i=t, J_i=j)$ are defined for whether event type $j$ occurs at time $t$; see Table~S1 of the Supplementary Material (SM). This setup allows for a conditional multinomial distribution of events. With $M=2$, we get   $\{\delta_{1it},\delta_{2it},1-\delta_{1it}-\delta_{2it}\}$ and the estimation of $(\alpha_{11},\ldots,\alpha_{1d},\bbeta_1^T)$ utilizes a collapsed log-likelihood  where $\delta_{2it}$ and $1-\delta_{1it}-\delta_{2it}$ are combined. This collapsed log-likelihood is tailored for analyzing cause $j=1$ using a binary regression model, with $\delta_{1it}$ serving  as the outcome variable, and is given by
$$
\log L_1(\alpha_{11},\ldots,\alpha_{1d},\bbeta_1) = \sum_{i=1}^n \sum_{t=1}^{X_i}\left[ \delta_{1it} \log \lambda_1(t|{\bf Z}_i)+(1-\delta_{1it})\log \{1-\lambda_1(t|{\bf Z}_i)\} \right] \, .
$$
Similarly, the collapsed log-likelihood for cause $j=2$  with $\delta_{2it}$ as the outcome becomes
$$
\log L_2(\alpha_{21},\ldots,\alpha_{2d},\bbeta_2) = \sum_{i=1}^n \sum_{t=1}^{X_i}\left[ \delta_{2it} \log \lambda_2(t|{\bf Z}_i)+(1-\delta_{2it})\log \{1-\lambda_2(t|{\bf Z}_i)\} \right] \, ,
$$
and one can fit the two models, separately.  In general, for $M$ competing events, 
the estimators of $(\alpha_{j1},\ldots,\alpha_{jd},\bbeta_j^T)$, are the respective values that maximize  
\begin{equation}\label{eq:logLj}
\log L_j(\alpha_{j1},\ldots,\alpha_{jd},\bbeta_j) = \sum_{i=1}^n \sum_{t=1}^{X_i}\left[ \delta_{jit} \log \lambda_j(t|{\bf Z}_i)+(1-\delta_{jit})\log \{1-\lambda_j(t|{\bf Z}_i)\} \right] 
\end{equation}
with $j=1,\ldots,M$. Namely, each maximization for event $j$ involves   $d + p$ parameters. \cite{lee_analysis_2018} showed that the estimators  are asymptotically  multivariate normally distributed and the covariance matrix can be consistently estimated.   Since $L$ does not separate into distinct components for each event type, optimizing each collapsed likelihood $L_j$ separately does not produce the same results as maximizing the entire likelihood across all parameters. This introduces a trade-off between computational simplicity and the potential loss of estimation efficiency. The authors also pointed out that standard generalized linear models (GLM) could be used for each $\log L_j$, and due to the Markov property ensuring conditional independence, the basic variance estimator from the GLM, which presumes independence, remains valid.

\subsection{The Proposed Approach}
 When applying penalized regression or screening analysis (i.e., performing separate regression  for each covariate)  with the above collapsed log-likelihoods, it is necessary to estimate both $(\alpha_{j1},\ldots,\alpha_{jd})$  and $\bbeta_j$ for each cause $j$, rather than only $\bbeta_j$. Our proposed procedure separates the estimation of  $(\alpha_{j1},\ldots,\alpha_{jd})$ and $\bbeta_j$ within each cause. This separation allows for focusing solely on estimating $\bbeta_j$ during the penalized regression or screening processes. Subsequently, $(\alpha_{j1},\ldots,\alpha_{jd})$ is consistently estimated using new estimating equations.

For separating the estimation of $(\alpha_{j1},\ldots,\alpha_{jd})$ and $\bbeta_j$ within each cause, we adopt  the conditional-logistic regression approach \citep{cox2018analysis,gail1981likelihood}. This involves analyzing the  expanded dataset. Let $\mathcal{N}_t$ be the set of all dummy observations with $\widetilde{X}$ equal to $t$ (see Table S1 of SM). A likelihood based on conditional-logistic regression is replacing Eq. (\ref{eq:logLj}), which stratifies the expanded dataset by $\widetilde{X}$ and conditions on the number of observed events within each stratum,  $\sum_{i \in \mathcal{N}_t} \delta_{jit}$. Specifically, define ${\bf d}_{jt}$ as a vector of 0s and 1s with a length equal to the cardinality of  $\mathcal{N}_t$, where $d_{jit}$ represents its components. Also, let $\mathcal{S}_{jt}$ be the set of all possible vectors ${\bf d}_{jt}$ such that $\sum_{i \in \mathcal{N}_t }d_{jit}=\sum_{i \in \mathcal{N}_t}\delta_{jit}$. Then,
the conditional likelihoods of the expanded data, stratified by $\widetilde{X}$ and given $\sum_{i \in \mathcal{N}_t} \delta_{jit}$, $t=1,\ldots,d$,
are given by
\begin{equation}\label{eq:betaj}
L_j^{\mathcal{C}}(\bbeta_j)  =  \prod_{t=1}^{d} \frac{\exp(\sum_{i \in \mathcal{N}_t} \delta_{jit} {\bf Z}_i^T \bbeta_j)}{\sum_{{\bf d}_{jt} \in \mathcal{S}_{jt}} \exp(\sum_{i \in \mathcal{N}_t} d_{jit} {\bf Z}_i^T \bbeta_j)} \,\,\,\,\,\,\,  j=1,\ldots,M \, .
\end{equation}
The estimators $\widehat{\bbeta}_j$ are the values of ${\bbeta}_j$ that maximize the conditional likelihoods. Clearly, $\exp(\alpha_{jt})$ in the numerator and denominator, within each $j$ and $t$, is canceled out. 

Eq. (\ref{eq:betaj}) resembles the partial likelihood from a Cox regression model when ties are present (see, for example, Eq. (8.4.3) of \cite{klein_survival_2003}), enabling the use of standard Cox-model routines for estimating $\bbeta_j$, $j=1,\ldots,M$.  In \texttt{R}, the \texttt{clogit}  function employs this strategy by creating necessary dummy variables and strata, then calling \texttt{coxph}. This function defaults to the Breslow approximation for conditional likelihood, with options for exact forms and other common tie approximations available.  The use of available Cox model routine for maximizing Eq. (\ref{eq:betaj}) is only a mathematical trick while Eq. (\ref{eq:logis}) still holds.

Leveraging the estimators  $\widehat{\bbeta}_j$, $j=1,\ldots,M$, we propose estimating $\alpha_{jt}$, $j=1,\ldots,M$, $t=1,\ldots,d$, through a series of $Md$ single-dimensional optimization algorithms applied to the original (i.e., non-expanded) dataset such that for each $(j,t)$, 
\begin{equation}\label{eq:alpha} 
        \widehat{\alpha}_{jt} = 
        \mbox{argmin}_{a} \left\{ \frac{1}{Y.(t)} \sum_{i=1}^n I(X_i \geq t)\frac{\exp(a+{\bf Z}_i^T\widehat{\bbeta}_j)}{1+\exp(a+{\bf Z}_i^T\widehat{\bbeta}_j)} - \frac{N_j(t)}{Y.(t)}\right\}^2
        \, ,
\end{equation}
where $Y.(t)=\sum_{i=1}^n I(X_i \geq t)$ and $N_j(t)=\sum_{i=1}^n I(X_i = t, J_i=j)$. Eq.~(\ref{eq:alpha}) involves minimizing the squared difference between the observed proportion of failures of type $j$ at time $t$, i.e., $N_j(t)/Y.(t)$, and the expected proportion of failures, as determined by Model (\ref{eq:logis}) and $\widehat{\bbeta}_j$. Since each $\alpha_{jt}$ is estimated separately, standard optimization routines like \texttt{nlminb} in  \texttt{R} or \texttt{minimize} of \texttt{scipy} in \texttt{python} are suitable for use.

In summary, the new proposed estimation procedure consists of the following two steps:
\begin{enumerate}
    \item Using the expanded dataset, estimate each  $\bbeta_j$ individually,  by maximizing Eq.~(\ref{eq:betaj}) using a stratified Cox routine, such as the \texttt{clogit} function in the \texttt{survival} R package, and get $\widehat{\bbeta}_j$, $j = 1,\ldots, M$.
    \item Using $\widehat{\bbeta}_j$, $j = 1,\ldots, M$, and the  non-expanded dataset, estimate each $\alpha_{jt}$, $j = 1,\ldots,M$, $t=1,\ldots,d$, separately, by Eq.~(\ref{eq:alpha}).
\end{enumerate}
The simulation results in Section \ref{sec:simul} show that the above two-step procedure performs well in terms of bias and provides similar standard errors to those of \cite{lee_analysis_2018}. 

The consistency and asymptotic normality of each $\widehat{\bbeta}_j$, $j=1,\ldots,M$, follow a similar argument of  \cite{lee_analysis_2018}. Namely, due to the Markov property, which includes conditional independence of the binary variables, the properties of the estimators and the naive variances' estimators from the conditional logistic regression approach above which assumes independence remain valid, as $n \rightarrow \infty$ and under finite fixed values of $d$ and $M$. The consistency and asymptotic normality of  $\widehat{\alpha}_{jt}$ are derived in Web Appendix B.

The proposed two-step estimation procedure can easily handle covariates or coefficients that change over time, ${\bf Z}(t)$ and $\bbeta_j(t)$, respectively. Similarly to  continuous survival time,  time-dependent covariates are coded  by breaking the individual's time into multiple time intervals, with one row of data for each interval. Hence, combining this data expansion step with the expansion described in Table S1 is straightforward. For time-dependent coefficients, $\bbeta_j(t)$, Eq. (\ref{eq:betaj}) is replaced by
$
L_j^{\mathcal{C}}(\bbeta_j(t))  =   \frac{\exp\{\sum_{i \in \mathcal{N}_t} \delta_{jit} {\bf Z}_i^T \bbeta_j(t)\}}{\sum_{{\bf d}_{jt} \in \mathcal{S}_t} \exp\{\sum_{i \in \mathcal{N}_t} d_{jit} {\bf Z}_i^T \bbeta_j(t)\}}  
$
with $j=1,\ldots,M \, , \, t=1,\ldots,d$.
Clearly, one can easily combine time-dependent covariate with time-dependent coefficients. Estimating $\alpha_{jt}$ with  time-dependent covariates or regression coefficients involves using ${\bf Z}(t)$ and $\widehat{\bbeta}_j(t)$ in the modified version of Eq.~(\ref{eq:alpha}).

\subsection{The Utility of the Proposed Approach}
Advancements in data collection technologies have greatly increased the number of potential predictors. Our method of separating the estimation of $\bbeta_j$ from $(\alpha_{j1},\ldots,\alpha_{jd})$ is particularly useful in dimension reduction and model selection. Below are two examples demonstrating the effectiveness of our two-step estimation procedure.

{\bf Example 1: Regularized regression.}  Penalized regression~\citep{hastie2009elements} methods  place a constraint on the size of the regression coefficients. We propose to apply penalized regression methods  in Lagrangian form based on Eq.~(\ref{eq:betaj}) by minimizing  
\begin{equation}\label{ref:betajregul}
-\log L_j^{\mathcal{C}}(\bbeta_j)  + \eta_j P(\bbeta_j)
\,\,\, , \,\,\, j=1,\ldots,M \, ,    
\end{equation}
where $P$ is a penalty function and $\eta_j>0$ is a shrinkage tuning parameter. For instance, in the $l_1$ penalty employed by lasso, $P(\bbeta_j)=\sum_{k=1}^p|\beta_{jk}|$. In the case of $l_2$ regularization for ridge regression, $P(\bbeta_j)=\sum_{k=1}^p\beta^2_{jk}$. Elastic net, on the other hand, involves an additional set of tuning parameters to balance  between lasso and ridge regression (see \cite{hastie2009elements} for additional penalty functions). Based on the proposed approach, any routine of regularized Cox regression model can be used for estimating  $\bbeta_j$, $j=1,\ldots,M$, based on  (\ref{ref:betajregul}) (e.g., \texttt{glmnet} of \texttt{R} or \texttt{CoxPHFitter} of \texttt{Python}).  Finally, $\alpha_{j1},\ldots,\alpha_{jd}$ are estimated only once the regularization step is completed and  models are selected. In contrast, penalized regression using the collapsed log-likelihood approach of \cite{lee_analysis_2018} requires minimizing $-\log L_j(\alpha_{j1},\ldots,\alpha_{jd},\bbeta_j) + \eta_j P(\bbeta_j)$, which necessitates estimating $\alpha_{j1},\ldots,\alpha_{jd}$. 

The tuning parameters $\eta_j$, $j=1,\ldots,M$, control the amount of regularization and their values play a crucial role. In our Python package, \texttt{PyDTS},  the values of $\eta_j$ are selected by K-fold cross validation while the criterion is to maximize the out-of-sample global area under the receiver operating characteristics curve (AUC). Appendix A provides the definitions and estimators of the area under the receiver operating characteristics curve and Brier score for discrete-survival data with competing risks and right censoring. This includes the cause-specific AUC and Brier score at each time $t$, $\mbox{AUC}_j(t)$ and $\mbox{BS}_{j}(t)$; integrated cause-specific AUC and Brier score, $\mbox{AUC}_j$ and $\mbox{BS}_{j}$; and global AUC and Brier score, $\mbox{AUC}$ and $\mbox{BS}$.

{\bf Example 2: Sure independent screening.} Under ultra-high dimension settings, most of the regularized methods suffer from the curse of dimensionality, high variance and over-fitting \citep{hastie2009elements,fan2012road}. To overcome these issues, the marginal screening technique, sure independent screening (SIS) has been shown to filter out many uninformative variables under an ordinary linear model with normal errors~\citep{fan2008sure}. Subsequently, penalized variable selection methods are often applied to the remaining variables. The key idea of the SIS procedure is to rank all predictors by using a utility measure between the response and each predictor and then to retain the top variables. The SIS procedure has been extended to various models and data types such as generalized linear models~\citep{fan2010sure}, additive models~\citep{fan2011nonparametric}, and Cox regression models~\citep{fan2010high,zhao2012principled,saldana2018sis}.  We focus on SIS and SIS followed by lasso (SIS-L)  \citep{fan2010high,saldana2018sis}  within the proposed two-step procedure. 

SIS involves fitting a marginal regression for each covariate by maximizing
\begin{equation}\label{eq:SIS}
L_j^{\mathcal{C}}(\beta_{jr}) \,\,\,\,\,\,\, j=1,\ldots,M, \,\,\,\, r=1,\ldots,p    
\end{equation} 
where $\bbeta_j=(\beta_{j1},\ldots,\beta_{jp})^T$. The SIS procedure subsequently assesses the importance of features by ranking them according to the magnitude of their marginal regression coefficients. Then,
the selected sets of variables are given by
$\widehat{\mathcal{M}}_{j,w_n} = \{1 \leq k \leq p \, : \, |\widehat{\bbeta}_{jk}| \geq w_n \}$,   $j=1,\ldots,M$, 
where $w_n$ is a threshold value. We adopt the data-driven threshold of \cite{saldana2018sis}. Given data of the form $\{X_i,\delta_i,J_i,{\bf Z}_i \, ; \, i=1,\dots,n\}$, a random permutation $\pi$ of $\{1,\ldots,n\}$ is used to decouple ${\bf Z}_i$ and
$(X_i,\delta_i,J_i)$ so that the resulting data $\{ X_i,\delta_i,J_i,{\bf Z}_{\pi(i)} \, ; \, i=1,\dots,n \}$ follow a model in which the covariates have no predicted power over the survival time of any event type. For the permuted data, we re-estimate individual regression coefficients 
and get $\widehat{\beta}^*_{jr}$. The data-driven threshold is defined by
$w_n = \max_{1\leq j \leq M, 1\leq k \leq p} |\widehat{\beta}^*_{jk}|$.
For SIS-L procedure, the lasso regularization is then added in the first step of our procedure applied to the set of covariates selected by SIS. In contrast to  (\ref{eq:SIS}), applying SIS or SIS-L with the collapsed log-likelihood approach requires maximizing
$L_j(\alpha_{j1},\ldots,\alpha_{jd},\beta_{jr})$,  $j=1,\ldots,M$, $r=1,\ldots,p$,   
which involves estimating $\alpha_{j1},\ldots,\alpha_{jd}$.

\section{Simulation Study}\label{sec:simul}
We evaluated our approach using a simulation study across 19 settings, detailed in Table~S2 of the SM, and compared the results with \cite{lee_analysis_2018}. The sampling process starts by selecting a vector of covariates ${\bf Z}$ for each individual. Based on the model, Eq. (\ref{eq:logis}), the event type is sampled according to the true probabilities $\Pr(J=j|{\bf Z})$. The event time is then sampled from $\Pr(T=t|J=j,{\bf Z})=\Pr(T=t,J=j|{\bf Z})/\Pr(J=j|{\bf Z})$, detailed in Section 2.1. For simulation settings 1-10, covariates were drawn from a standard uniform distribution. Parameters for Settings 1-2 include $\alpha_{1t} = -1.4 + 0.4 \log t$ and $\alpha_{2t} = -1.3 + 0.4\log t$ for $t=1,\ldots,7$, with $\bbeta_1 = -0.7 (\log 0.8, \log 3, \log 3, \log 2.5, \log 2)$, and $\bbeta_{2} = -0.6 (\log 1, \log 3, \log 4, \log 3, \log 2)$. Censoring times followed a discrete uniform distribution with a probability of 0.02 for each $t=1,\ldots,7$. For Settings 3-4, parameters were set to $\alpha_{1t} = -2.0 - 0.2 \log t$ and $\alpha_{2t} = -2.2 - 0.2\log t$, $t=1,\ldots,30$, with $\bbeta$ values the same as in Settings 1-2. Censoring times were sampled with a probability of 0.01 for each $t$.

\begin{figure}[ht!]
    \centering
    \includegraphics[width=\textwidth]{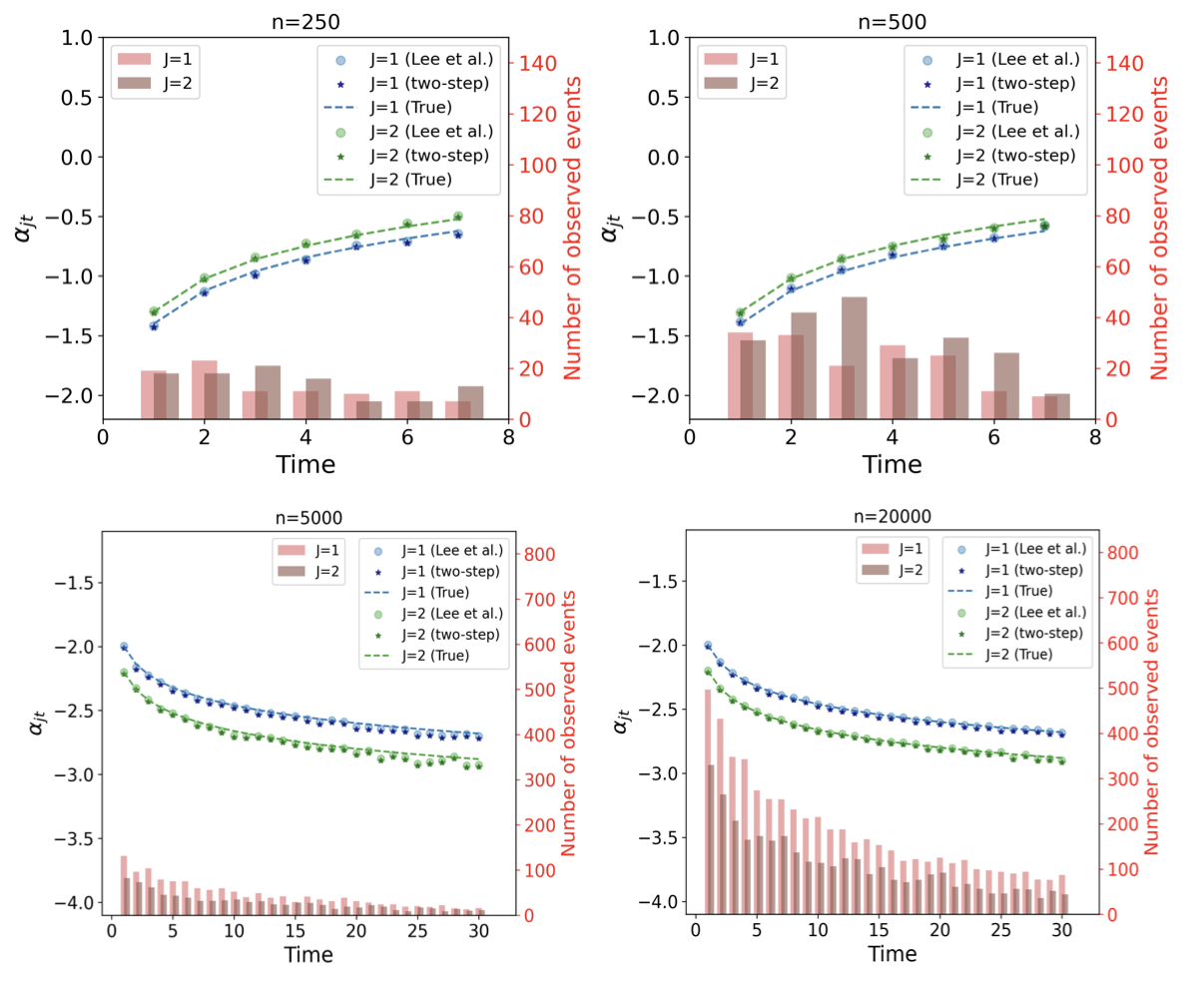}
    \caption{Simulation results of two competing events. Results of ${\alpha}_{jt}$. Each panel is based on a different sample size. Number of observed events are shown in red and brown bars for event types $j=1$ and $j=2$, respectively. True values and mean of estimates are in blue and green for $j=1$ and $j=2$. True values are shown in dashed lines, mean of estimates based on Lee et al. (2018) and the proposed two-step approach denoted by circles and diamonds, respectively.}
    \label{fig:different_n_alpha_results}
\end{figure}

 Table \ref{tab:methods_comparison_table}  and Fig. \ref{fig:different_n_alpha_results} summarise the results of $\bbeta_{j}$ and $\alpha_{jt}$, respectively, for two competing risks. Results with other sample sizes and three competing risks are provided in Web Appendix D and Web Appendix E.  Evidently, the method of \cite{lee_analysis_2018} and the proposed method perform similarly in terms of bias and standard errors. In addition, the empirical coverage rates of 95\% Wald-type confidence intervals for each regression coefficient, based on the proposed approach, are reasonably close to 95\%.

The aim of Settings 11--16 is to showcase how lasso regularization is integrated into our two-step procedure for feature selection. In Settings 11--13 $p=100$ covariates were considered, and only five of them are with non-zero values.  Two settings of zero-mean normally distributed  covariates were considered: (i) independent covariates, each with variance 0.4;  (ii) the following covariances were updated in setting (i) $Cov(Z_1, Z_9) = 0.1$, $Cov(Z_2, Z_{10}) = 0.3$, $Cov(Z_4, Z_8) = -0.3$, and $Cov(Z_5, Z_{12}) = -0.1$. In order to get appropriate survival probabilities based on Eq.~(\ref{eq:logis}), covariates were truncated to be within $[-1.5,1.5]$. The parameters of the model were set to be  $\alpha_{1t} = -3.4 - 0.1 \log t$, $\alpha_{2t} = -3.4 - 0.2\log t$, $t = 1, \ldots , 15$. The first five components of $\bbeta_1$ and $\bbeta_2$ were set to be $(1.2,1.5,-1,-0.3,-1.2)$ and $(-1.2,1,1,-1,1.4)$, respectively, and the rest of the coefficients were set to zero.


Based on one simulated dataset of Setting 11 (see Figure~S5 of the SM) and the selected values of $\eta_j$, the means and  standard deviations (SD) based on the 5-fold integrated cause-specific $\widehat{\mbox{AUC}}_j$ were $\widehat{\mbox{AUC}}_1=0.796$ (SD=0.007) and $\widehat{\mbox{AUC}}_2=0.803$ (SD=0.007), with a mean global $\widehat{\mbox{AUC}}=0.8$ (SD=0.003). The mean global AUC of the non-regularized procedure was $\widetilde{\mbox{AUC}}=0.795$ (SD=0.002). Looking at this specific example, we observe a substantial reduction in the number of covariates selected by the lasso penalty, without a significant change in the discrimination performance as measured by the AUC. The mean integrated cause-specific Brier Scores were $\widehat{\mbox{BS}}_1=0.045$ (SD=0.002) and $\widehat{\mbox{BS}}_2=0.044$ (SD=0.003), with a mean global Brier Score $\widehat{\mbox{BS}}=0.044$ (SD=0.002). 
Similar results were observed for the one simulated dataset of Setting 12 (see Web Appendix F). 

Setting 13 is similar to Setting 12, but with 100 repetitions. It shows that the means of true- and false-positive discoveries for each event type, $\mbox{TP}_j$ and $\mbox{FP}_j$, $j=1,2$, under the selected values of $\eta_j$ were $\mbox{TP}_1=4.99$, $\mbox{FP}_1=0.01$, $\mbox{TP}_2=5$, and $\mbox{FP}_2=0$. The results indicate that the correct model was selected in all 100 repetitions, with a single exception for $j=1$.  
Similar results were observed with smaller sample size of $n=500$ (see Web Appendix F, Settings 14--16).
Web Appendix C provides a detailed description of Settings 17–19, demonstrating the excellent performance of integrating screening methods into the two-step procedure.

\section{MIMIC Data Analysis - Length of Hospital Stay in ICU}
\label{sec:mimic}

Although the MIMIC dataset records admission and discharge times to the minute, it is advisable to use daily units for survival analysis, because times within a day are more influenced by hospital procedures than by patients' health status. The analysis includes 25,170 ICU admissions with three competing events: discharge to home ($J=1$, 69.0\%), transfer to another medical facility ($J=2$, 21.4\%), and in-hospital death ($J=3$, 6.1\%). 
The  analysis is restricted to admissions classified as ``emergency'', with a distinction between direct emergency and emergency ward (EW). 
Emergency admission history is included by two covariates: the number of previous emergency admissions (admissions number), and a dummy variable indicating whether the previous admission ended within 30 days prior to the last one (recent admission). Additional covariates included in the analysis are: year of admission (available in resolution of three years); standardized age at admission; a binary variable indicating night admission (between 20:00 to 8:00); ethnicity (Asian, Black, Hispanic, White, Other); and lab test results (normal or abnormal) performed upon arrival and with results within the first 24 hours of admission.  Note that it is common to include initial laboratory test results when predicting  hospital length of stay \citep{almeida2024hospital}.  The analysis includes 36 covariates in total. Web Appendix G summarizes the covariates' distribution. 

\begin{figure}[ht!]
    \centering
    \includegraphics[width=0.85\textwidth]{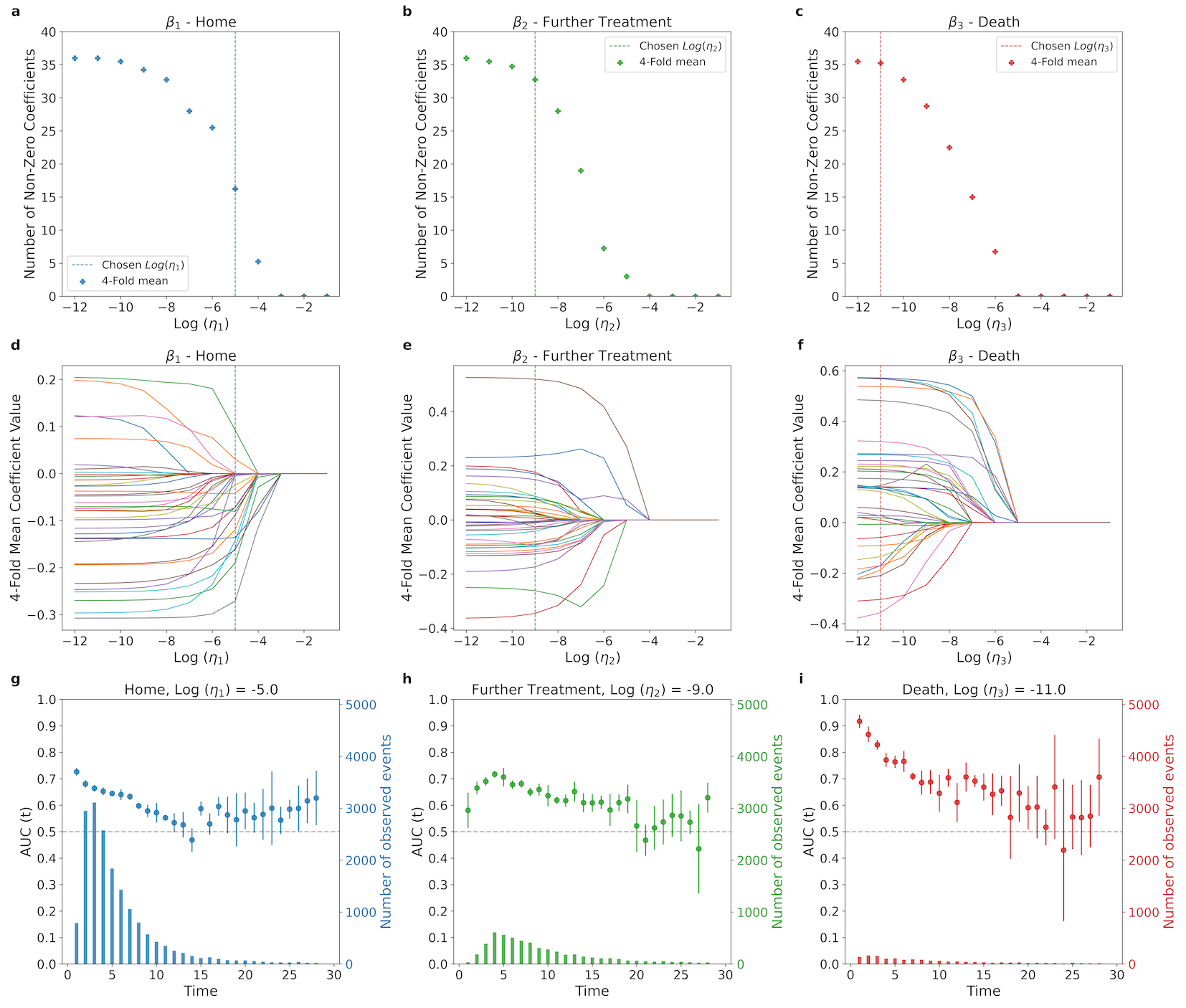}
    \includegraphics[width=0.45\textwidth]{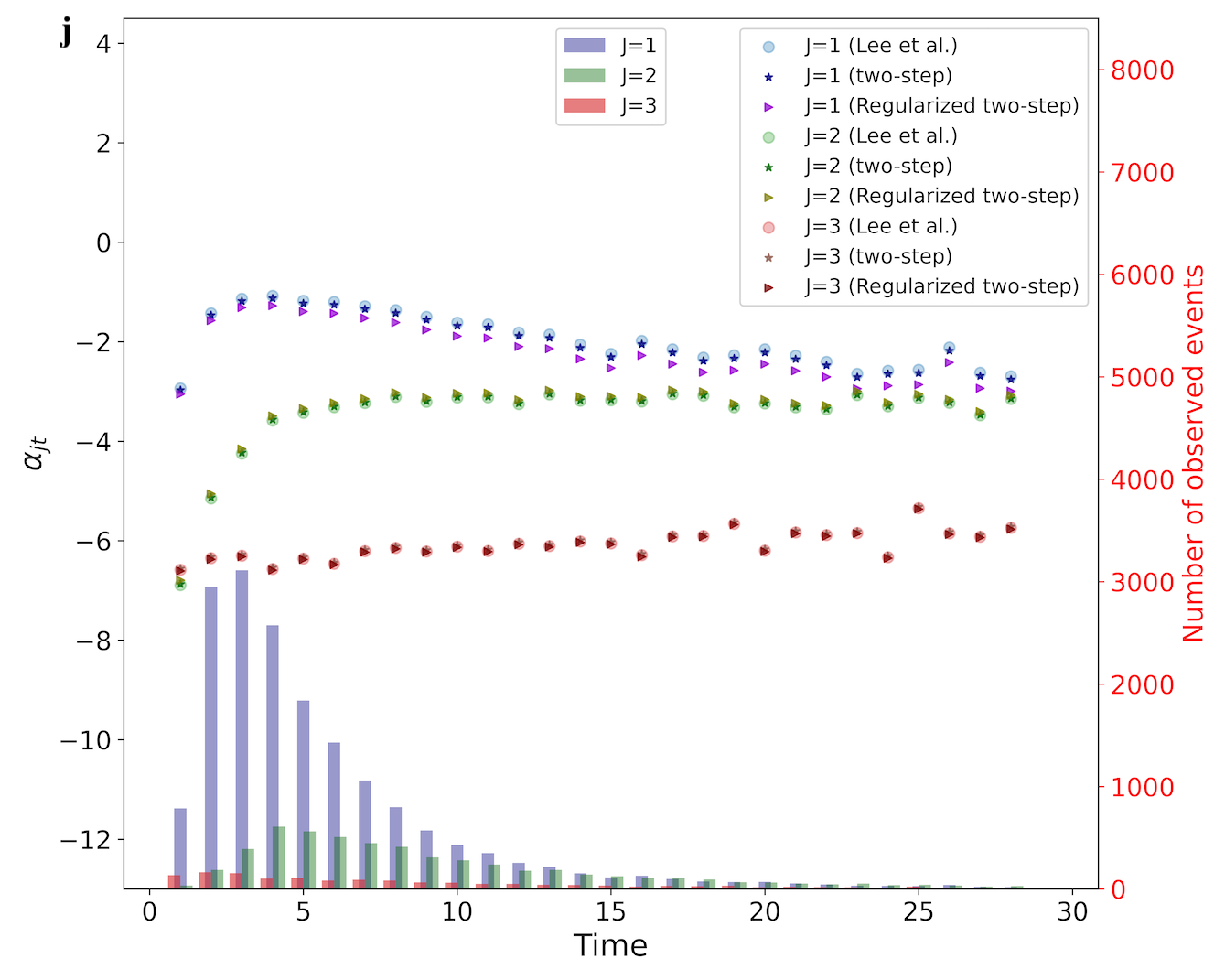}
    \caption{MIMIC dataset - LOS analysis. Regularized regression with 4-fold CV. The selected values of $\eta_j$ are shown in dashed-dotted lines on panels \textbf{a-f}. \textbf{a-c.} Number of non-zero coefficients for $j=1,2,3$. \textbf{d-f.} The estimated coefficients, as a function of $\eta_j$, $j=1,2,3$. \textbf{g-i.} Mean (and SD bars) of the 4 folds $\widehat{\mbox{AUC}}_j(t)$, $j=1,2,3$,  for the selected values  $\log \eta_1=-5$, $\log \eta_2=-9$ and $\log \eta_3=-11$. The number of observed events of each type is shown by bars. \textbf{j.} Results of estimated $\alpha_{jt}$ by the method of Lee et al. (2018) (circle), the  proposed two-step approach (stars) with no regularization and the proposed approach with lasso (left triangular). Numbers of observed events are shown in blue bars for home discharge ($j=1$), in green bars for further treatment ($j=2$), and in red bars for in-hospital death ($j=3$). lasso estimates are based on  $\log \eta_1=-5$, $\log \eta_2=-9$ and $\log \eta_3=-11$.}
    \label{fig:reg_mimic_results}
\end{figure}

Three methods were considered: \cite{lee_analysis_2018}, the proposed two-step approach, and the proposed two-step approach with lasso. For the latter,  the selection of $\eta_j$, $j=1,2,3$, were carried out using 4-fold cross validation, and by maximizing the out-of-sample global AUC. $\log \eta_j$ was allowed to vary from -12 to -1, in steps of 1. The resulting selected values of $\log \eta_j$, $j=1,2,3$, were -5, -9 and -11. The results of the three procedures are presented in Tables~\ref{table:los-j1}--\ref{table:los-j3} and Figure~\ref{fig:reg_mimic_results}. The parameters' estimates were similar between \cite{lee_analysis_2018}'s approach and the two-step procedure without regularization, as expected. Computation time was also similar between \cite{lee_analysis_2018}'s approach and the two-step procedure without regularization with estimation time of 29.5 seconds and 22.1 seconds, respectively.

The global AUCs of the proposed approach without and with lasso penalty were highly similar,
$\widehat{\mbox{AUC}}=0.649$ (SD=0.003) and 
$\widehat{\mbox{AUC}}=0.651$ (SD=0.003). By adding lasso regularization, the number of predictors for each event type was reduced (see last column of Tables~\ref{table:los-j1}--\ref{table:los-j3}), but the corresponding estimators for $\alpha_{jt}$ remained highly similar. 

The estimates for $\mbox{AUC}_j(t)$ typically range from 0.5 to 0.8 for discharges to home or further treatment, and are  higher for death within the first three days of hospitalization.
The integrated cause-specific AUCs were $\widehat{\mbox{AUC}}_1=0.642$ (SD=0.002), $\widehat{\mbox{AUC}}_2=0.655$ (SD=0.012), and $\widehat{\mbox{AUC}}_3=0.740$ (SD=0.006), with a global $\widehat{\mbox{AUC}}=0.651$ (SD=0.003).  The integrated cause-specific Brier Scores were $\widehat{\mbox{BS}}_1=0.105$ (SD=0.002), $\widehat{\mbox{BS}}_2=0.042$ (SD=0.001), and $\widehat{\mbox{BS}}_3=0.010$ (SD=0.001), with a global Brier Score of $\widehat{\mbox{BS}}=0.085$ (SD=0.001). Additional discussion of the results is provided in Web Appendix G. 

\section{Discussion}
\label{sec:discussion}
This work provides a new estimation procedure for a semi-parametric logit-link survival model of discrete time  with competing events. Our current deviation from \cite{lee_analysis_2018} involves a simplification by segregating the estimation procedures for $\alpha_{jt}$ and $\bbeta_j$. 
Our approach is valid when using both the logit- and log-link functions; however, it does not hold under the complementary log-log model. Our current software uses the logit link. 

The hazard models considered in \cite{tutz2016modeling}, \cite{most2016variable} and \cite{schmid2021competing} are of the form
$
\lambda^*_{j}(t|{\bf Z}) = \frac{\exp(\alpha^*_{jt}+{\bf Z}^T \bbeta^*_j)}{1+\sum_{j'=1}^M\exp(\alpha^*_{j't}+{\bf Z}^T \bbeta^*_{j'})}
\,\,\, j=1,\ldots,M \, .
$
Namely, the hazard model $\lambda^*_{j}$ is a function  not only of the parameters associated with the $j$th competing event but also of the parameters related to  all other event types. In contrast, the hazard function 
$\lambda_j$, adopted by \cite{allison_discrete-time_1982}, \cite{lee_analysis_2018}, \cite{wu2022analysis} and in this work, is a function only of the parameters of the $j$th competing event. Both models, $\lambda_j$ and $\lambda_j^*$, are valid and were presented by \cite{allison_discrete-time_1982}. However, as discussed by \cite{allison_discrete-time_1982}, models in the spirit of $\lambda_j$ provide a natural and direct analogy  to the cause-specific hazard function in the context of continuous survival time. Because the discrete-time likelihood cannot be factored into separate components for each of the $M$ types of events, \cite{allison_discrete-time_1982} considered a more tractable formulation. In particular, he explored the generalization of the logistic model $\lambda^*_{j}$ which was later adopted by \cite{tutz2016modeling}, \cite{most2016variable} and \cite{schmid2021competing}.

In Web Appendix H, we show that although computation times for the two methods are comparable at lower values of $d$, our proposed method becomes more efficient as $d$ increases. Furthermore, during tests on a system with 16GB RAM, \cite{lee_analysis_2018}'s method experienced memory errors at relatively low values of $d$, while our two-step procedure ran smoothly without any issues.

\section*{Data and Code Availability Statement}
The estimation procedures and simulation study were implemented in Python using the \texttt{PyDTS} package \citep{meir_pydts_2022}. An example of our approach implemented in \texttt{R} is also available. Codes are available  at \url{https://github.com/tomer1812/pydts/} and  \url{https://github.com/tomer1812/DiscreteTimeSurvivalPenalization}. The MIMIC dataset is accessible at \url{https://physionet.org/content/mimiciv/2.0/} and subjected to credentials.


\section*{Acknowledgements}
T.M. is supported by the Israeli Council for Higher Education (Vatat) fellowship in data science via the Technion; M.G. work was supported by the ISF 767/21 grant and Malag competitive grant in data science (DS).

\bibliographystyle{plainnat}
\bibliography{refs}

\begin{thebibliography}{32}
\providecommand{\natexlab}[1]{#1}
\providecommand{\url}[1]{\texttt{#1}}
\expandafter\ifx\csname urlstyle\endcsname\relax
  \providecommand{\doi}[1]{doi: #1}\else
  \providecommand{\doi}{doi: \begingroup \urlstyle{rm}\Url}\fi

\bibitem[Adhikari et~al.(2010)Adhikari, Fowler, Bhagwanjee, and
  Rubenfeld]{adhikari_critical_2010}
NKJ Adhikari, RA~Fowler, S~Bhagwanjee, and GD~Rubenfeld.
\newblock Critical care and the global burden of critical illness in adults.
\newblock \emph{The Lancet}, 376\penalty0 (9749):\penalty0 1339--1346, oct
  2010.
\newblock ISSN 01406736.
\newblock \doi{10.1016/S0140-6736(10)60446-1}.

\bibitem[Allison(1982)]{allison_discrete-time_1982}
PD~Allison.
\newblock Discrete-time methods for the analysis of event histories.
\newblock \emph{Sociological Methodology}, 13:\penalty0 61--98, 1982.
\newblock ISSN 00811750, 14679531.
\newblock URL \url{http://www.jstor.org/stable/270718}.

\bibitem[Almeida et~al.(2024)Almeida, Brito~Correia, Borges, and
  Bernardino]{almeida2024hospital}
Guilherme Almeida, Fernanda Brito~Correia, Ana~Rosa Borges, and Jorge
  Bernardino.
\newblock Hospital length-of-stay prediction using machine learning
  algorithms—a literature review.
\newblock \emph{Applied Sciences}, 14\penalty0 (22):\penalty0 10523, 2024.

\bibitem[Awad et~al.(2017)Awad, Bader–El–Den, and
  McNicholas]{awad_patient_2017}
A~Awad, M~Bader–El–Den, and J~McNicholas.
\newblock Patient length of stay and mortality prediction: {A} survey.
\newblock \emph{Health Services Management Research}, 30\penalty0 (2):\penalty0
  105--120, may 2017.
\newblock ISSN 0951-4848, 1758-1044.
\newblock \doi{10.1177/0951484817696212}.

\bibitem[Bazick et~al.(2011)Bazick, Chang, Mahadevappa, Gibbons, and
  Christopher]{bazick_red_2011}
HS~Bazick, D~Chang, K~Mahadevappa, FK~Gibbons, and KB~Christopher.
\newblock Red cell distribution width and all-cause mortality in critically ill
  patients*:.
\newblock \emph{Critical Care Medicine}, 39\penalty0 (8):\penalty0 1913--1921,
  aug 2011.
\newblock ISSN 0090-3493.
\newblock \doi{10.1097/CCM.0b013e31821b85c6}.
\newblock URL \url{http://journals.lww.com/00003246-201108000-00008}.

\bibitem[Cox(1972)]{cox_regression_1972}
DR~Cox.
\newblock Regression models and life-tables.
\newblock \emph{Journal of the Royal Statistical Society: Series B
  (Methodological)}, 34\penalty0 (2):\penalty0 187--220, 1972.
\newblock ISSN 00359246.
\newblock \doi{10.1111/j.2517-6161.1972.tb00899.x}.

\bibitem[Cox(2018)]{cox2018analysis}
DR~Cox.
\newblock \emph{Analysis of binary data}.
\newblock Routledge, 2018.

\bibitem[Fan and Lv(2008)]{fan2008sure}
J~Fan and J~Lv.
\newblock Sure independence screening for ultrahigh dimensional feature space.
\newblock \emph{Journal of the Royal Statistical Society: Series B (Statistical
  Methodology)}, 70\penalty0 (5):\penalty0 849--911, 2008.
\newblock \doi{10.1111/j.1467-9868.2008.00674.x}.

\bibitem[Fan and Song(2010)]{fan2010sure}
J~Fan and R~Song.
\newblock Sure independence screening in generalized linear models with
  np-dimensionality.
\newblock \emph{The Annals of Statistics}, 38\penalty0 (6):\penalty0
  3567--3604, 2010.

\bibitem[Fan et~al.(2010)Fan, Feng, and Wu]{fan2010high}
J~Fan, Y~Feng, and Y~Wu.
\newblock High-dimensional variable selection for cox’s proportional hazards
  model.
\newblock \emph{Institute of Mathematical Statistics}, 6:\penalty0 70--86,
  2010.

\bibitem[Fan et~al.(2011)Fan, Feng, and Song]{fan2011nonparametric}
J~Fan, Y~Feng, and R~Song.
\newblock Nonparametric independence screening in sparse ultra-high-dimensional
  additive models.
\newblock \emph{Journal of the American Statistical Association}, 106\penalty0
  (494):\penalty0 544--557, 2011.

\bibitem[Fan et~al.(2012)Fan, Feng, and Tong]{fan2012road}
J~Fan, Y~Feng, and X~Tong.
\newblock A road to classification in high dimensional space: the regularized
  optimal affine discriminant.
\newblock \emph{Journal of the Royal Statistical Society: Series B (Statistical
  Methodology)}, 74\penalty0 (4):\penalty0 745--771, 2012.

\bibitem[Gail et~al.(1981)Gail, Lubin, and Rubinstein]{gail1981likelihood}
MH~Gail, JH~Lubin, and LV~Rubinstein.
\newblock Likelihood calculations for matched case-control studies and survival
  studies with tied death times.
\newblock \emph{Biometrika}, \penalty0 (3):\penalty0 703--707, 1981.

\bibitem[Goldberger et~al.(2000)Goldberger, Amaral, Glass, Hausdorff, Ivanov,
  Mark, and et~al.]{goldberger_physiobank_2000}
AL~Goldberger, LAN Amaral, L~Glass, JM~Hausdorff, PC~Ivanov, RG~Mark, and
  et~al.
\newblock {PhysioBank}, {PhysioToolkit}, and {PhysioNet}: {Components} of a
  {New} {Research} {Resource} for {Complex} {Physiologic} {Signals}.
\newblock \emph{Circulation}, 101\penalty0 (23):\penalty0 e215--e220, jun 2000.
\newblock ISSN 0009-7322, 1524-4539.
\newblock \doi{10.1161/01.CIR.101.23.e215}.
\newblock URL \url{https://www.ahajournals.org/doi/10.1161/01.CIR.101.23.e215}.

\bibitem[Hastie et~al.(2009)Hastie, Tibshirani, and
  Friedman]{hastie2009elements}
T~Hastie, R~Tibshirani, and JH~Friedman.
\newblock \emph{The elements of statistical learning: data mining, inference,
  and prediction}.
\newblock Springer, 2009.

\bibitem[Johnson et~al.(2022)Johnson, Bulgarelli, Pollard, Horng, Celi, and
  Mark]{johnson_mimic-iv_2022}
A~Johnson, L~Bulgarelli, T~Pollard, S~Horng, LA~Celi, and R~Mark.
\newblock {MIMIC}-{IV} (version 2.0).
\newblock \emph{PhysioNet}, pages 49--55, jun 2022.
\newblock \doi{https://doi.org/10.13026/7vcr-e114}.

\bibitem[Kalbfleisch and Prentice(2011)]{kalbfleisch_statistical_2011}
JD~Kalbfleisch and RL~Prentice.
\newblock \emph{The Statistical Analysis of Failure Time Data}.
\newblock Wiley, 2nd edition, 2011.
\newblock ISBN 978-1-118-03123-0.

\bibitem[Klein and Moeschberger(2003)]{klein_survival_2003}
JP~Klein and ML~Moeschberger.
\newblock \emph{Survival Analysis}.
\newblock Springer, 2003.
\newblock ISBN 978-0-387-95399-1.

\bibitem[Lee et~al.(2018)Lee, Feuer, and Fine]{lee_analysis_2018}
M~Lee, EJ~Feuer, and JP~Fine.
\newblock On the analysis of discrete time competing risks data.
\newblock \emph{Biometrics}, 74\penalty0 (4):\penalty0 1468--1481, 2018.
\newblock ISSN 0006-341X, 1541-0420.
\newblock \doi{10.1111/biom.12881}.

\bibitem[Lequertier et~al.(2021)Lequertier, Wang, Fondrevelle, Augusto, and
  Duclos]{lequertier2021hospital}
V~Lequertier, T~Wang, J~Fondrevelle, V~Augusto, and A~Duclos.
\newblock Hospital length of stay prediction methods: A systematic review.
\newblock \emph{Medical Care}, 59\penalty0 (10):\penalty0 929--938, 2021.

\bibitem[Meir et~al.(2022)Meir, Gutman, and Gorfine]{meir_pydts_2022}
T~Meir, R~Gutman, and M~Gorfine.
\newblock Pydts: A python package for discrete-time survival (regularized)
  regression with competing risks.
\newblock \emph{arXiv}, 2022.
\newblock \doi{10.48550/ARXIV.2204.05731}.
\newblock URL \url{https://arxiv.org/abs/2204.05731}.

\bibitem[Meynaar et~al.(2013)Meynaar, Knook, Coolen, Le, Bos, Van Der~Dijs, von
  Lindern, and Steyerberg]{meynaar2013red}
IA~Meynaar, AH~Knook, S~Coolen, H~Le, MM~Bos, F~Van Der~Dijs, Marieke von
  Lindern, and EW~Steyerberg.
\newblock Red cell distribution width as predictor for mortality in critically
  ill patients.
\newblock \emph{Neth J Med}, 71\penalty0 (9):\penalty0 488--493, 2013.

\bibitem[M{\"o}st et~al.(2016)M{\"o}st, P{\"o}{\ss}necker, and
  Tutz]{most2016variable}
S~M{\"o}st, W~P{\"o}{\ss}necker, and G~Tutz.
\newblock Variable selection for discrete competing risks models.
\newblock \emph{Quality \& Quantity}, 50:\penalty0 1589--1610, 2016.

\bibitem[Saldana and Feng(2018)]{saldana2018sis}
DF~Saldana and Y~Feng.
\newblock Sis: An r package for sure independence screening in
  ultrahigh-dimensional statistical models.
\newblock \emph{Journal of Statistical Software}, 83\penalty0 (2):\penalty0
  1--25, 2018.

\bibitem[Schmid and Berger(2021)]{schmid2021competing}
M~Schmid and M~Berger.
\newblock Competing risks analysis for discrete time-to-event data.
\newblock \emph{Wiley Interdisciplinary Reviews: Computational Statistics},
  13\penalty0 (5):\penalty0 e1529, 2021.

\bibitem[Tsiatis(2006)]{tsiatis2006semiparametric}
Anastasios~A Tsiatis.
\newblock \emph{Semiparametric theory and missing data}, volume~4.
\newblock Springer, 2006.

\bibitem[Tutz et~al.(2016)Tutz, Schmid, et~al.]{tutz2016modeling}
G~Tutz, M~Schmid, et~al.
\newblock \emph{Modeling discrete time-to-event data}.
\newblock Springer, 2016.

\bibitem[Van~der Vaart(2000)]{van2000asymptotic}
Aad~W Van~der Vaart.
\newblock \emph{Asymptotic statistics}.
\newblock Cambridge university press, 2000.

\bibitem[Wernly et~al.(2018)Wernly, Lichtenauer, Vellinga, Boerma, Ince, Kelm,
  and Jung]{wernly_blood_2018}
B~Wernly, M~Lichtenauer, NAR Vellinga, EC~Boerma, C~Ince, M~Kelm, and C~Jung.
\newblock Blood urea nitrogen ({BUN}) independently predicts mortality in
  critically ill patients admitted to {ICU}: {A} multicenter study.
\newblock \emph{Clinical Hemorheology and Microcirculation}, 69\penalty0
  (1-2):\penalty0 123--131, may 2018.
\newblock ISSN 13860291, 18758622.
\newblock \doi{10.3233/CH-189111}.
\newblock URL
  \url{https://www.medra.org/servlet/aliasResolver?alias=iospress&doi=10.3233/CH-189111}.

\bibitem[Wu et~al.(2022)Wu, He, Shi, Schaubel, and Kalbfleisch]{wu2022analysis}
W~Wu, K~He, X~Shi, DE~Schaubel, and JD~Kalbfleisch.
\newblock Analysis of hospital readmissions with competing risks.
\newblock \emph{Statistical Methods in Medical Research}, 31\penalty0
  (11):\penalty0 2189--2200, 2022.
\newblock \doi{10.1177/09622802221115879}.

\bibitem[Zhao and Li(2012)]{zhao2012principled}
SD~Zhao and Y~Li.
\newblock Principled sure independence screening for cox models with
  ultra-high-dimensional covariates.
\newblock \emph{Journal of multivariate analysis}, 105\penalty0 (1):\penalty0
  397--411, 2012.

\bibitem[Zhong et~al.(2021)Zhong, Gao, Luo, Zheng, Tu, and
  Xue]{zhong_serum_2021}
J~Zhong, J~Gao, JC~Luo, JL~Zheng, GW~Tu, and Y~Xue.
\newblock Serum creatinine as a predictor of mortality in patients readmitted
  to the intensive care unit after cardiac surgery: a retrospective cohort
  study in {China}.
\newblock \emph{Journal of Thoracic Disease}, 13\penalty0 (3):\penalty0
  1728--1736, mar 2021.
\newblock ISSN 20721439, 20776624.
\newblock \doi{10.21037/jtd-20-3205}.
\newblock URL \url{https://jtd.amegroups.com/article/view/49824/html}.

\end{thebibliography}

\clearpage

\begin{table}[h!]
    \centering
        \caption{Simulation results of two competing events. Results of Lee et al. (2018) include mean and estimated standard error (Est SE). Results of the proposed two-step approach include mean, estimated SE, empirical SE (Emp SE) and empirical coverage rate (CR) of 95\% Wald-type confidence interval.}
    \label{tab:methods_comparison_table}
\begin{tabular}{llrrrrrrr}
\toprule
      &            &  True  & \multicolumn{2}{c}{Lee et al.} & \multicolumn{4}{c}{Two-Step} \\
 $n$     &    $\beta_{jk}$    &  Value      &   Mean & Est SE & Mean & Est SE & Emp SE &  CR \\
\midrule
250 & $\beta_{11}$ &  0.156 &      0.138 &        0.390 &    0.137 &        0.389 &        0.375 &         0.965 \\
    & $\beta_{12}$ & -0.769 &     -0.751 &        0.395 &   -0.745 &        0.393 &        0.399 &         0.945 \\
    & $\beta_{13}$ & -0.769 &     -0.817 &        0.395 &   -0.811 &        0.393 &        0.378 &         0.965 \\
    & $\beta_{14}$ & -0.641 &     -0.642 &        0.395 &   -0.637 &        0.393 &        0.409 &         0.950 \\
    & $\beta_{15}$ & -0.485 &     -0.496 &        0.393 &   -0.492 &        0.391 &        0.425 &         0.925 \\
    & $\beta_{21}$ &  0.000 &     -0.002 &        0.380 &   -0.002 &        0.378 &        0.357 &         0.960 \\
    & $\beta_{22}$ & -0.659 &     -0.704 &        0.384 &   -0.698 &        0.383 &        0.394 &         0.950 \\
    & $\beta_{23}$ & -0.832 &     -0.849 &        0.385 &   -0.842 &        0.383 &        0.378 &         0.955 \\
    & $\beta_{24}$ & -0.659 &     -0.675 &        0.384 &   -0.669 &        0.382 &        0.406 &         0.945 \\
    & $\beta_{25}$ & -0.416 &     -0.451 &        0.382 &   -0.447 &        0.381 &        0.402 &         0.940 \\
\hline
500 & $\beta_{11}$ &  0.156 &      0.133 &        0.273 &    0.132 &        0.273 &        0.270 &         0.925 \\
    & $\beta_{12}$ & -0.769 &     -0.795 &        0.276 &   -0.791 &        0.276 &        0.295 &         0.945 \\
    & $\beta_{13}$ & -0.769 &     -0.815 &        0.278 &   -0.812 &        0.277 &        0.294 &         0.945 \\
    & $\beta_{14}$ & -0.641 &     -0.642 &        0.275 &   -0.640 &        0.275 &        0.260 &         0.965 \\
    & $\beta_{15}$ & -0.485 &     -0.472 &        0.274 &   -0.470 &        0.273 &        0.258 &         0.975 \\
    & $\beta_{21}$ &  0.000 &      0.005 &        0.265 &    0.005 &        0.265 &        0.254 &         0.955 \\
    & $\beta_{22}$ & -0.659 &     -0.681 &        0.268 &   -0.678 &        0.267 &        0.277 &         0.925 \\
    & $\beta_{23}$ & -0.832 &     -0.855 &        0.269 &   -0.852 &        0.269 &        0.268 &         0.950 \\
    & $\beta_{24}$ & -0.659 &     -0.634 &        0.267 &   -0.631 &        0.267 &        0.274 &         0.940 \\
    & $\beta_{25}$ & -0.416 &     -0.415 &        0.266 &   -0.414 &        0.265 &        0.272 &         0.940 \\
\hline
5,000  & $\beta_{11}$ &  0.223 &      0.227 &        0.094 &    0.225 &        0.093 &        0.104 &         0.940 \\
      & $\beta_{12}$ & -1.099 &     -1.093 &        0.096 &   -1.082 &        0.095 &        0.104 &         0.920 \\
      & $\beta_{13}$ & -1.099 &     -1.102 &        0.096 &   -1.090 &        0.095 &        0.105 &         0.935 \\
      & $\beta_{14}$ & -0.916 &     -0.914 &        0.095 &   -0.904 &        0.094 &        0.092 &         0.955 \\
      & $\beta_{15}$ & -0.693 &     -0.701 &        0.095 &   -0.694 &        0.094 &        0.099 &         0.940 \\
      & $\beta_{21}$ & -0.000 &      0.004 &        0.121 &    0.004 &        0.120 &        0.119 &         0.945 \\
      & $\beta_{22}$ & -1.099 &     -1.091 &        0.124 &   -1.083 &        0.123 &        0.129 &         0.925 \\
      & $\beta_{23}$ & -1.386 &     -1.402 &        0.125 &   -1.393 &        0.125 &        0.137 &         0.920 \\
      & $\beta_{24}$ & -1.099 &     -1.109 &        0.124 &   -1.101 &        0.123 &        0.135 &         0.925 \\
      & $\beta_{25}$ & -0.693 &     -0.704 &        0.122 &   -0.698 &        0.121 &        0.120 &         0.945 \\
\hline
20,000 & $\beta_{11}$ &  0.223 &      0.220 &        0.047 &    0.217 &        0.047 &        0.046 &         0.935 \\
      & $\beta_{12}$ & -1.099 &     -1.099 &        0.048 &   -1.088 &        0.048 &        0.044 &         0.965 \\
      & $\beta_{13}$ & -1.099 &     -1.098 &        0.048 &   -1.087 &        0.048 &        0.046 &         0.940 \\
      & $\beta_{14}$ & -0.916 &     -0.920 &        0.048 &   -0.910 &        0.047 &        0.041 &         0.980 \\
      & $\beta_{15}$ & -0.693 &     -0.690 &        0.047 &   -0.682 &        0.047 &        0.046 &         0.945 \\
      & $\beta_{21}$ & -0.000 &      0.003 &        0.060 &    0.003 &        0.060 &        0.065 &         0.930 \\
      & $\beta_{22}$ & -1.099 &     -1.095 &        0.062 &   -1.088 &        0.061 &        0.066 &         0.940 \\
      & $\beta_{23}$ & -1.386 &     -1.394 &        0.063 &   -1.385 &        0.062 &        0.057 &         0.980 \\
      & $\beta_{24}$ & -1.099 &     -1.096 &        0.062 &   -1.089 &        0.061 &        0.061 &         0.950 \\
      & $\beta_{25}$ & -0.693 &     -0.700 &        0.061 &   -0.695 &        0.061 &        0.056 &         0.970 \\
\bottomrule
\end{tabular}
\end{table}


\setstretch{1.0}

\begin{table} 
\centering
\caption{MIMIC dataset - LOS analysis: Estimated regression coefficients of event type discharge to home, $J=1$.}
\label{table:los-j1}
\begin{tabular}{llccc}
\toprule
{} & {} &     Lee et al. &        Two-Step & Two-Step \& lasso \\
{} & {} &  Estimate (SE) &   Estimate (SE) &    Estimate (SE) \\
\midrule
Admissions Number  &         2 &   0.000 (0.024) &   0.003 (0.022) &   {\bf 0.000 (0.000)} \\
{} &        3+ &  -0.032 (0.023) &  -0.027 (0.022) &   {\bf 0.000 (0.000)} \\
Anion Gap            &  Abnormal &  -0.137 (0.032) &  -0.128 (0.030) &   {\bf 0.000 (0.000)} \\
Bicarbonate          &  Abnormal &  -0.208 (0.021) &  -0.194 (0.020) &   -0.119 (0.019) \\
Calcium Total        &  Abnormal &  -0.291 (0.020) &  -0.270 (0.019) &   -0.190 (0.018) \\
Chloride             &  Abnormal &  -0.148 (0.024) &  -0.137 (0.023) &   -0.071 (0.021) \\
Creatinine           &  Abnormal &  -0.103 (0.024) &  -0.098 (0.023) &   -0.072 (0.021) \\
Direct Emergency     &       Yes &  -0.011 (0.026) &  -0.014 (0.024) &   {\bf 0.000 (0.000)} \\
Ethnicity      &     Black &   0.006 (0.046) &   0.009 (0.042) &   {\bf 0.000 (0.000)} \\
{}   &  Hispanic &   0.132 (0.053) &   0.120 (0.048) &   {\bf 0.000 (0.000)} \\
{}      &     Other &  -0.162 (0.051) &  -0.146 (0.047) &   {\bf 0.000 (0.000)} \\
{}      &     White &  -0.031 (0.041) &  -0.026 (0.038) &   {\bf 0.000 (0.000)} \\
Glucose              &  Abnormal &  -0.215 (0.018) &  -0.192 (0.016) &   -0.088 (0.016) \\
Hematocrit           &  Abnormal &  -0.042 (0.032) &  -0.037 (0.029) &   -0.042 (0.029) \\
Hemoglobin           &  Abnormal &  -0.080 (0.033) &  -0.071 (0.030) &   -0.081 (0.030) \\
Insurance    &  Medicare &   0.138 (0.039) &   0.125 (0.036) &   {\bf 0.000 (0.000)} \\
{}      &     Other &   0.219 (0.036) &   0.200 (0.033) &    0.030 (0.016) \\
MCH                  &  Abnormal &  -0.002 (0.023) &  -0.002 (0.022) &   {\bf 0.000 (0.000)} \\
MCHC                 &  Abnormal &  -0.128 (0.019) &  -0.116 (0.018) &   -0.003 (0.017) \\
MCV                  &  Abnormal &  -0.048 (0.026) &  -0.045 (0.024) &   {\bf 0.000 (0.000)} \\
Magnesium            &  Abnormal &  -0.080 (0.030) &  -0.074 (0.028) &   {\bf 0.000 (0.000)} \\
Marital Status     &   Married &   0.224 (0.032) &   0.205 (0.030) &    0.093 (0.016) \\
{}       &    Single &  -0.087 (0.033) &  -0.079 (0.031) &   {\bf 0.000 (0.000)} \\
{}      &   Widowed &   0.026 (0.040) &   0.020 (0.037) &   {\bf 0.000 (0.000)} \\
Night Admission      &       Yes &   0.081 (0.017) &   0.075 (0.016) &   {\bf 0.000 (0.000)} \\
Phosphate            &  Abnormal &  -0.052 (0.019) &  -0.048 (0.018) &   {\bf 0.000 (0.000)} \\
Platelet Count       &  Abnormal &  -0.068 (0.019) &  -0.062 (0.018) &   {\bf 0.000 (0.000)} \\
Potassium            &  Abnormal &  -0.103 (0.032) &  -0.095 (0.030) &   {\bf 0.000 (0.000)} \\
RDW                  &  Abnormal &  -0.327 (0.021) &  -0.308 (0.020) &   -0.271 (0.019) \\
Recent Admission     &       Yes &  -0.262 (0.035) &  -0.247 (0.033) &   -0.001 (0.027) \\
Red Blood Cells      &  Abnormal &  -0.089 (0.027) &  -0.078 (0.024) &   -0.024 (0.025) \\
Sex                  &    Female &  -0.007 (0.018) &  -0.006 (0.016) &   {\bf 0.000 (0.000)} \\
Sodium               &  Abnormal &  -0.312 (0.030) &  -0.297 (0.029) &   -0.142 (0.026) \\
Standardized Age     &           &  -0.260 (0.011) &  -0.234 (0.010) &   -0.162 (0.009) \\
Urea Nitrogen        &  Abnormal &  -0.148 (0.022) &  -0.139 (0.020) &   -0.136 (0.020) \\
White Blood Cells    &  Abnormal &  -0.276 (0.018) &  -0.252 (0.016) &   -0.159 (0.016) \\
\bottomrule
\end{tabular}
\end{table}

\begin{table} 
\centering
\caption{MIMIC dataset - LOS analysis: Estimated regression coefficients of event type discharged to another facility, $J=2$.}
\label{table:los-j2}
\begin{tabular}{llccc}
\toprule
{} &  {} &    Lee et al. &        Two-Step & Two-Step \& lasso \\
{} &  {} & Estimate (SE) &   Estimate (SE) &    Estimate (SE) \\
\midrule
Admissions Number   &         2 &   0.108 (0.041) &   0.107 (0.040) &    0.087 (0.038) \\
{}  &        3+ &   0.194 (0.037) &   0.190 (0.036) &    0.169 (0.034) \\
Anion Gap            &  Abnormal &  -0.006 (0.048) &  -0.006 (0.047) &    {\bf 0.000 (0.002)} \\
Bicarbonate          &  Abnormal &  -0.121 (0.033) &  -0.117 (0.032) &   -0.110 (0.032) \\
Calcium Total        &  Abnormal &  -0.098 (0.031) &  -0.094 (0.031) &   -0.088 (0.030) \\
Chloride             &  Abnormal &   0.016 (0.036) &   0.015 (0.035) &    {\bf 0.000 (0.002)} \\
Creatinine           &  Abnormal &  -0.199 (0.036) &  -0.191 (0.035) &   -0.173 (0.035) \\
Direct Emergency     &       Yes &  -0.373 (0.052) &  -0.363 (0.050) &   -0.345 (0.050) \\
Ethnicity       &     Black &   0.084 (0.090) &   0.079 (0.088) &    0.028 (0.086) \\
{}   &  Hispanic &  -0.068 (0.111) &  -0.070 (0.108) &   -0.088 (0.106) \\
{}     &     Other &   0.026 (0.099) &   0.022 (0.097) &   -0.006 (0.095) \\
{}      &     White &   0.144 (0.082) &   0.138 (0.081) &    0.094 (0.079) \\
Glucose              &  Abnormal &  -0.138 (0.031) &  -0.132 (0.030) &   -0.126 (0.030) \\
Hematocrit           &  Abnormal &   0.038 (0.057) &   0.039 (0.055) &    0.032 (0.055) \\
Hemoglobin           &  Abnormal &   0.018 (0.062) &   0.015 (0.060) &    0.005 (0.059) \\
Insurance    &  Medicare &   0.237 (0.075) &   0.230 (0.074) &    0.238 (0.073) \\
{}     &     Other &  -0.094 (0.074) &  -0.091 (0.072) &   -0.081 (0.072) \\
MCH                  &  Abnormal &   0.042 (0.038) &   0.040 (0.037) &    0.019 (0.031) \\
MCHC                 &  Abnormal &  -0.010 (0.031) &  -0.011 (0.030) &    {\bf 0.000 (0.003)} \\
MCV                  &  Abnormal &  -0.020 (0.041) &  -0.019 (0.039) &    {\bf 0.000 (0.003)} \\
Magnesium            &  Abnormal &  -0.039 (0.048) &  -0.038 (0.047) &   -0.025 (0.046) \\
Marital Status      &   Married &  -0.254 (0.054) &  -0.249 (0.053) &   -0.262 (0.052) \\
{}      &    Single &   0.209 (0.054) &   0.200 (0.053) &    0.176 (0.052) \\
{}      &   Widowed &   0.175 (0.058) &   0.163 (0.056) &    0.149 (0.056) \\
Night Admission      &       Yes &   0.056 (0.029) &   0.054 (0.028) &    0.047 (0.028) \\
Phosphate            &  Abnormal &  -0.042 (0.033) &  -0.040 (0.032) &   -0.034 (0.031) \\
Platelet Count       &  Abnormal &  -0.130 (0.032) &  -0.125 (0.031) &   -0.118 (0.031) \\
Potassium            &  Abnormal &   0.042 (0.048) &   0.042 (0.047) &    0.023 (0.047) \\
RDW                  &  Abnormal &  -0.107 (0.033) &  -0.104 (0.032) &   -0.093 (0.031) \\
Recent Admission     &       Yes &  -0.021 (0.051) &  -0.023 (0.049) &    {\bf 0.000 (0.004)} \\
Red Blood Cells      &  Abnormal &   0.083 (0.052) &   0.079 (0.050) &    0.073 (0.050) \\
Sex                  &    Female &   0.090 (0.031) &   0.088 (0.030) &    0.078 (0.030) \\
Sodium               &  Abnormal &  -0.056 (0.042) &  -0.056 (0.041) &   -0.039 (0.038) \\
Standardized Age     &           &   0.536 (0.021) &   0.525 (0.021) &    0.519 (0.021) \\
Urea Nitrogen        &  Abnormal &   0.100 (0.035) &   0.095 (0.034) &    0.077 (0.034) \\
White Blood Cells    &  Abnormal &  -0.107 (0.029) &  -0.103 (0.028) &   -0.099 (0.028) \\
\bottomrule
\end{tabular}
\end{table}

\begin{table} 
\centering
\caption{MIMIC dataset - LOS analysis: Estimated regression coefficients of event type in-hospital death, $J=3$.}
\label{table:los-j3}
\begin{tabular}{llccc}
\toprule
{} & {} &     Lee et al. &        Two-Step & Two-Step \& lasso \\
{} & {} &  Estimate (SE) &   Estimate (SE) &    Estimate (SE) \\
\midrule
Admissions Number   &         2 &   0.147 (0.074) &   0.147 (0.073) &    0.140 (0.074) \\
{}  &        3+ &   0.142 (0.069) &   0.140 (0.068) &    0.134 (0.068) \\
Anion Gap            &  Abnormal &   0.582 (0.064) &   0.573 (0.064) &    0.571 (0.064) \\
Bicarbonate          &  Abnormal &   0.543 (0.056) &   0.537 (0.056) &    0.535 (0.056) \\
Calcium Total        &  Abnormal &   0.204 (0.054) &   0.204 (0.054) &    0.203 (0.054) \\
Chloride             &  Abnormal &   0.147 (0.059) &   0.143 (0.058) &    0.142 (0.058) \\
Creatinine           &  Abnormal &   0.273 (0.067) &   0.271 (0.067) &    0.271 (0.067) \\
Direct Emergency     &       Yes &  -0.318 (0.096) &  -0.311 (0.095) &   -0.302 (0.095) \\
Ethnicity       &     Black &  -0.236 (0.140) &  -0.235 (0.139) &   -0.203 (0.140) \\
{}    &  Hispanic &  -0.395 (0.183) &  -0.393 (0.181) &   -0.351 (0.181) \\
{}      &     Other &   0.145 (0.147) &   0.133 (0.145) &    0.155 (0.146) \\
{}      &     White &  -0.156 (0.123) &  -0.157 (0.122) &   -0.130 (0.123) \\
Glucose              &  Abnormal &   0.215 (0.064) &   0.212 (0.063) &    0.208 (0.063) \\
Hematocrit           &  Abnormal &  -0.198 (0.108) &  -0.194 (0.107) &   -0.165 (0.108) \\
Hemoglobin           &  Abnormal &   0.024 (0.122) &   0.023 (0.121) &    0.003 (0.121) \\
Insurance    &  Medicare &  -0.224 (0.136) &  -0.225 (0.135) &   -0.171 (0.138) \\
{}      &     Other &  -0.242 (0.133) &  -0.240 (0.132) &   -0.188 (0.135) \\
MCH                  &  Abnormal &  -0.066 (0.070) &  -0.066 (0.069) &   -0.057 (0.069) \\
MCHC                 &  Abnormal &   0.027 (0.056) &   0.029 (0.055) &    0.027 (0.055) \\
MCV                  &  Abnormal &   0.060 (0.072) &   0.061 (0.071) &    0.055 (0.071) \\
Magnesium            &  Abnormal &   0.329 (0.073) &   0.324 (0.072) &    0.320 (0.072) \\
Marital Status      &   Married &   0.156 (0.102) &   0.154 (0.101) &    0.127 (0.061) \\
{}       &    Single &   0.026 (0.107) &   0.027 (0.106) &   {\bf 0.000 (0.008)} \\
{}     &   Widowed &   0.047 (0.115) &   0.048 (0.114) &    0.020 (0.084) \\
Night Admission      &       Yes &  -0.096 (0.053) &  -0.093 (0.052) &   -0.089 (0.052) \\
Phosphate            &  Abnormal &   0.178 (0.056) &   0.176 (0.055) &    0.174 (0.055) \\
Platelet Count       &  Abnormal &   0.235 (0.054) &   0.232 (0.054) &    0.229 (0.054) \\
Potassium            &  Abnormal &   0.227 (0.072) &   0.221 (0.071) &    0.221 (0.071) \\
RDW                  &  Abnormal &   0.492 (0.058) &   0.486 (0.058) &    0.483 (0.058) \\
Recent Admission     &       Yes &   0.250 (0.083) &   0.242 (0.082) &    0.242 (0.082) \\
Red Blood Cells      &  Abnormal &   0.142 (0.105) &   0.140 (0.104) &    0.130 (0.104) \\
Sex                  &    Female &  -0.011 (0.057) &  -0.008 (0.057) &   -0.005 (0.057) \\
Sodium               &  Abnormal &   0.276 (0.064) &   0.270 (0.063) &    0.268 (0.063) \\
Standardized Age     &           &   0.580 (0.041) &   0.574 (0.040) &    0.568 (0.040) \\
Urea Nitrogen        &  Abnormal &   0.141 (0.070) &   0.141 (0.070) &    0.141 (0.070) \\
White Blood Cells    &  Abnormal &   0.579 (0.056) &   0.571 (0.056) &    0.568 (0.055) \\
\bottomrule
\end{tabular}
\end{table}

\clearpage

\renewcommand{\thefigure}{S\arabic{figure} }
\renewcommand{\thetable}{S\arabic{table} }
\setcounter{figure}{0}
\setcounter{table}{0}

\phantomsection
\section*{Web Appendix A: Continuous-Time versus Discrete-Time Analyses}
\begin{figure}[!ht]
    \centering
    \includegraphics[width=\textwidth]{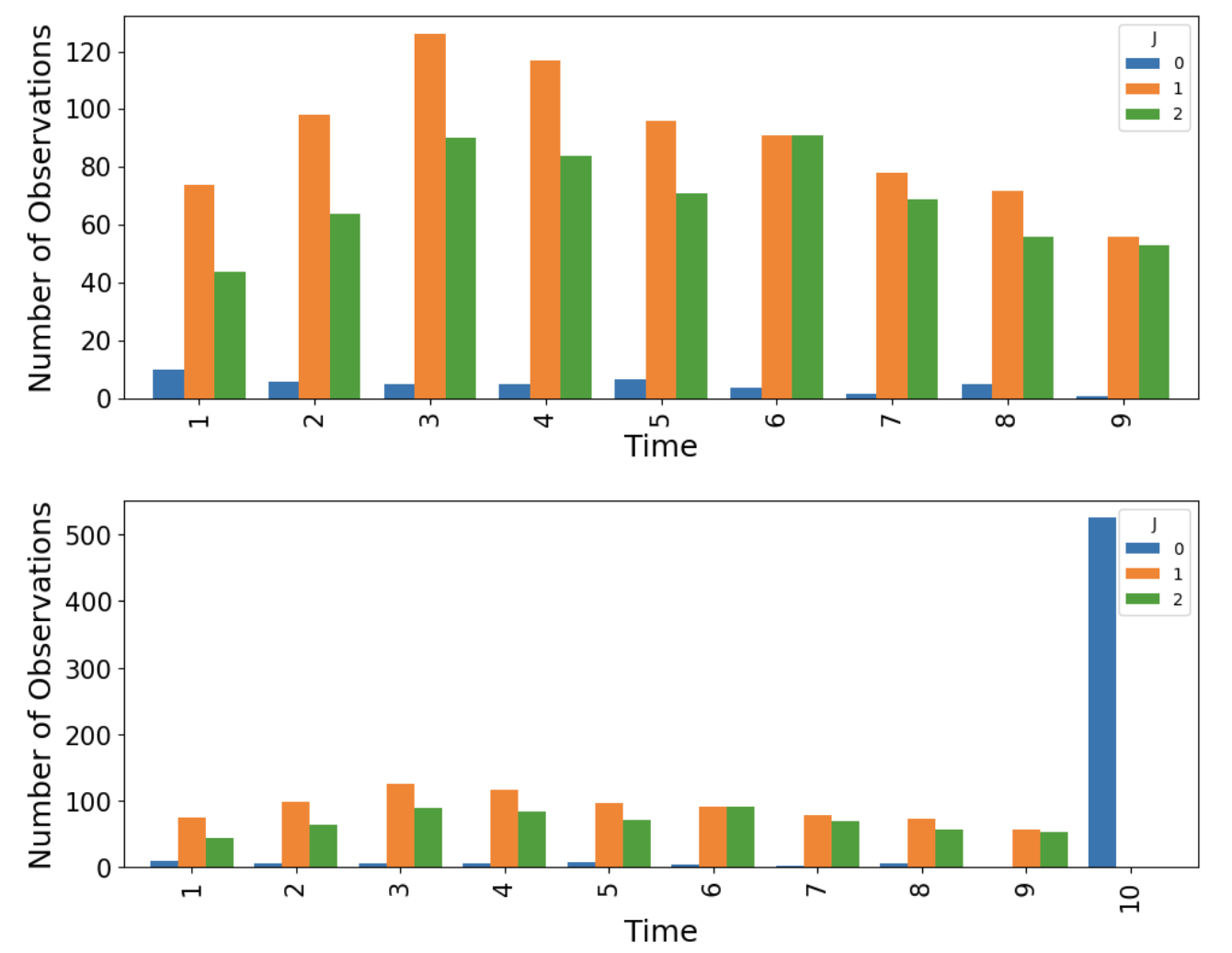}
    \caption{Summary of one simulated dataset of Section S1: $d=9$, $n = 2,000$, $M = 2$. }\label{fig:single_sample_d_9}
\end{figure}

Here we  compare the proposed discrete analysis method against a naive approach using the standard partial likelihood method and Breslow estimator from the R function \texttt{coxph}. The dataset consists of 2,000 observations, each with 5 covariates drawn from a standard uniform distribution. Two  competing events were considered and 9 time points. The parameter values are set as follows: $\alpha_{1t} = -1.9 + 0.6 \log t$ and $\alpha_{2t} = -2.5 + 0.6 \log t$ for $t = 1, \dots, 9$,  $\bbeta_1 = -0.7(\log 0.8, \log 3, \log 3, \log 2.5, \log 2)^T$ and $\bbeta_2 = -0.7(\log 1, \log 3, \log 4, \log 3, \log 2)^T$. 
Censoring times are randomly sampled from a discrete uniform distribution with a probability of 0.005 at each time $t$. Results from one dataset are illustrated in Figure \ref{fig:single_sample_d_9}.

The results, based on 500 repetitions, are summarized  in  Table \ref{tab:continuous_discrete_comparison_table_bias}. 
It is evident that the biases of the estimators of the baseline hazards, ${\alpha}_{jt}$, based on the naive approach—using any of the tie correction methods (Breslow, Efron, or Exact)—are substantial. In contrast, the proposed approach yields practically unbiased results.

\begin{figure}[!ht]
    \centering
    \includegraphics[width=\textwidth]{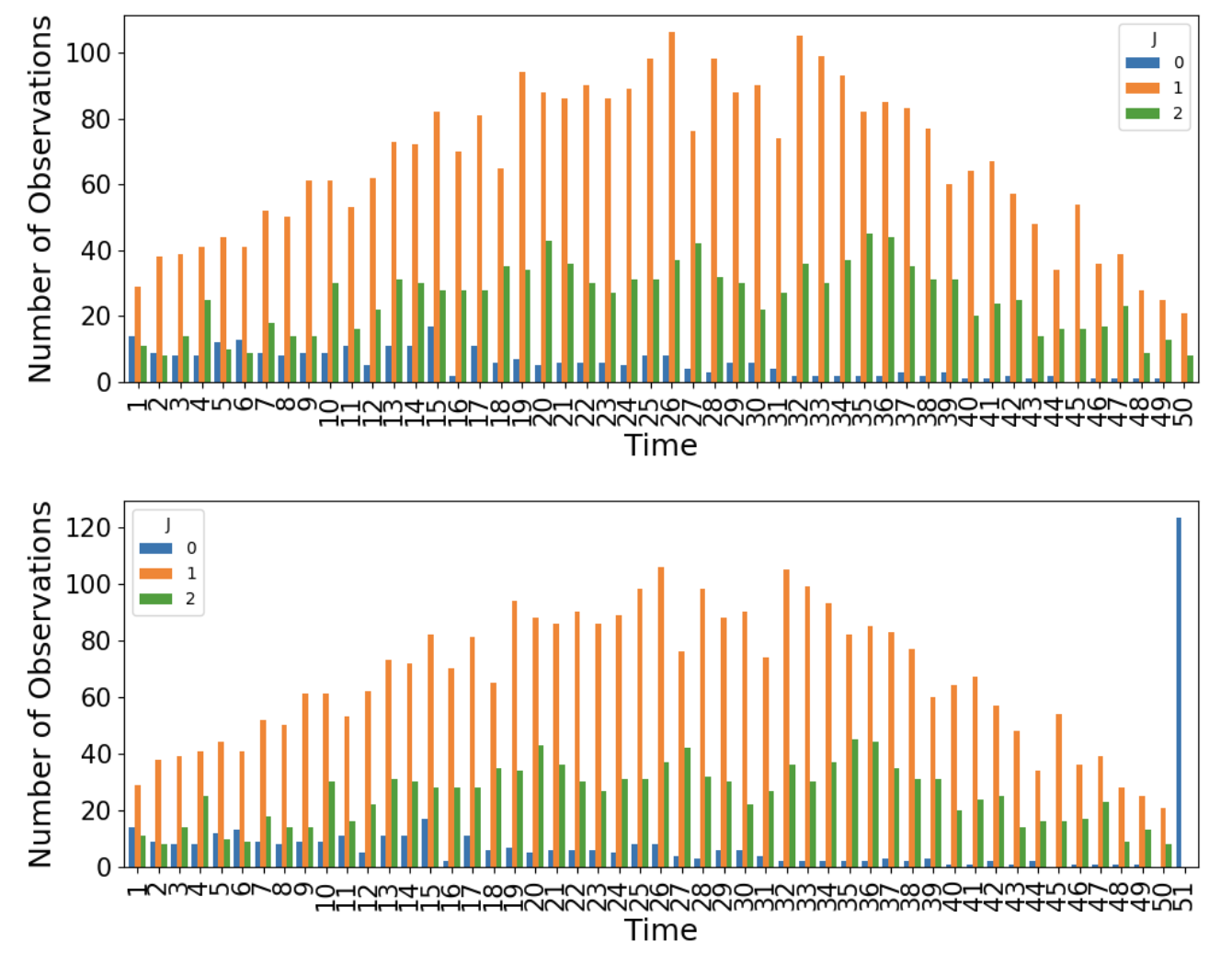}
    \caption{Summary of one simulated dataset of Section S1: $d=50$, $n = 5,000$, $M = 2$. }\label{fig:single_sample_d_50}
\end{figure}

We repeat the analysis with $d=50$ time points, 5,000, observations and $p = 5$ standard uniform covariates.  The parameter values are set as follows: $\alpha_{1t} = -4 + 0.07 t$, $\alpha_{2t} = -5.3 + 0.07 t$, $t = 1, \dots, 50$,  
$$
\bbeta_1 = -0.5(\log 0.8, \log 3, \log 3, \log 2.5, \log 2)^T
$$ and 
$$
\bbeta_2=-0.5(0,\log 3, \log 4, \log 3, \log2)^T \, .
$$ 
Results are presented in Figure \ref{fig:single_sample_d_50} and Tables \ref{tab:continuous_discrete_comparison_table_bias_large_d_risk_1}--\ref{tab:continuous_discrete_comparison_table_bias_large_d_risk_2}. 
Evidently, similar findings are observed with large $d$.
 The substantial bias of the naive approach stems from the inappropriate use of the Breslow estimator for baseline hazard functions. Theoretically, the Breslow estimator is justifiable with competing events only when the likelihood function can be decomposed into distinct components for each event type, a condition not met by the discrete-time regression model considered in our paper.

\phantomsection
\section*{Web Appendix B: Asymptotic Results}\label{appx:asymptotics}
Let us assume that as $n \rightarrow \infty$, both $d$ and $M$ are finite fixed values, and the vectors of covariates ${\bf Z}$ are bounded. Additionally, it is  assumed that 
$\Pr(Y_{.}(t) \geq 1)>0 $ for all $t=1,\ldots,d$. 
We denote the true parameters values as $\bbeta^o_j$ and $\alpha_{jt^o}$. Assume that $\alpha_{jt}$, $j=1,\ldots,M$, $t=1,\ldots,d$, lie in a compact convex set $\mathcal{A}$ that includes an open neighborhood around each true value $\alpha^o_{jt}$. For each $(j,t)$, $j=1,\dots,M$, $t=1,\ldots,d$, minimizing
    \begin{equation*}
         \left\{ \frac{1}{Y.(t)} \sum_{i=1}^n I(X_i \geq t)\frac{\exp(a+{\bf Z}_i^T\widehat{\bbeta}_j)}{1+\exp(a+{\bf Z}_i^T\widehat{\bbeta}_j)} - \frac{N_j(t)}{Y.(t)}\right\}^2
    \end{equation*}
as a function of $a$ is equivalent to solving
\begin{eqnarray*}
  && 2 \left\{ \frac{1}{Y.(t)} \sum_{i=1}^n I(X_i \geq t)\frac{\exp(a+{\bf Z}_i^T\widehat{\bbeta}_j)}{1+\exp(a+{\bf Z}_i^T\widehat{\bbeta}_j)} - \frac{N_j(t)}{Y.(t)}\right\} \\
  && \hspace{0.4cm} \times \frac{\partial}{\partial a} 
\frac{1}{Y.(t)} \sum_{i=1}^n I(X_i \geq t)\frac{\exp(a+{\bf Z}_i^T\widehat{\bbeta}_j)}{1+\exp(a+{\bf Z}_i^T\widehat{\bbeta}_j)} = 0 
\end{eqnarray*}
or alternatively solving 
$$
\frac{1}{n}\sum_{i=1}^n \left\{ I(X_i \geq t)\frac{\exp(a+{\bf Z}_i^T\widehat{\bbeta}_j)}{1+\exp(a+{\bf Z}_i^T\widehat{\bbeta}_j)} - I(X_i=t, J_i=j) \right\} = 0 \, .
$$
Define 
$$
U_n(\alpha_{jt}) = \frac{1}{n}\sum_{i=1}^n \left\{ I(X_i \geq t)\frac{\exp(\alpha_{jt}+{\bf Z}_i^T\widehat{\bbeta}_j)}{1+\exp(\alpha_{jt}+{\bf Z}_i^T\widehat{\bbeta}_j)} - I(X_i=t, J_i=j) \right\}
$$
and 
$$
\widetilde{U}_n(\alpha_{jt}) = \frac{1}{n}\sum_{i=1}^n \left\{ I(X_i \geq t)\frac{\exp(\alpha_{jt}+{\bf Z}_i^T {\bbeta}^o_j)}{1+\exp(\alpha_{jt}+{\bf Z}_i^T {\bbeta}^o_j)} - I(X_i=t, J_i=j) \right\} \, .
$$
Then, 
$$
U_n(\alpha_{jt}) - \widetilde{U}_n(\alpha_{jt})  = \frac{1}{n}\sum_{i=1}^n 
I(X_i \geq t)
\left\{ \frac{\exp(\alpha_{jt}+{\bf Z}_i^T\widehat{\bbeta}_j)}{1+\exp(\alpha_{jt}a+{\bf Z}_i^T\widehat{\bbeta}_j)} - 
\frac{\exp(\alpha_{jt}+{\bf Z}_i^T {\bbeta}^o_j)}{1+\exp(\alpha_{jt}+{\bf Z}_i^T {\bbeta}^o_j)} \right\}   \, .
$$
Taking a first-order Taylor expansion about $\bbeta^o_j$ gives
$$
U_n(\alpha_{jt}) - \widetilde{U}_n(\alpha_{jt})  = 
\frac{1}{n}\sum_{i=1}^n I(X_i \geq t) \frac{\partial}{\partial \bbeta^o_j} 
\frac{\exp(\alpha_{jt}+{\bf Z}_i^T {\bbeta}^o_j)}{1+\exp(\alpha_{jt}+{\bf Z}_i^T {\bbeta}^o_j)} (\widehat{\bbeta}_j - \bbeta^o_j) + o_p(1) \, .
$$
Since $||\widehat{\bbeta}_j - \bbeta^o_j||_2 = o_p(1)$ as $n \rightarrow \infty$, where $||\cdot||_2$ denotes the $l_2$ norm, and since $U_n(\alpha_{jt}) - \widetilde{U}_n(\alpha_{jt})$ is continuous in $\widehat{\bbeta}_j$, then by the continuous mapping theorem and Slutsky theorem, 
$\sup_{ \alpha_{jt} \in \mathcal{A} } |U_n(\alpha_{jt}) - \widetilde{U}_n(\alpha_{jt})| =o_p(1)$. Finally, since 
$$
E(T_i=t, J_i=j|T_i \geq t) =  \frac{\exp(\alpha^o_{jt}+{\bf Z}_i^T {\bbeta}^o_j)}{1+\exp(\alpha^o_{jt}+{\bf Z}_i^T {\bbeta}^o_j)} \, ,
$$
consistency of each $\widehat{\alpha}_{jt}$, as $n \rightarrow \infty$, follows by standard theory of moment estimators \citep{van2000asymptotic} applied for  $\widetilde{U}_n(\alpha_{jt})$. For the asymptotic normality, write
$$
0 = U_n(\widehat{\alpha}_{jt}) = \widetilde{U}_n(\alpha^o_{jt}) + \left\{U_n(\widehat{\alpha}_{jt}) - \widetilde{U}_n(\widehat{\alpha}_{jt})\right\} +
\left\{\widetilde{U}_n(\widehat{\alpha}_{jt}) - \widetilde{U}_n(\alpha^o_{jt}) \right\}
$$
and in the following we consider each of the terms of the right-hand side of the equation.

We can write $\widetilde{U}_n(\alpha_{jt}^o) = n^{-1} \sum_{i=1}^n \xi_{ijt}$ where
$$
\xi_{ijt} =  I(X_i \geq t)\frac{\exp(\alpha^o_{jt}+{\bf Z}_i^T {\bbeta}^o_j)}{1+\exp(\alpha^o_{jt}+{\bf Z}_i^T {\bbeta}^o_j)} - I(X_i=t, J_i=j) \, .
$$
Thus, $\widetilde{U}_n(\alpha^o_{jt})$ is the mean of the iid mean-zero random variables $\xi_{ijt}$. It hance follows from the central limit theorem that $n^{1/2}U(\alpha_{jt}^o)$ is asymptotically mean-zero normal. To estimate the variance, let $\widehat{\xi}_{i}$ be the counterpart of $\xi_i$ with estimates of $\bbeta_j$ and $\alpha_{jt}$ substituted for the true values. Then, the empirical estimator of the variance is given by
$$
V_{1jt} = n^{-1} \sum_{i=1}^n \widehat{\xi}_i^2 \, .
$$
First order Taylor expansion of $U_n(\widehat{\alpha}_{jt})$ about $\bbeta_j^o$ gives
\begin{eqnarray*}
n^{1/2} \{ U_n(\widehat{\alpha}_{jt}) - \widetilde{U}_n(\widehat{\alpha}_{jt}) \} &=&   n^{-1/2} 
\sum_{i=1}^n I(X_i \geq t) D_{ijt}(\widehat{\alpha}_{jt},\bbeta_j^o) (\widehat{\bbeta}_{j}-\bbeta^o_j) +o_p(1) \\
& = & n^{-1/2} 
\sum_{i=1}^n I(X_i \geq t) D_{ijt}({\alpha}^o_{jt},\bbeta_j^o) (\widehat{\bbeta}_{j}-\bbeta^o_j) +o_p(1) \\
& = & n^{-1/2}  D_{.jt}({\alpha}^o_{jt},\bbeta_j^o) (\widehat{\bbeta}_{j}-\bbeta^o_j) +o_p(1) 
\end{eqnarray*}
where
$$
D_{ijt}({\alpha}_{jt},\bbeta_j^o) =\frac{\partial}{\partial \bbeta_j^o} \frac{\exp({\alpha}_{jt} + {\bf Z}_i^T \bbeta_j^o)}{1+ \exp({\alpha}_{jt} + {\bf Z}_i^T \bbeta_j^o)}
$$
and 
$D_{.jt}({\alpha}_{jt},\bbeta_j^o) = \sum_{i=1}^n I(X_i \geq t) D_{ijt}({\alpha}_{jt},\bbeta_j^o)$. Therefore, $n^{1/2} \{ U_n(\widehat{\alpha}_{jt}) - \widetilde{U}_n(\widehat{\alpha}_{jt}) \}$ is asymptotically mean-zero normal with covariance matrix that can be consistently estimated by 
$$
V_{2jt} = n^{-1} \sum_{i=1}^n I(X_i \geq t) D_{ijt}(\widehat{\alpha}_{jt},\widehat{\bbeta}_j) \widehat{var}(\widehat{\bbeta}_j) \, .
$$ 

First order Taylor expansion of $\widetilde{U}_n(\widehat{\alpha}_{jt})$ about $\alpha_{jt}^o$ gives
$$
n^{1/2} \{ \widetilde{U}_n(\widehat{\alpha}_{jt}) - \widetilde{U}_n({\alpha}^o_{jt}) \} =
n^{-1/2} \sum_{i=1}^n I(X_i \geq t) A_{ijt}(\alpha_{jt}^o,\bbeta_j^o)(\widehat{\alpha}_{jt}-\alpha^o_{jt}) + o_p(1)
$$
where
$$
A_{ijt}({\alpha}^o_{jt},\bbeta_j^o) =\frac{\partial}{\partial \alpha_{jt}^o} \frac{\exp({\alpha}^o_{jt} + {\bf Z}_i^T \bbeta_j^o)}{1+ \exp({\alpha}^o_{jt} + {\bf Z}_i^T \bbeta_j^o)} \, .
$$
Let $A_{.jt}({\alpha}^o_{jt},\bbeta_j^o)=\sum_{i=1}^n I(X_i \geq t)A_{ijt}({\alpha}^o_{jt},\bbeta_j^o)$. Then, combining the results above we get $n^{1/2}(\widehat{\alpha}_{jt}-\alpha_{jt}^o)$ is asymptotically zero-mean normally distributed. For the variance of $\widehat{\alpha}_{jt}$ we write
$$
\widetilde{U}(\alpha_{jt}^o) + \{ U_n(\widehat{\alpha}_{jt} - \widetilde{U}_n(\widehat{\alpha}_{jt})\} = n^{-1} \sum_{i=1}^n (\xi_{ijt} + \psi_{ij}) + o_p(1)
$$
where $\psi_{ij}$ is the asymptotic representation of $\widehat{\bbeta}_j$ \citep{tsiatis2006semiparametric} since $\widehat{\bbeta}_j$ is a regular asymptotically linear estimator, namely, 
$$
n^{1/2} (\widehat{\bbeta}_j - \bbeta_j^o) = n^{-1/2} \sum_{i=1}^n \psi_{ij} + o_p(1) \, .
 $$
Therefore, the variance matrix can be consistently estimated by 
$
A^{-2}_{.jt}(\widehat{\alpha}_{jt},\widehat{\bbeta}_j)(V_{1jt}+V_{2jt}+V_{3jt})
$
where $V_{3jt}=2/n\sum_{i=1}^n\widehat{\xi}_{ijt}+\widehat{\phi}_{ij}$.

\phantomsection
\section*{Web Appendix C: SIS -  Additional Simulation Results}
\label{appx:PSIS}

The simulated datasets (Setting 17--19) consist of $n=1,000$ observations and $p=15,000$ covariates. Each covariate is a zero-mean normally distributed with variance 1. Three settings were considered: with independent covariates ($\rho=0$), and with correlated covariates such that  $Cov(Z_{il}, Z_{ih})=\rho^{|l-h|}$ and $\rho=0.5, 0.9$, following a similar approach as \cite{zhao2012principled}. To ensure appropriate survival probabilities, covariates were truncated to be within $[-3, 3]$.  We considered $M=2$ competing events with $d=8$. The first five components of $\bbeta_1$ and $\bbeta_2$ were set to be non-zero, and the remaining coefficients set to zero. The non-zero parameters $\beta_{1k}$, $k=1, \ldots, 5$, took on the values of $-0.7, -0.6, 0.8, 0.7, -0.8$ while $\beta_{2k}$, $k=1, \ldots, 5$,  had values of $0.7, 0.8, -0.8, -0.6, -0.7$. Additionally, $\alpha_{1t} = -3.2 + 0.3  \log t$ and $\alpha_{2t} = -3.3 + 0.4  \log t$.

For SIS-L, the lasso parameters $\eta_1$ and $\eta_2$ were tuned using a grid search and 3-fold cross-validation, where $\log \eta_1, \log \eta_2$ ranged between -12 to -2 with a step size of 0.5. The selected $\eta_1, \eta_2$ maximize the global-AUC. The simulation results are summarised in Tables \ref{tab:psis_etas}-\ref{tab:psis_metrics}.

The mean (SE) of the data-driven thresholds, $w_n$, of the SIS procedure were $0.224 \; (0.015)$, $0.224 \; (0.017)$, and $0.230 \; (0.018)$, for $\rho$ values of 0, 0.5, and 0.9, respectively. The means and SEs of the selected regularization parameters of the SIS-L are shown in Table \ref{tab:psis_etas}. The size of the selected models, the false positive (FP) and false negative (FN) are summarized in Table \ref{tab:psis_fp_fn}. As expected, higher values of $\rho$ result in higher number of FPs. Additionally, adding lasso regularization resulted in similar or reduced mean selected-model size and the mean FP. Both methods resulted with similar performance measures, as shown in Table \ref{tab:psis_metrics}. Adding lasso after the SIS, allowed us to retain a smaller set of covariates, while maintaining similar performances.


\phantomsection
\section*{Web Appendix D: Simulation Results of Two Competing Events - Additional Sample Sizes}
\label{appx:twocompt1}
\begin{figure}[!ht]
    \centering
    \includegraphics[width=\textwidth]{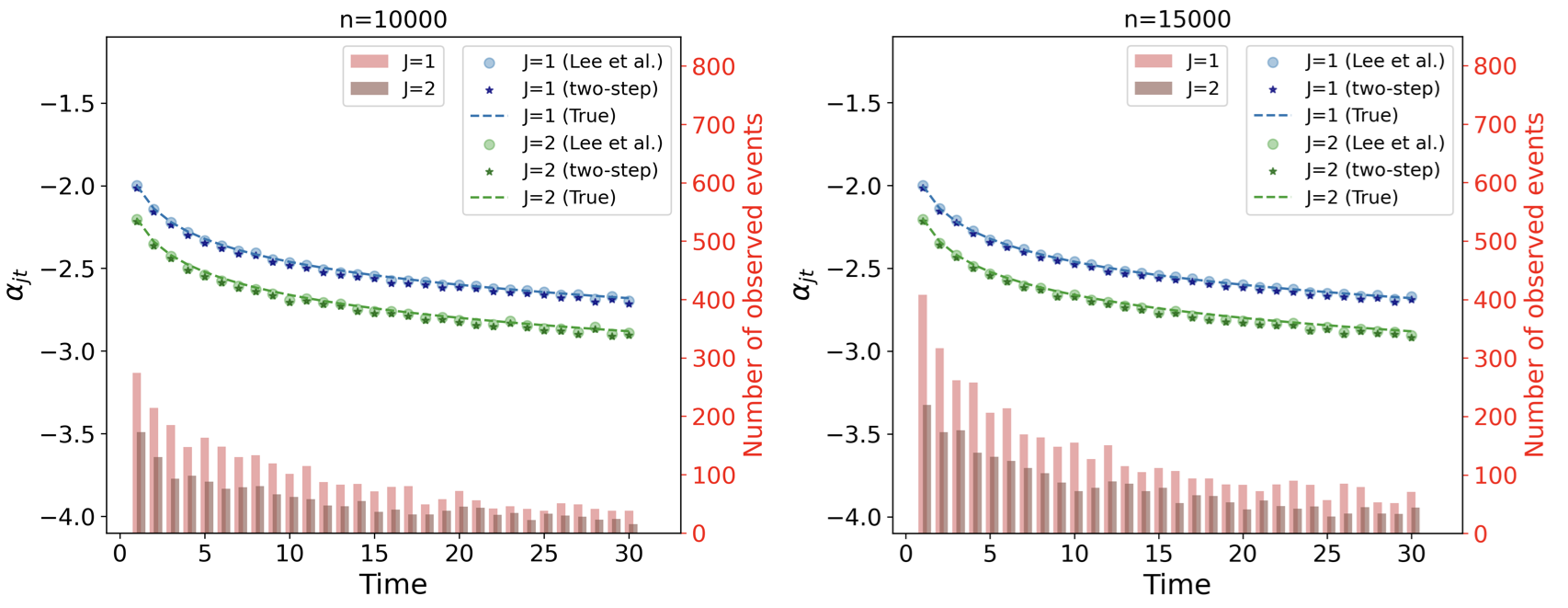}
    \caption{Simulation results of two competing events. Results of ${\alpha}_{jt}$. Each panel is based on a different sample size. Number of observed events are shown in blue, green and red bars for $j=1,2$, respectively. True values and mean of estimates are in blue, green and red for $j=1,2$. True values are shown in dashed lines, mean of estimates based on Lee et al. and the proposed two-step approach denoted by circles and diamonds, respectively.}
    \label{fig:alpha_sim_SM}
\end{figure}

We considered sample sizes of $n=10,000$ and 15,000. The vector of covariates ${\bf Z}$ is of $p=5$ dimension, and each covariate was sampled from a standard uniform distribution. For each observation, based on the sampled covariates ${\bf Z}$ and the true model of Eq.(1), the event type was sampled, and then the failure time, with $d=30$. 
The parameters' values of Settings 3-4 were set to be $\alpha_{1t} = -2.0 - 0.2 \log t$, $\alpha_{2t} = -2.2 - 0.2\log t$, $t=1,\ldots,30$, 
 $\bbeta_1 = -(\log 0.8, \log 3, \log 3, \log 2.5, \log 2)^T$, and 
 $\bbeta_{2} = -(\log 1, \log 3, \log 4, \log 3, \log 2)^T$. 
 The censoring times were sampled from a discrete uniform distribution with probability 0.01 at each $t$. The simulation results are based on 200 repetitions of each setting. Results are shown in Figure~\ref{fig:alpha_sim_SM} and Table~\ref{tab:methods_comparison_table_SM}.

\phantomsection
\section*{Web Appendix E: Simulation Results of Three Competing Events}
\label{appx:threecompt1}
\begin{figure}[!ht]
    \centering
    \includegraphics[width=\textwidth]{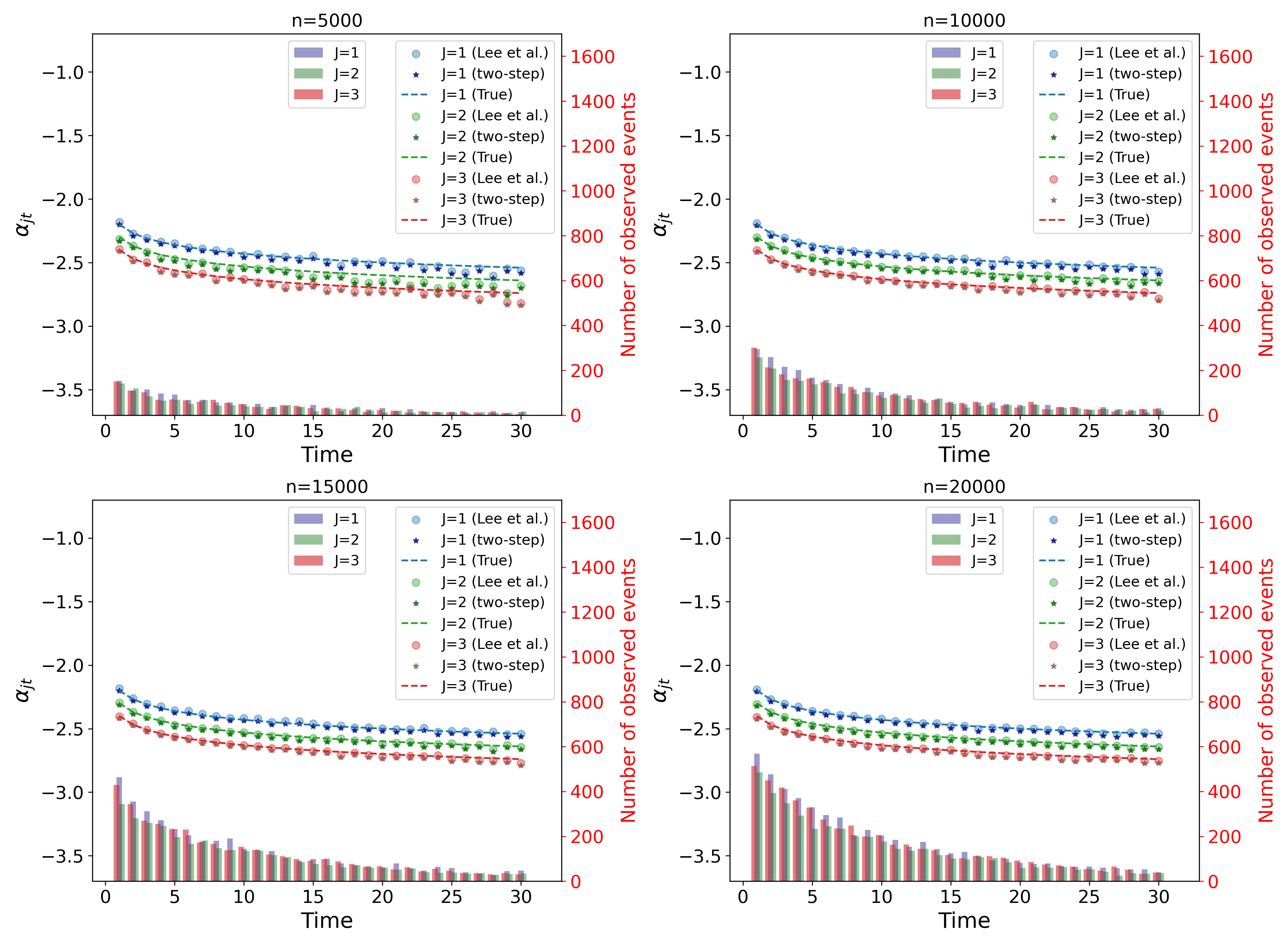}
    \caption{Simulation results of three competing events. Results of ${\alpha}_{jt}$. Each panel is based on a different sample size. Number of observed events are shown in blue, green and red bars for $j=1,2$ and 3, respectively. True values and mean of estimates are in blue, green and red for $j=1,2$ and 3. True values are shown in dashed lines, mean of estimates based on Lee et al. and the proposed two-step approach denoted by circles and diamonds, respectively.}
    \label{fig:different_n_alpha_results-j3}
\end{figure}

We considered sample sizes of $n=5,000,10,000,15,000$ and $20,000$. The  vector of covariates ${\bf Z}$ is of $p=5$ dimension, and each covariate was sampled from a standard uniform distribution. For each observation, based on the sampled covariates ${\bf Z}$ and the true model of Eq.(1), the event type was sampled, and then the failure time, with $d=30$. 
The parameters' values were set to be $\alpha_{1t} = -2.2 - 0.1 \log t$, $\alpha_{2t} = -2.3 - 0.1\log t$, and $\alpha_{3t} = -2.4 - 0.1\log t$ $t=1,\ldots,30$, $\bbeta_1 = -(\log 2.5, \log 1.5, \log 0.8, \log 3, \log 2)$, $\bbeta_2 = -(\log 0.8, \log 3, \log 2.8, \log 2.2, \log 1.5)$, and $\bbeta_{3} = -(\log 1.8, \log 0.8, \log 2.5, \log 1.2, \log 3)$. Finally, the censoring times were sampled from a discrete uniform distribution with probability 0.01 at each $t \leq 30$. The simulation results are based on 200 repetitions of each setting. Results are shown in Figure~\ref{fig:different_n_alpha_results-j3} and Tables~\ref{tab:methods_comparison_table_j3}-\ref{tab:methods_comparison_table_j3b}.

\phantomsection
\section*{Web Appendix F: Lasso - Additional Simulation Results}
Figure \ref{fig:reg_corr_sim_results} demonstrates the results of the regularization parameters $\eta_j$,  $j= 1,2$, of one simulated dataset under Setting 12. Based on the one simulated dataset of Setting 12  and the selected values of $\eta_j$: $\widehat{\mbox{AUC}}_1=0.796$ (SD=0.007), $\widehat{\mbox{AUC}}_2=0.801$ (SD=0.008), $\widehat{\mbox{AUC}}=0.799$ (SD=0.005),  and $\widetilde{\mbox{AUC}}=0.794$ (SD=0.005). The mean Brier Scores were $\widehat{\mbox{BS}}_1=0.046$ (SD=0.002), $\widehat{\mbox{BS}}_2=0.043$ (SD=0.003), and $\widehat{\mbox{BS}}=0.045$ (SD=0.001). 

To demonstrate the performance of the proposed approach with lasso regularization in small sample sizes, we repeat the same sampling procedure as in Settings 11--13, but with sample size of $n=500$ observations and $d=10$ times. The parameters of the model were set to be  $\alpha_{1t} = -4.4 + 0.3 t$, $\alpha_{2t} = -4.3 + 0.3 t$, $t = 1, \ldots , 10$. The first five out of $p=35$ components of $\bbeta_1$ and $\bbeta_2$ were set to be $(1.2,1.5,-1,-0.3,-1.2)$ and $(-1.2,-1,1.4,1,1)$, respectively, and the rest of the coefficients were set to zero. 

Figure \ref{fig:reg_sim_results_small_sample} demonstrates the results of the regularization parameters $\eta_j$,  $j= 1,2$, of one simulated dataset under Setting 14. Based on the one simulated dataset of Setting 14  and the selected values of $\eta_j$: $\widehat{\mbox{AUC}}_1=0.746$ (SD=0.013), $\widehat{\mbox{AUC}}_2=0.726$ (SD=0.024), $\widehat{\mbox{AUC}}=0.767$ (SD=0.039),  and $\widetilde{\mbox{AUC}}=0.706$ (SD=0.017). The mean Brier Scores were $\widehat{\mbox{BS}}_1=0.112$ (SD=0.005), $\widehat{\mbox{BS}}_2=0.109$ (SD=0.014), and $\widehat{\mbox{BS}}=0.114$ (SD=0.009).

Figure \ref{fig:reg_sim_results_small_sample_corr} demonstrates the results of the regularization parameters $\eta_j$,  $j= 1,2$, of one simulated dataset under Setting 15. Based on the one simulated dataset of Setting 15  and the selected values of $\eta_j$: $\widehat{\mbox{AUC}}_1=0.758$ (SD=0.011), $\widehat{\mbox{AUC}}_2=0.751$ (SD=0.009), $\widehat{\mbox{AUC}}=0.765$ (SD=0.026),  and $\widetilde{\mbox{AUC}}=0.724$ (SD=0.006). The mean Brier Scores were $\widehat{\mbox{BS}}_1=0.105$ (SD=0.009), $\widehat{\mbox{BS}}_2=0.104$ (SD=0.019), and $\widehat{\mbox{BS}}=0.104$ (SD=0.001).

Setting 16 is similar to Setting 14, but with 100 repetitions. It shows that the means of true- and false-positive discoveries for each event type, $\mbox{TP}_j$ and $\mbox{FP}_j$, $j=1,2$, under the selected values of $\eta_j$ were $\mbox{TP}_1=4.34$, $\mbox{FP}_1=2.56$, $\mbox{TP}_2=4.99$, and $\mbox{FP}_2=2.0$. These findings indicate that even with a small sample size, the proposed grid search and  AUC-based selection of $\eta_j$, $j=1,2$, successfully identified the 9 out of 10 non-zero parameters in all 100 repetitions. However, the smaller parameter in $j=1$ was not always selected. Additionally, most of the 30 parameters with true value of zero were excluded from the final model, leaving only a small number of false positives.

\phantomsection
\section*{Web Appendix G: MIMIC Data Analysis - Additional Discussion of the Results}\label{appx:MIMIC}

The estimated coefficients for lab tests in the discharge-to-home ($j=1$) model were all negative, consistent with the expected result that abnormal test results at admission reduce the hazard of home discharge. Older age and recent admission were also found to reduce this hazard, while being married and having Medicare or ``other'' insurance increased it. Female gender, admission number, direct emergency admission, and night admission had a relatively small impact on this hazard. lasso regularization excluded several features from the model, including admissions number, night admission, direct emergency admission, ethnicity, Medicare insurance, single or widowed status, sex, and certain lab tests (Anion Gap, MCH, MCV, Magnesium, Phosphate, Platelet count, and Potassium).

The hazard of being discharged for further treatment ($j=2$) is primarily increased by admissions number, White ethnicity, Medicare insurance, single or widowed marital status, and older age. Direct emergency admission and being married decrease the hazard. Most lab test results had a minor impact on the hazard, except for white blood cell count, RDW, platelet count, glucose, creatinine, and bicarbonate, which reduced the hazard of being discharged for further treatment when abnormal. lasso regularization excluded only a few lab tests (Anion Gap, Chloride, MCHC, and MCV) and recent admission. The main factors that increased the hazard and were included in the model were admissions number, single or widowed marital status, Medicare insurance, and older age, while direct emergency admission, being married, and abnormal results of bicarbonate, creatinine, glucose, and platelet count decreased the hazard.

The hazard of in-hospital death ($j=3$) had the lowest number of observed events, resulting in noisier estimators, especially in later times. The lasso penalty had only a minor effect in terms of the number of excluded features. Lab test results that increased the hazard of in-hospital death were abnormal Anion Gap, Bicarbonate, Creatinine, Magnesium, White Blood Cells, RDW, and Sodium. Some of these lab test results had already been identified as predictors of in-hospital mortality in previous studies \citep{zhong_serum_2021, wernly_blood_2018, meynaar2013red, bazick_red_2011}. Other lab test results that increased the hazard of in-hospital death were abnormal Calcium total, Chloride, Glucose, Phosphate, Platelet Count, Potassium, Urea Nitrogen, and Red Blood Cells. Admissions number, ``other'' ethnicity, married status, recent admission, and older age also increased the hazard of in-hospital death. Direct emergency admission, black, Hispanic, or white ethnicity, and Medicare or ``other'' insurance types decreased the hazard of in-hospital death.

\phantomsection
\section*{Web Appendix H: Comparison of Computation Time}
For demonstrating the reduction in computation time as a function of $d$, a sample size of $n=20,000$ observations was considered with $p=10$ covariates, two competing events,  various values of $d$, and 10 repetitions for each $d$. 
Furthermore,  $\alpha_{1t} = -2.5 - 0.3 \log t$, $\alpha_{2t} = -2.8 - 0.3\log t$, 
$$\bbeta_{1} = -0.5  (\log 0.8, \log 3, \log 3, \log 2.5, \log 4, \log 1, \log 3, \log 2, \log 2, \log 3)^T \, ,$$ 
and 
$$\bbeta_{2} = -0.5  (\log 1, \log 3, \log 2, \log 1, \log 4, \log 3, \log 4, \log 3, \log 3, \log 2)^T \, .$$
The median and the interquartile range of the computation times are presented in Figure \ref{fig:fitting-time-comparison}. These results are based on a single CPU out of 40 CPUs server of type Intel Xeon Silver 4114 CPU @ 2.20GHz X 2 and 377GB RAM. Evidently, under low values of $d$, the computation times of the two approaches are comparable. However, as $d$ increases, the advantage of the proposed approach, in terms of computation time, increases as well.   
Moreover,  when running this analysis using hardware with 16GB RAM,  the estimation procedure of Lee et al. reached a memory error at a low value of $d$, while the two-step procedure was completed successfully. 


\clearpage

\begin{table}[h!]
	\centering
	\caption{Original and expanded datasets with $M=2$ competing events \citep{lee_analysis_2018}}\label{tbl:expanded}
		\begin{tabular}{cccc cccccc}
			\toprule
	\multicolumn{4}{c}{Original Data} & \multicolumn{6}{c}{Expanded Data}  \\
 \cmidrule(lr){1-4} \cmidrule(lr){5-10} \\ [-6pt]
	$i$ & $X_i$ & $\delta_i$ & $Z_i$ & $i$ & $\widetilde{X}_i$ & $\delta_{1it}$ &  $\delta_{2it}$ & 
 $1-\delta_{1it}-\delta_{2it}$ & $Z_i$ \\
			\hline
   1 & 2 & 1 & $Z_1$ & 1 & 1 & 0 & 0 & 1 &  $Z_1$ \\
     &   &   &       & 1 & 2 & 1 & 0 & 0 &  $Z_1$ \\
   2 & 3 & 2 & $Z_2$ & 2 & 1 & 0 & 0 & 1 &  $Z_2$ \\
     &   &   &       & 2 & 2 & 0 & 0 & 1 &  $Z_2$ \\    
     &   &   &       & 2 & 3 & 0 & 1 & 0 &  $Z_2$ \\ 
   3 & 3 & 0 & $Z_3$ & 3 & 1 & 0 & 0 & 1 &  $Z_3$ \\
     &   &   &       & 3 & 2 & 0 & 0 & 1 & $Z_3$ \\
     &   &   &       & 3 & 3 & 0 & 0 & 1 & $Z_3$ \\
     \bottomrule
		\end{tabular}
  \vspace*{2cm}
\end{table}

\clearpage

\begin{sidewaystable}[h!]
\setlength{\tabcolsep}{3pt}
\centering
\caption{Overview of simulations settings: number of repetitions is denoted by Rep, Dist stands for the covariates' distribution, where ``indep'' and ``dep'' denote independent and dependent covariates, respectively, ``unif'' and ``normal'' denote standard uniform and normal distributions.}\label{tbl:settings-summary}
\begin{tabular}{ccc ccccc cccc cc c}
\hline
      \multicolumn{3}{c}{}            & \multicolumn{5}{c}{Setting Details} & \multicolumn{4}{c}{\% of Cen \& Event Type} & \multicolumn{2}{c}{Estimation Method}     &                \\
      \cmidrule(lr){4-8} \cmidrule(lr){9-12} \cmidrule(lr){13-14}\\ [-6pt]
No. & Objective      & Rep & $n$              & $p$ & Dist  & $M$ & $d$ & $J=0$ & $J=1$ & $J=2$ & $J=3$  & Lee et al.    & two-step & Results           \\
\hline
1   & Performance    & 200         & 250           & 5  & indep, unif  & 2 & 7 & 23.3 & 37.1 & 39.6 & -  & Yes                & Yes                     & Tbl 1, Fig~1 \\
2   & Performance    & 200         & 500           & 5 & indep, unif   & 2 & 7 & 23.4 & 37.3 & 39.3 & -  & Yes                & Yes                     & Tbl 1, Fig~1 \\
3   & Performance    & 200         & 5000            & 5  & indep, unif & 2 & 30  & 55.5 & 27.8 & 16.7 & -   & Yes                & Yes                  & Tbl 1, Fig~1 \\
4   & Performance    & 200         & 20000           & 5 & indep, unif  & 2 & 30 & 55.5 & 27.8 & 16.7 & -  & Yes                & Yes                    & Tbl  1, Fig~1 \\
5   & Performance    & 200         & 10000           & 5 & indep, unif  & 2 & 30 & 55.5 & 27.8 & 16.7 & -   & Yes                & Yes               & Tbl S9, Fig S3 \\
6   & Performance    & 200         & 15000           & 5 & indep, unif & 2 & 30 & 55.5 & 27.8 & 16.7 & -  & Yes                & Yes                     & Tbl S9, Fig S3 \\
7   & Performance    & 200         & 5000            & 5  & indep, unif  & 3 & 30 & 25.7 & 26.6 & 22.1 & 25.6  & Yes                & Yes                     & Tbl S10, Fig S4  \\
8   & Performance    & 200         & 10000           & 5 & indep, unif   & 3 & 30 & 25.7 & 26.6 & 22.1 & 25.6  & Yes                & Yes                     & Tbl S10, Fig S4  \\
9   & Performance    & 200         & 15000           & 5 & indep, unif   & 3 & 30 & 25.7 & 26.6 & 22.1 & 25.6  & Yes                & Yes                    & Tbl S11, Fig S4  \\
10   & Performance    & 200         & 20000           & 5 & indep, unif   & 3 & 30 & 25.7 & 26.6 & 22.1 & 25.6  & Yes                & Yes                   & Tbl S11, Fig S4  \\
11  & Regularization & 1           & 10000           & 100 & indep, normal  & 2 & 15 & 34.5 & 34.5 & 31.0 & -   & No                       & Yes              & Fig~S5      \\
12  & Regularization & 1           & 10000           & 100 & dep, normal & 2 & 15 & 36.0 & 32.6 & 31.4 & -  & No                      & Yes              & Fig S6    \\
13  & Regularization & 100         & 10000           & 100 & indep, normal & 2 & 15 & 35.0 & 33.8 & 31.2 & - & No                       & Yes              & Section~3   \\
14  & Regularization & 1           & 500           & 35 & indep, normal  & 2 & 10 & 12.4 & 41.8 & 45.8 & -   & No                       & Yes              & Fig~S7      \\
15  & Regularization & 1           & 500           & 35 & dep, normal & 2 & 10 & 12.4 & 40.8 & 46.8 & -  & No                      & Yes              & Fig S8    \\
16  & Regularization & 100         & 500           & 35 & indep, normal & 2 & 10 & 12.4 & 33.8 & 31.2 & - & No                       & Yes              & Web Appendix F   \\
17  & SIS, SIS-L & 100           & 1000        & 15000 & indep, normal & 2 & 8 & 24.1 & 37.7 & 38.2 & - & No                       & Yes              & Web Appendix C\\
18  & SIS, SIS-L & 100           & 1000        & 15000 & dep, normal & 2 & 8 & 24.5 & 36.5 & 39.0 & - & No                       & Yes              & Web Appendix C \\
19  & SIS, SIS-L & 100           & 1000        & 15000 & dep, normal & 2 & 8 & 31.4 & 33.0 & 35.6 & - & No                       & Yes              & Web Appendix C\\
\hline
\end{tabular}
\end{sidewaystable}

\clearpage

\begin{table}
    \centering
        \caption{Simulation results of continuous time versus discrete time analyses.  $n=2,000$, $p=5$, $d=9$, and $M=2$. The results are the mean bias and empirical standard error (SE) over 500 repetitions.}
    \label{tab:continuous_discrete_comparison_table_bias}
\vspace{0.4cm}
\small
\begin{tabular}{cccccccccc}
\toprule
& \multicolumn{7}{c}{Naive Continuous-Time Analysis} & \multicolumn{2}{c}{The Proposed} \\
{} & {} & \multicolumn{2}{c}{Breslow} & \multicolumn{2}{c}{Efron} & \multicolumn{2}{c}{Exact} & \multicolumn{2}{c}{Approach} \\
{} &   True &        Bias &    SE &      Bias &    SE &      Bias &    SE &     Bias &    SE \\
\midrule
$\alpha_{11}$ & -1.900 &      -0.034 & 0.008 &     0.058 & 0.008 &     0.096 & 0.009 &   -0.006 & 0.008 \\
$\alpha_{12}$ & -1.484 &       0.025 & 0.008 &     0.141 & 0.009 &     0.168 & 0.009 &    0.005 & 0.007 \\
$\alpha_{13}$ & -1.241 &       0.065 & 0.009 &     0.203 & 0.009 &     0.221 & 0.010 &    0.004 & 0.007 \\
$\alpha_{14}$ & -1.068 &       0.087 & 0.009 &     0.243 & 0.010 &     0.254 & 0.010 &   -0.009 & 0.007 \\
$\alpha_{15}$ & -0.934 &       0.127 & 0.009 &     0.303 & 0.010 &     0.308 & 0.011 &   -0.006 & 0.007 \\
$\alpha_{16}$ & -0.825 &       0.167 & 0.010 &     0.363 & 0.012 &     0.364 & 0.012 &   -0.003 & 0.008 \\
$\alpha_{17}$ & -0.732 &       0.215 & 0.011 &     0.433 & 0.013 &     0.429 & 0.013 &    0.006 & 0.008 \\
$\alpha_{18}$ & -0.652 &       0.240 & 0.012 &     0.478 & 0.015 &     0.472 & 0.015 &   -0.004 & 0.008 \\
$\alpha_{19}$ & -0.582 &       0.280 & 0.013 &     0.541 & 0.016 &     0.532 & 0.016 &   -0.001 & 0.008 \\
$\beta_{11}$  &  0.156 &      -0.014 & 0.005 &    -0.007 & 0.005 &     0.002 & 0.005 &    0.002 & 0.005 \\
$\beta_{12}$  & -0.769 &       0.072 & 0.004 &     0.035 & 0.005 &    -0.006 & 0.005 &   -0.006 & 0.005 \\
$\beta_{13}$  & -0.769 &       0.074 & 0.005 &     0.037 & 0.005 &    -0.004 & 0.005 &   -0.004 & 0.005 \\
$\beta_{14}$  & -0.641 &       0.062 & 0.005 &     0.031 & 0.005 &    -0.003 & 0.006 &   -0.003 & 0.006 \\
$\beta_{15}$  & -0.485 &       0.056 & 0.005 &     0.033 & 0.005 &     0.007 & 0.005 &    0.007 & 0.005 \\
$\alpha_{21}$ & -2.500 &      -0.003 & 0.011 &     0.045 & 0.012 &     0.068 & 0.012 &    0.005 & 0.011 \\
$\alpha_{22}$ & -2.084 &       0.033 & 0.011 &     0.089 & 0.011 &     0.107 & 0.012 &    0.011 & 0.010 \\
$\alpha_{23}$ & -1.841 &       0.059 & 0.011 &     0.123 & 0.012 &     0.137 & 0.012 &    0.011 & 0.010 \\
$\alpha_{24}$ & -1.668 &       0.058 & 0.012 &     0.129 & 0.012 &     0.141 & 0.012 &   -0.010 & 0.010 \\
$\alpha_{25}$ & -1.534 &       0.099 & 0.012 &     0.177 & 0.013 &     0.186 & 0.013 &    0.006 & 0.010 \\
$\alpha_{26}$ & -1.425 &       0.102 & 0.013 &     0.185 & 0.013 &     0.193 & 0.013 &   -0.009 & 0.011 \\
$\alpha_{27}$ & -1.332 &       0.146 & 0.014 &     0.236 & 0.014 &     0.243 & 0.015 &    0.009 & 0.011 \\
$\alpha_{28}$ & -1.252 &       0.146 & 0.014 &     0.241 & 0.015 &     0.248 & 0.015 &   -0.007 & 0.011 \\
$\alpha_{29}$ & -1.182 &       0.183 & 0.015 &     0.284 & 0.016 &     0.291 & 0.016 &    0.004 & 0.012 \\
$\beta_{21}$  &  0.000 &      -0.012 & 0.007 &    -0.012 & 0.007 &    -0.012 & 0.007 &   -0.012 & 0.007 \\
$\beta_{22}$  & -0.769 &       0.029 & 0.007 &     0.010 & 0.008 &    -0.010 & 0.008 &   -0.010 & 0.008 \\
$\beta_{23}$  & -0.970 &       0.046 & 0.007 &     0.023 & 0.007 &    -0.002 & 0.008 &   -0.002 & 0.008 \\
$\beta_{24}$  & -0.769 &       0.040 & 0.007 &     0.022 & 0.007 &     0.002 & 0.008 &    0.002 & 0.008 \\
$\beta_{25}$  & -0.485 &       0.014 & 0.007 &     0.002 & 0.007 &    -0.010 & 0.007 &   -0.010 & 0.007 \\
\bottomrule
\end{tabular}
\end{table}

\clearpage

\begin{table}
    \centering
        \caption{Simulation results of continuous time versus discrete time analyses. $n=5,000$, $p=5$, $d=50$, $M=2$ and $j=1$. The results are the mean bias and empirical standard error (SE) over 500 repetitions.}
    \label{tab:continuous_discrete_comparison_table_bias_large_d_risk_1}
\vspace{0.4cm}
\footnotesize
\begin{tabular}{cccccccccc}
\toprule
   &    & \multicolumn{6}{c}{Naive Continuous-Time Analysis} & \multicolumn{2}{c}{The Proposed} \\ 
{} & {} & \multicolumn{2}{c}{Breslow} & \multicolumn{2}{c}{Efron} & \multicolumn{2}{c}{Exact} & \multicolumn{2}{c}{Approach} \\
Parameter &  True &       Bias &    SE &      Bias &    SE &      Bias &    SE &     Bias &    SE \\
\midrule
$\alpha_{11}$  & -3.930 &      -0.059 & 0.007 &    -0.034 & 0.007 &    -0.015 & 0.007 &   -0.026 & 0.007 \\
$\alpha_{12}$  & -3.860 &      -0.047 & 0.007 &    -0.021 & 0.008 &    -0.003 & 0.008 &   -0.015 & 0.007 \\
$\alpha_{13}$  & -3.790 &      -0.037 & 0.007 &    -0.010 & 0.008 &     0.008 & 0.008 &   -0.005 & 0.007 \\
$\alpha_{14}$  & -3.720 &      -0.044 & 0.008 &    -0.017 & 0.008 &     0.000 & 0.008 &   -0.013 & 0.007 \\
$\alpha_{15}$  & -3.650 &      -0.041 & 0.007 &    -0.014 & 0.007 &     0.003 & 0.007 &   -0.011 & 0.007 \\
$\alpha_{16}$  & -3.580 &      -0.039 & 0.007 &    -0.012 & 0.007 &     0.005 & 0.007 &   -0.010 & 0.007 \\
$\alpha_{17}$  & -3.510 &      -0.034 & 0.007 &    -0.005 & 0.007 &     0.011 & 0.007 &   -0.005 & 0.007 \\
$\alpha_{18}$  & -3.440 &      -0.049 & 0.007 &    -0.020 & 0.007 &    -0.004 & 0.007 &   -0.021 & 0.006 \\
$\alpha_{19}$  & -3.370 &      -0.033 & 0.007 &    -0.003 & 0.007 &     0.012 & 0.007 &   -0.007 & 0.006 \\
$\alpha_{110}$ & -3.300 &      -0.022 & 0.006 &     0.008 & 0.006 &     0.023 & 0.006 &    0.002 & 0.006 \\
$\alpha_{111}$ & -3.230 &      -0.028 & 0.006 &     0.003 & 0.006 &     0.018 & 0.006 &   -0.004 & 0.006 \\
$\alpha_{112}$ & -3.160 &      -0.026 & 0.006 &     0.006 & 0.006 &     0.020 & 0.006 &   -0.004 & 0.006 \\
$\alpha_{113}$ & -3.090 &      -0.035 & 0.006 &    -0.003 & 0.006 &     0.011 & 0.006 &   -0.015 & 0.006 \\
$\alpha_{114}$ & -3.020 &      -0.004 & 0.006 &     0.030 & 0.006 &     0.043 & 0.006 &    0.015 & 0.006 \\
$\alpha_{115}$ & -2.950 &      -0.018 & 0.006 &     0.017 & 0.006 &     0.029 & 0.006 &   -0.001 & 0.006 \\
$\alpha_{116}$ & -2.880 &      -0.024 & 0.006 &     0.012 & 0.006 &     0.023 & 0.006 &   -0.008 & 0.006 \\
$\alpha_{117}$ & -2.810 &      -0.017 & 0.006 &     0.019 & 0.006 &     0.030 & 0.006 &   -0.004 & 0.006 \\
$\alpha_{118}$ & -2.740 &      -0.018 & 0.006 &     0.020 & 0.006 &     0.030 & 0.006 &   -0.007 & 0.006 \\
$\alpha_{119}$ & -2.670 &      -0.011 & 0.006 &     0.028 & 0.006 &     0.037 & 0.006 &   -0.003 & 0.006 \\
$\alpha_{120}$ & -2.600 &      -0.020 & 0.006 &     0.021 & 0.006 &     0.028 & 0.006 &   -0.014 & 0.006 \\
$\alpha_{121}$ & -2.530 &      -0.006 & 0.006 &     0.036 & 0.006 &     0.042 & 0.006 &   -0.003 & 0.006 \\
$\alpha_{122}$ & -2.460 &      -0.004 & 0.006 &     0.040 & 0.006 &     0.045 & 0.006 &   -0.004 & 0.005 \\
$\alpha_{123}$ & -2.390 &       0.007 & 0.006 &     0.053 & 0.006 &     0.057 & 0.006 &    0.003 & 0.006 \\
$\alpha_{124}$ & -2.320 &       0.002 & 0.006 &     0.049 & 0.006 &     0.052 & 0.006 &   -0.005 & 0.005 \\
$\alpha_{125}$ & -2.250 &       0.008 & 0.006 &     0.057 & 0.006 &     0.058 & 0.006 &   -0.003 & 0.005 \\
$\alpha_{126}$ & -2.180 &       0.011 & 0.005 &     0.062 & 0.006 &     0.062 & 0.006 &   -0.004 & 0.005 \\
$\alpha_{127}$ & -2.110 &       0.017 & 0.006 &     0.071 & 0.006 &     0.069 & 0.006 &   -0.003 & 0.005 \\
$\alpha_{128}$ & -2.040 &       0.031 & 0.006 &     0.088 & 0.006 &     0.084 & 0.006 &    0.006 & 0.005 \\
$\alpha_{129}$ & -1.970 &       0.031 & 0.006 &     0.089 & 0.006 &     0.084 & 0.006 &   -0.000 & 0.005 \\
$\alpha_{130}$ & -1.900 &       0.032 & 0.006 &     0.094 & 0.006 &     0.087 & 0.006 &   -0.004 & 0.005 \\
$\alpha_{131}$ & -1.830 &       0.041 & 0.006 &     0.106 & 0.006 &     0.097 & 0.006 &   -0.002 & 0.005 \\
$\alpha_{132}$ & -1.760 &       0.048 & 0.006 &     0.116 & 0.007 &     0.104 & 0.007 &   -0.002 & 0.006 \\
$\alpha_{133}$ & -1.690 &       0.054 & 0.006 &     0.126 & 0.006 &     0.112 & 0.006 &   -0.003 & 0.005 \\
$\alpha_{134}$ & -1.620 &       0.050 & 0.007 &     0.125 & 0.007 &     0.109 & 0.007 &   -0.014 & 0.006 \\
$\alpha_{135}$ & -1.550 &       0.082 & 0.006 &     0.162 & 0.007 &     0.142 & 0.006 &    0.006 & 0.006 \\
$\alpha_{136}$ & -1.480 &       0.078 & 0.006 &     0.162 & 0.007 &     0.139 & 0.007 &   -0.007 & 0.006 \\
$\alpha_{137}$ & -1.410 &       0.086 & 0.007 &     0.176 & 0.008 &     0.149 & 0.007 &   -0.009 & 0.006 \\
$\alpha_{138}$ & -1.340 &       0.111 & 0.007 &     0.207 & 0.008 &     0.176 & 0.007 &    0.002 & 0.006 \\
$\alpha_{139}$ & -1.270 &       0.119 & 0.008 &     0.221 & 0.008 &     0.186 & 0.008 &   -0.003 & 0.006 \\
$\alpha_{140}$ & -1.200 &       0.130 & 0.008 &     0.239 & 0.009 &     0.200 & 0.009 &   -0.006 & 0.007 \\
$\alpha_{141}$ & -1.130 &       0.135 & 0.008 &     0.250 & 0.009 &     0.207 & 0.008 &   -0.014 & 0.007 \\
$\alpha_{142}$ & -1.060 &       0.168 & 0.009 &     0.293 & 0.010 &     0.243 & 0.009 &   -0.003 & 0.007 \\
$\alpha_{143}$ & -0.990 &       0.183 & 0.010 &     0.318 & 0.011 &     0.261 & 0.010 &   -0.006 & 0.008 \\
$\alpha_{144}$ & -0.920 &       0.202 & 0.010 &     0.347 & 0.012 &     0.284 & 0.011 &   -0.009 & 0.008 \\
$\alpha_{145}$ & -0.850 &       0.233 & 0.011 &     0.390 & 0.012 &     0.318 & 0.011 &   -0.004 & 0.008 \\
$\alpha_{146}$ & -0.780 &       0.264 & 0.012 &     0.436 & 0.014 &     0.354 & 0.013 &   -0.003 & 0.009 \\
$\alpha_{147}$ & -0.710 &       0.283 & 0.013 &     0.470 & 0.015 &     0.379 & 0.014 &   -0.011 & 0.009 \\
$\alpha_{148}$ & -0.640 &       0.315 & 0.015 &     0.522 & 0.018 &     0.417 & 0.016 &   -0.015 & 0.010 \\
$\alpha_{149}$ & -0.570 &       0.365 & 0.017 &     0.600 & 0.021 &     0.475 & 0.018 &   -0.010 & 0.011 \\
$\alpha_{150}$ & -0.500 &       0.412 & 0.019 &     0.684 & 0.026 &     0.532 & 0.021 &   -0.013 & 0.012 \\
\hline
$\beta_{11}$   &  0.112 &      -0.006 & 0.002 &    -0.004 & 0.003 &    -0.001 & 0.003 &   -0.001 & 0.003 \\
$\beta_{12}$   & -0.549 &       0.030 & 0.002 &     0.016 & 0.002 &     0.001 & 0.003 &    0.001 & 0.003 \\
$\beta_{13}$   & -0.549 &       0.024 & 0.002 &     0.010 & 0.003 &    -0.005 & 0.003 &   -0.005 & 0.003 \\
$\beta_{14}$   & -0.458 &       0.028 & 0.002 &     0.016 & 0.003 &     0.004 & 0.003 &    0.004 & 0.003 \\
$\beta_{15}$   & -0.347 &       0.018 & 0.002 &     0.009 & 0.003 &    -0.001 & 0.003 &   -0.001 & 0.003 \\
\bottomrule
\end{tabular}
\end{table}

\clearpage

\begin{table}
    \centering
        \caption{Simulation results of continuous time versus discrete time analyses. $n=5,000$, $p=5$, $d=50$, $M=2$ and $j=2$. The results are the mean bias and empirical standard error (SE) over 500 repetitions.}
    \label{tab:continuous_discrete_comparison_table_bias_large_d_risk_2}
\vspace{0.4cm}
\footnotesize
\begin{tabular}{cccccccccc}
\toprule
   &    & \multicolumn{6}{c}{Naive Continuous-Time Analysis} & \multicolumn{2}{c}{The Proposed} \\ 
   {} & {} & \multicolumn{2}{c}{Breslow} & \multicolumn{2}{c}{Efron} & \multicolumn{2}{c}{Exact} & \multicolumn{2}{c}{Approach} \\
Parameter & True &      Bias &    SE &      Bias &    SE &      Bias &    SE &     Bias &    SE \\
\midrule
$\alpha_{21}$  & -5.230 &      -0.047 & 0.016 &    -0.040 & 0.016 &    -0.034 & 0.016 &   -0.037 & 0.016 \\
$\alpha_{22}$  & -5.160 &      -0.049 & 0.016 &    -0.041 & 0.016 &    -0.036 & 0.016 &   -0.039 & 0.016 \\
$\alpha_{23}$  & -5.090 &      -0.079 & 0.017 &    -0.072 & 0.017 &    -0.066 & 0.017 &   -0.070 & 0.017 \\
$\alpha_{24}$  & -5.020 &      -0.063 & 0.015 &    -0.055 & 0.015 &    -0.050 & 0.015 &   -0.054 & 0.015 \\
$\alpha_{25}$  & -4.950 &      -0.056 & 0.015 &    -0.048 & 0.015 &    -0.043 & 0.015 &   -0.047 & 0.015 \\
$\alpha_{26}$  & -4.880 &      -0.047 & 0.015 &    -0.039 & 0.015 &    -0.034 & 0.015 &   -0.039 & 0.014 \\
$\alpha_{27}$  & -4.810 &      -0.036 & 0.014 &    -0.028 & 0.014 &    -0.023 & 0.014 &   -0.028 & 0.014 \\
$\alpha_{28}$  & -4.740 &      -0.052 & 0.014 &    -0.043 & 0.014 &    -0.038 & 0.014 &   -0.044 & 0.014 \\
$\alpha_{29}$  & -4.670 &      -0.028 & 0.013 &    -0.020 & 0.013 &    -0.015 & 0.013 &   -0.021 & 0.013 \\
$\alpha_{210}$ & -4.600 &      -0.017 & 0.013 &    -0.009 & 0.013 &    -0.004 & 0.013 &   -0.010 & 0.013 \\
$\alpha_{211}$ & -4.530 &      -0.015 & 0.012 &    -0.007 & 0.013 &    -0.002 & 0.012 &   -0.009 & 0.012 \\
$\alpha_{212}$ & -4.460 &      -0.029 & 0.014 &    -0.020 & 0.014 &    -0.016 & 0.014 &   -0.023 & 0.014 \\
$\alpha_{213}$ & -4.390 &      -0.055 & 0.013 &    -0.045 & 0.013 &    -0.041 & 0.013 &   -0.049 & 0.013 \\
$\alpha_{214}$ & -4.320 &      -0.008 & 0.012 &     0.002 & 0.012 &     0.006 & 0.012 &   -0.002 & 0.012 \\
$\alpha_{215}$ & -4.250 &      -0.021 & 0.012 &    -0.012 & 0.012 &    -0.008 & 0.012 &   -0.017 & 0.012 \\
$\alpha_{216}$ & -4.180 &      -0.039 & 0.012 &    -0.029 & 0.012 &    -0.025 & 0.012 &   -0.034 & 0.012 \\
$\alpha_{217}$ & -4.110 &      -0.026 & 0.012 &    -0.016 & 0.012 &    -0.012 & 0.012 &   -0.023 & 0.012 \\
$\alpha_{218}$ & -4.040 &      -0.017 & 0.012 &    -0.007 & 0.012 &    -0.003 & 0.012 &   -0.014 & 0.012 \\
$\alpha_{219}$ & -3.970 &      -0.028 & 0.012 &    -0.017 & 0.012 &    -0.014 & 0.012 &   -0.026 & 0.012 \\
$\alpha_{220}$ & -3.900 &       0.004 & 0.011 &     0.015 & 0.011 &     0.018 & 0.011 &    0.005 & 0.011 \\
$\alpha_{221}$ & -3.830 &      -0.017 & 0.011 &    -0.006 & 0.011 &    -0.003 & 0.011 &   -0.017 & 0.011 \\
$\alpha_{222}$ & -3.760 &      -0.007 & 0.011 &     0.005 & 0.011 &     0.007 & 0.011 &   -0.008 & 0.011 \\
$\alpha_{223}$ & -3.690 &      -0.032 & 0.011 &    -0.020 & 0.011 &    -0.018 & 0.011 &   -0.033 & 0.011 \\
$\alpha_{224}$ & -3.620 &      -0.016 & 0.011 &    -0.004 & 0.011 &    -0.002 & 0.011 &   -0.019 & 0.011 \\
$\alpha_{225}$ & -3.550 &      -0.013 & 0.011 &    -0.001 & 0.012 &     0.001 & 0.011 &   -0.018 & 0.011 \\
$\alpha_{226}$ & -3.480 &      -0.007 & 0.011 &     0.006 & 0.011 &     0.007 & 0.011 &   -0.013 & 0.011 \\
$\alpha_{227}$ & -3.410 &      -0.000 & 0.011 &     0.013 & 0.011 &     0.014 & 0.011 &   -0.007 & 0.011 \\
$\alpha_{228}$ & -3.340 &      -0.015 & 0.011 &    -0.001 & 0.012 &    -0.000 & 0.011 &   -0.023 & 0.011 \\
$\alpha_{229}$ & -3.270 &       0.009 & 0.012 &     0.023 & 0.012 &     0.023 & 0.012 &   -0.002 & 0.011 \\
$\alpha_{230}$ & -3.200 &       0.004 & 0.011 &     0.019 & 0.011 &     0.019 & 0.011 &   -0.008 & 0.011 \\
$\alpha_{231}$ & -3.130 &      -0.005 & 0.011 &     0.011 & 0.012 &     0.010 & 0.011 &   -0.018 & 0.011 \\
$\alpha_{232}$ & -3.060 &      -0.014 & 0.012 &     0.002 & 0.012 &     0.001 & 0.012 &   -0.030 & 0.011 \\
$\alpha_{233}$ & -2.990 &       0.003 & 0.012 &     0.020 & 0.012 &     0.018 & 0.012 &   -0.015 & 0.011 \\
$\alpha_{234}$ & -2.920 &      -0.003 & 0.012 &     0.015 & 0.012 &     0.013 & 0.012 &   -0.023 & 0.012 \\
$\alpha_{235}$ & -2.850 &      -0.002 & 0.012 &     0.016 & 0.012 &     0.013 & 0.012 &   -0.025 & 0.012 \\
$\alpha_{236}$ & -2.780 &       0.005 & 0.012 &     0.023 & 0.013 &     0.020 & 0.013 &   -0.022 & 0.012 \\
$\alpha_{237}$ & -2.710 &       0.015 & 0.012 &     0.035 & 0.012 &     0.031 & 0.012 &   -0.014 & 0.012 \\
$\alpha_{238}$ & -2.640 &       0.025 & 0.014 &     0.045 & 0.014 &     0.041 & 0.014 &   -0.009 & 0.013 \\
$\alpha_{239}$ & -2.570 &      -0.015 & 0.014 &     0.006 & 0.014 &     0.002 & 0.014 &   -0.050 & 0.013 \\
$\alpha_{240}$ & -2.500 &       0.021 & 0.014 &     0.044 & 0.014 &     0.038 & 0.014 &   -0.020 & 0.013 \\
$\alpha_{241}$ & -2.430 &       0.015 & 0.015 &     0.039 & 0.015 &     0.032 & 0.015 &   -0.030 & 0.014 \\
$\alpha_{242}$ & -2.360 &       0.026 & 0.015 &     0.050 & 0.015 &     0.043 & 0.015 &   -0.025 & 0.014 \\
$\alpha_{243}$ & -2.290 &       0.040 & 0.015 &     0.066 & 0.015 &     0.058 & 0.015 &   -0.016 & 0.014 \\
$\alpha_{244}$ & -2.220 &       0.032 & 0.016 &     0.059 & 0.017 &     0.050 & 0.016 &   -0.030 & 0.015 \\
$\alpha_{245}$ & -2.150 &       0.014 & 0.018 &     0.041 & 0.018 &     0.032 & 0.018 &   -0.053 & 0.016 \\
$\alpha_{246}$ & -2.080 &       0.010 & 0.019 &     0.039 & 0.020 &     0.028 & 0.019 &   -0.064 & 0.018 \\
$\alpha_{247}$ & -2.010 &       0.050 & 0.020 &     0.081 & 0.020 &     0.069 & 0.020 &   -0.035 & 0.018 \\
$\alpha_{248}$ & -1.940 &       0.064 & 0.020 &     0.097 & 0.020 &     0.084 & 0.020 &   -0.029 & 0.017 \\
$\alpha_{249}$ & -1.870 &       0.115 & 0.020 &     0.150 & 0.021 &     0.135 & 0.021 &    0.007 & 0.018 \\
$\alpha_{250}$ & -1.800 &       0.170 & 0.021 &     0.208 & 0.021 &     0.192 & 0.021 &    0.045 & 0.018 \\
\hline
$\beta_{21}$   &  0.000 &       0.000 & 0.005 &     0.000 & 0.005 &     0.000 & 0.005 &    0.000 & 0.005 \\
$\beta_{22}$   & -0.549 &      -0.001 & 0.005 &    -0.005 & 0.005 &    -0.008 & 0.005 &   -0.008 & 0.005 \\
$\beta_{23}$   & -0.693 &       0.008 & 0.005 &     0.003 & 0.005 &    -0.001 & 0.005 &   -0.001 & 0.005 \\
$\beta_{24}$   & -0.549 &       0.002 & 0.005 &    -0.002 & 0.005 &    -0.006 & 0.005 &   -0.006 & 0.005 \\
$\beta_{25}$   & -0.347 &       0.006 & 0.005 &     0.004 & 0.006 &     0.001 & 0.006 &    0.001 & 0.006 \\
\bottomrule
\end{tabular}
\end{table}

\clearpage

\begin{table}[h!]
    \centering
    \caption{Simulation results of SIS-L using the first step of the proposed two-step approach. Results include the mean and standard error of the chosen regularization parameters $ \log \eta_1$ and $\log \eta_2$. }
    \label{tab:psis_etas}
\begin{tabular}{lrrrrrr}
\toprule
$\rho$ & \multicolumn{2}{c}{0.0} & \multicolumn{2}{c}{0.5} & \multicolumn{2}{c}{0.9} \\
{} &  Mean &     SE &  Mean &     SE &  Mean &     SE \\
\midrule
$\eta_1$ & -4.323 &  1.262 & -5.735 &  1.120 & -7.285 &  1.142 \\
$\eta_2$ & -4.505 &  1.711 & -5.140 &  0.714 & -6.540 &  1.002 \\
\bottomrule
\end{tabular}
\end{table}


\begin{table}[h!]
\setlength{\tabcolsep}{3pt}
    \centering
    \caption{Simulation results of SIS and SIS-L procedures using the first step of the proposed two-step approach. Results  include mean and SE of the selected-model size (Size), false positive (FP), and false negative (FN). }
    \label{tab:psis_fp_fn}
\begin{tabular}{llrrrrrrrrrrrr}
\toprule
      &     & \multicolumn{2}{c}{Size} & \multicolumn{2}{c}{$\mbox{FP}_1$} & \multicolumn{2}{c}{$\mbox{FN}_1$} & \multicolumn{2}{c}{Size} & \multicolumn{2}{c}{$\mbox{FP}_2$} & \multicolumn{2}{c}{$\mbox{FN}_2$} \\
      &     &              Mean &    SE &          Mean &    SE &          Mean &    SE &              Mean &    SE &          Mean &    SE &          Mean &    SE \\
{} & $\rho$ &                   &       &               &       &               &       &                   &       &               &       &               &       \\
\midrule
SIS & 0.0 &              5.56 &  0.88 &          0.56 &  0.88 &          0.00 &  0.00 &              5.49 &  0.95 &          0.52 &  0.93 &          0.02 &  0.14 \\
      & 0.5 &              5.50 &  1.50 &          0.78 &  1.33 &          0.28 &  0.51 &              6.56 &  1.05 &          1.67 &  0.96 &          0.11 &  0.31 \\
      & 0.9 &             10.84 &  3.82 &          6.06 &  3.58 &          0.22 &  0.60 &             14.03 &  2.98 &         10.47 &  2.76 &          1.44 &  0.50 \\
\midrule
SIS-L & 0.0 &              4.29 &  2.40 &          0.40 &  0.71 &          1.11 &  2.09 &              4.25 &  2.43 &          0.38 &  0.82 &          1.13 &  2.08 \\
      & 0.5 &              5.48 &  1.51 &          0.76 &  1.33 &          0.28 &  0.51 &              5.69 &  1.09 &          0.80 &  0.99 &          0.11 &  0.31 \\
      & 0.9 &              8.02 &  3.12 &          3.34 &  2.80 &          0.32 &  0.86 &              7.21 &  3.18 &          3.83 &  2.87 &          1.62 &  0.81 \\
\bottomrule
\end{tabular}
\end{table}


\begin{table}[h!]
\setlength{\tabcolsep}{3pt}
    \centering
    \caption{Simulation results of SIS and SIS-L procedures using the proposed two-step approach. Results include mean and SE of selected models metrics: AUC, BS, $\mbox{AUC}_1$, $\mbox{AUC}_2$, $\mbox{BS}_1$, and $\mbox{BS}_1$. }
    \label{tab:psis_metrics}
\begin{tabular}{llcccccccccccc}
\toprule
    &        & \multicolumn{2}{c}{AUC} & \multicolumn{2}{c}{BS} & \multicolumn{2}{c}{$\mbox{AUC}_1$} & \multicolumn{2}{c}{$\mbox{AUC}_2$} & \multicolumn{2}{c}{$\mbox{BS}_1$} & \multicolumn{2}{c}{$\mbox{BS}_2$} \\
 $\rho$   &        &   Mean &     SE &   Mean &     SE &           Mean &     SE &           Mean &     SE &          Mean &     SE &          Mean &     SE \\
\midrule
0.0 & SIS &  0.786 &  0.002 &  0.082 &  0.000 &          0.789 &  0.002 &          0.784 &  0.002 &         0.080 &  0.001 &         0.083 &  0.000 \\
    & SIS-L &  0.789 &  0.002 &  0.084 &  0.000 &          0.792 &  0.002 &          0.786 &  0.001 &         0.083 &  0.001 &         0.085 &  0.000 \\
\midrule
0.5 & SIS &  0.774 &  0.001 &  0.082 &  0.001 &          0.755 &  0.001 &          0.793 &  0.003 &         0.081 &  0.000 &         0.082 &  0.001 \\
    & SIS-L &  0.780 &  0.001 &  0.082 &  0.001 &          0.760 &  0.001 &          0.799 &  0.003 &         0.081 &  0.000 &         0.082 &  0.001 \\
\midrule
0.9 & SIS &  0.689 &  0.002 &  0.073 &  0.000 &          0.658 &  0.002 &          0.717 &  0.002 &         0.071 &  0.001 &         0.074 &  0.000 \\
    & SIS-L &  0.708 &  0.002 &  0.073 &  0.000 &          0.673 &  0.002 &          0.740 &  0.002 &         0.071 &  0.001 &         0.073 &  0.000 \\
\bottomrule
\end{tabular}
\end{table}

\clearpage

\begin{table}[h!]
    \centering
        \caption{Simulation results of two competing events. Results of Lee et al. (2018) include mean and estimated standard error (Est SE). Results of the proposed two-step approach include mean, estimated SE, empirical SE (Emp SE) and empirical coverage rate (CR) of 95\% Wald-type confidence interval.}
    \label{tab:methods_comparison_table_SM}
\begin{tabular}{llrrrrrrr}
\toprule
      &            &  True  & \multicolumn{2}{c}{Lee et al.} & \multicolumn{4}{c}{Two-Step} \\
 $n$     &    $\beta_{jk}$    &  Value      &   Mean & Est SE & Mean & Est SE & Emp SE &  CR \\
\midrule
10,000 & $\beta_{11}$ &  0.223 &      0.226 &        0.067 &    0.224 &        0.066 &        0.059 &         0.985 \\
      & $\beta_{12}$ & -1.099 &     -1.100 &        0.068 &   -1.088 &        0.067 &        0.070 &         0.915 \\
      & $\beta_{13}$ & -1.099 &     -1.097 &        0.068 &   -1.086 &        0.067 &        0.065 &         0.960 \\
      & $\beta_{14}$ & -0.916 &     -0.922 &        0.067 &   -0.912 &        0.067 &        0.064 &         0.945 \\
      & $\beta_{15}$ & -0.693 &     -0.688 &        0.067 &   -0.681 &        0.066 &        0.064 &         0.955 \\
      & $\beta_{21}$ & -0.000 &      0.007 &        0.085 &    0.007 &        0.085 &        0.085 &         0.955 \\
      & $\beta_{22}$ & -1.099 &     -1.107 &        0.087 &   -1.100 &        0.087 &        0.091 &         0.930 \\
      & $\beta_{23}$ & -1.386 &     -1.389 &        0.088 &   -1.380 &        0.088 &        0.084 &         0.965 \\
      & $\beta_{24}$ & -1.099 &     -1.085 &        0.087 &   -1.077 &        0.087 &        0.080 &         0.965 \\
      & $\beta_{25}$ & -0.693 &     -0.690 &        0.086 &   -0.685 &        0.085 &        0.084 &         0.945 \\
\hline
15,000 & $\beta_{11}$ &  0.223 &      0.225 &        0.054 &    0.222 &        0.054 &        0.057 &         0.945 \\
      & $\beta_{12}$ & -1.099 &     -1.103 &        0.055 &   -1.092 &        0.055 &        0.059 &         0.940 \\
      & $\beta_{13}$ & -1.099 &     -1.100 &        0.055 &   -1.089 &        0.055 &        0.054 &         0.950 \\
      & $\beta_{14}$ & -0.916 &     -0.925 &        0.055 &   -0.916 &        0.054 &        0.051 &         0.970 \\
      & $\beta_{15}$ & -0.693 &     -0.686 &        0.055 &   -0.679 &        0.054 &        0.052 &         0.950 \\
      & $\beta_{21}$ & -0.000 &     -0.000 &        0.070 &   -0.000 &        0.069 &        0.066 &         0.955 \\
      & $\beta_{22}$ & -1.099 &     -1.092 &        0.071 &   -1.085 &        0.071 &        0.064 &         0.955 \\
      & $\beta_{23}$ & -1.386 &     -1.381 &        0.072 &   -1.372 &        0.072 &        0.067 &         0.955 \\
      & $\beta_{24}$ & -1.099 &     -1.101 &        0.071 &   -1.094 &        0.071 &        0.076 &         0.940 \\
      & $\beta_{25}$ & -0.693 &     -0.689 &        0.070 &   -0.684 &        0.070 &        0.071 &         0.945 \\
\bottomrule
\end{tabular}
\end{table}

\clearpage

\begin{table}[h!]
    \centering
            \caption{Simulation results of three competing events. Results of Lee et al. include mean and estimated standard error (Est SE). Results of the proposed two-step approach include mean, estimated SE, empirical SE (Emp SE) and empirical coverage rate (CR) of 95\% Wald-type confidence interval.}
             \label{tab:methods_comparison_table_j3}
\begin{tabular}{llrrrrrrr}
\toprule
      &            &  True  & \multicolumn{2}{c}{Lee et al.} & \multicolumn{4}{c}{Two-Step} \\
   $n$  &    $\beta_{jk}$    &  Value      &   Mean & Est SE & Mean & Est SE & Emp SE &  CR \\
      \midrule
5,000  & $\beta_{11}$ & -0.916 &     -0.929 &        0.098 &   -0.917 &        0.097 &        0.090 &         0.965 \\
      & $\beta_{12}$ & -0.405 &     -0.410 &        0.097 &   -0.404 &        0.096 &        0.097 &         0.950 \\
      & $\beta_{13}$ &  0.223 &      0.224 &        0.098 &    0.221 &        0.097 &        0.098 &         0.935 \\
      & $\beta_{14}$ & -1.099 &     -1.110 &        0.098 &   -1.095 &        0.097 &        0.110 &         0.900 \\
      & $\beta_{15}$ & -0.693 &     -0.697 &        0.098 &   -0.687 &        0.096 &        0.099 &         0.935 \\
      & $\beta_{21}$ &  0.223 &      0.246 &        0.107 &    0.243 &        0.105 &        0.109 &         0.945 \\
      & $\beta_{22}$ & -1.099 &     -1.094 &        0.108 &   -1.081 &        0.107 &        0.105 &         0.945 \\
      & $\beta_{23}$ & -1.030 &     -1.035 &        0.107 &   -1.023 &        0.106 &        0.111 &         0.940 \\
      & $\beta_{24}$ & -0.788 &     -0.783 &        0.107 &   -0.773 &        0.105 &        0.103 &         0.950 \\
      & $\beta_{25}$ & -0.405 &     -0.416 &        0.107 &   -0.411 &        0.105 &        0.095 &         0.965 \\
      & $\beta_{31}$ & -0.588 &     -0.592 &        0.099 &   -0.584 &        0.098 &        0.106 &         0.940 \\
      & $\beta_{32}$ &  0.223 &      0.237 &        0.099 &    0.234 &        0.098 &        0.098 &         0.970 \\
      & $\beta_{33}$ & -0.916 &     -0.925 &        0.100 &   -0.913 &        0.099 &        0.107 &         0.930 \\
      & $\beta_{34}$ & -0.182 &     -0.183 &        0.099 &   -0.180 &        0.098 &        0.110 &         0.905 \\
      & $\beta_{35}$ & -1.099 &     -1.108 &        0.100 &   -1.094 &        0.099 &        0.099 &         0.945 \\
\hline
10,000 & $\beta_{11}$ & -0.916 &     -0.922 &        0.069 &   -0.909 &        0.068 &        0.067 &         0.965 \\
      & $\beta_{12}$ & -0.405 &     -0.412 &        0.068 &   -0.406 &        0.068 &        0.067 &         0.950 \\
      & $\beta_{13}$ &  0.223 &      0.221 &        0.069 &    0.218 &        0.068 &        0.064 &         0.955 \\
      & $\beta_{14}$ & -1.099 &     -1.093 &        0.069 &   -1.078 &        0.068 &        0.067 &         0.925 \\
      & $\beta_{15}$ & -0.693 &     -0.691 &        0.069 &   -0.681 &        0.068 &        0.070 &         0.940 \\
      & $\beta_{21}$ &  0.223 &      0.222 &        0.075 &    0.220 &        0.075 &        0.080 &         0.945 \\
      & $\beta_{22}$ & -1.099 &     -1.104 &        0.076 &   -1.091 &        0.075 &        0.076 &         0.965 \\
      & $\beta_{23}$ & -1.030 &     -1.030 &        0.076 &   -1.019 &        0.075 &        0.074 &         0.940 \\
      & $\beta_{24}$ & -0.788 &     -0.789 &        0.075 &   -0.780 &        0.075 &        0.070 &         0.950 \\
      & $\beta_{25}$ & -0.405 &     -0.412 &        0.075 &   -0.407 &        0.074 &        0.077 &         0.920 \\
      & $\beta_{31}$ & -0.588 &     -0.587 &        0.070 &   -0.579 &        0.069 &        0.069 &         0.945 \\
      & $\beta_{32}$ &  0.223 &      0.231 &        0.070 &    0.228 &        0.069 &        0.067 &         0.945 \\
      & $\beta_{33}$ & -0.916 &     -0.923 &        0.071 &   -0.912 &        0.070 &        0.065 &         0.965 \\
      & $\beta_{34}$ & -0.182 &     -0.175 &        0.070 &   -0.173 &        0.069 &        0.069 &         0.940 \\
      & $\beta_{35}$ & -1.099 &     -1.116 &        0.071 &   -1.102 &        0.070 &        0.069 &         0.955 \\
\bottomrule
\end{tabular}
\end{table}

\clearpage

\begin{table}[h!]
    \centering
            \caption{Simulation results of three competing events. Results of Lee et al. include mean and estimated standard error (Est SE). Results of the proposed two-step approach include mean, estimated SE, empirical SE (Emp SE) and empirical coverage rate (CR) of 95\% Wald-type confidence interval.}
    \label{tab:methods_comparison_table_j3b}
\begin{tabular}{llrrrrrrr}
\toprule
      &            &  True  & \multicolumn{2}{c}{Lee et al.} & \multicolumn{4}{c}{Two-Step} \\
  $n$    &    $\beta_{jk}$    &  Value      &   Mean & Est SE & Mean & Est SE & Emp SE &  CR \\
\midrule
15,000 & $\beta_{11}$ & -0.916 &     -0.921 &        0.057 &   -0.909 &        0.056 &        0.059 &         0.940 \\
      & $\beta_{12}$ & -0.405 &     -0.407 &        0.056 &   -0.402 &        0.055 &        0.055 &         0.945 \\
      & $\beta_{13}$ &  0.223 &      0.219 &        0.056 &    0.216 &        0.056 &        0.055 &         0.940 \\
      & $\beta_{14}$ & -1.099 &     -1.099 &        0.057 &   -1.085 &        0.056 &        0.055 &         0.960 \\
      & $\beta_{15}$ & -0.693 &     -0.706 &        0.056 &   -0.697 &        0.055 &        0.053 &         0.945 \\
      & $\beta_{21}$ &  0.223 &      0.224 &        0.062 &    0.222 &        0.061 &        0.061 &         0.940 \\
      & $\beta_{22}$ & -1.099 &     -1.095 &        0.062 &   -1.083 &        0.062 &        0.064 &         0.935 \\
      & $\beta_{23}$ & -1.030 &     -1.029 &        0.062 &   -1.017 &        0.061 &        0.059 &         0.950 \\
      & $\beta_{24}$ & -0.788 &     -0.789 &        0.062 &   -0.780 &        0.061 &        0.062 &         0.950 \\
      & $\beta_{25}$ & -0.405 &     -0.413 &        0.061 &   -0.408 &        0.061 &        0.064 &         0.935 \\
      & $\beta_{31}$ & -0.588 &     -0.587 &        0.057 &   -0.580 &        0.056 &        0.053 &         0.955 \\
      & $\beta_{32}$ &  0.223 &      0.220 &        0.057 &    0.217 &        0.057 &        0.057 &         0.945 \\
      & $\beta_{33}$ & -0.916 &     -0.916 &        0.058 &   -0.905 &        0.057 &        0.057 &         0.935 \\
      & $\beta_{34}$ & -0.182 &     -0.179 &        0.057 &   -0.177 &        0.057 &        0.055 &         0.955 \\
      & $\beta_{35}$ & -1.099 &     -1.102 &        0.058 &   -1.088 &        0.057 &        0.056 &         0.950 \\
\hline
20,000 & $\beta_{11}$ & -0.916 &     -0.917 &        0.049 &   -0.905 &        0.048 &        0.047 &         0.945 \\
      & $\beta_{12}$ & -0.405 &     -0.407 &        0.048 &   -0.402 &        0.048 &        0.047 &         0.960 \\
      & $\beta_{13}$ &  0.223 &      0.226 &        0.049 &    0.223 &        0.048 &        0.045 &         0.960 \\
      & $\beta_{14}$ & -1.099 &     -1.098 &        0.049 &   -1.084 &        0.048 &        0.048 &         0.940 \\
      & $\beta_{15}$ & -0.693 &     -0.702 &        0.049 &   -0.692 &        0.048 &        0.047 &         0.955 \\
      & $\beta_{21}$ &  0.223 &      0.220 &        0.053 &    0.218 &        0.053 &        0.049 &         0.975 \\
      & $\beta_{22}$ & -1.099 &     -1.090 &        0.054 &   -1.078 &        0.053 &        0.049 &         0.950 \\
      & $\beta_{23}$ & -1.030 &     -1.032 &        0.054 &   -1.020 &        0.053 &        0.053 &         0.950 \\
      & $\beta_{24}$ & -0.788 &     -0.789 &        0.053 &   -0.780 &        0.053 &        0.050 &         0.980 \\
      & $\beta_{25}$ & -0.405 &     -0.400 &        0.053 &   -0.395 &        0.053 &        0.051 &         0.970 \\
      & $\beta_{31}$ & -0.588 &     -0.588 &        0.049 &   -0.581 &        0.049 &        0.049 &         0.960 \\
      & $\beta_{32}$ &  0.223 &      0.230 &        0.050 &    0.228 &        0.049 &        0.052 &         0.950 \\
      & $\beta_{33}$ & -0.916 &     -0.910 &        0.050 &   -0.898 &        0.049 &        0.050 &         0.935 \\
      & $\beta_{34}$ & -0.182 &     -0.181 &        0.050 &   -0.178 &        0.049 &        0.050 &         0.940 \\
      & $\beta_{35}$ & -1.099 &     -1.103 &        0.050 &   -1.090 &        0.049 &        0.046 &         0.950 \\
\bottomrule
\end{tabular}
\end{table}

\clearpage

\begin{table}[h!]
\setlength{\tabcolsep}{3pt}
\centering
\caption{MIMIC dataset. Summary of covariates of overall sample, among censored observations, and by event type: in-hospital death (Death), discharged to another medical facility (Another Facility), and discharge to home (Home). }
\label{table:los-tableone}
\small
\begin{tabular}{llccccc}
\toprule
                          &      &  &  &\multicolumn{3}{c}{Event Type} \\
                          &      &                       Overall &     Censored &          Death & Another Facility &           Home \\
\midrule
$n$ &  &                         25170 &          894 &          1540 &              5379 &          17357 \\
Sex  (\%) & Female &                  12291 (48.8) &   373 (41.7) &    695 (45.1) &       2865 (53.3) &    8358 (48.2) \\
                          & Male &                  12879 (51.2) &   521 (58.3) &    845 (54.9) &       2514 (46.7) &    8999 (51.8) \\
Age, mean (SD) &      &                   64.1 (17.9) &  58.4 (16.5) &   72.7 (14.5) &       73.3 (15.7) &    60.8 (17.6) \\
Race  (\%) & Asian &                    1035 (4.1) &     27 (3.0) &      76 (4.9) &         165 (3.1) &      767 (4.4) \\
                          & Black &                   3543 (14.1) &   154 (17.2) &    197 (12.8) &        741 (13.8) &    2451 (14.1) \\
                          & Hispanic &                    1326 (5.3) &     53 (5.9) &      53 (3.4) &         180 (3.3) &     1040 (6.0) \\
                          & White &                  17595 (69.9) &   595 (66.6) &   1072 (69.6) &       3977 (73.9) &   11951 (68.9) \\
                        & Other &                    1671 (6.6) &     65 (7.3) &     142 (9.2) &         316 (5.9) &     1148 (6.6) \\
Insurance (\%) & Medicaid &                    1423 (5.7) &     86 (9.6) &      66 (4.3) &         222 (4.1) &     1049 (6.0) \\
                          & Medicare &                  10609 (42.1) &   316 (35.3) &    843 (54.7) &       3253 (60.5) &    6197 (35.7) \\
                          & Other &                  13138 (52.2) &   492 (55.0) &    631 (41.0) &       1904 (35.4) &   10111 (58.3) \\
Marital Status (\%) & Divorced &                    2043 (8.1) &    94 (10.5) &     121 (7.9) &         464 (8.6) &     1364 (7.9) \\
                          & Married &                  11289 (44.9) &   329 (36.8) &    751 (48.8) &       1853 (34.4) &    8356 (48.1) \\
                          & Single &                   8414 (33.4) &   403 (45.1) &    386 (25.1) &       1729 (32.1) &    5896 (34.0) \\
                          & Widowed &                   3424 (13.6) &     68 (7.6) &    282 (18.3) &       1333 (24.8) &    1741 (10.0) \\
Direct Emergency (\%) & No &                  22398 (89.0) &   790 (88.4) &   1413 (91.8) &       4924 (91.5) &   15271 (88.0) \\
                          & Yes &                   2772 (11.0) &   104 (11.6) &     127 (8.2) &         455 (8.5) &    2086 (12.0) \\
Night Admission (\%) & No &                  11604 (46.1) &   404 (45.2) &    736 (47.8) &       2414 (44.9) &    8050 (46.4) \\
                          & Yes &                  13566 (53.9) &   490 (54.8) &    804 (52.2) &       2965 (55.1) &    9307 (53.6) \\
\makecell[l]{Previous Admission \\ this Month (\%)} & No &                  23138 (91.9) &   795 (88.9) &   1318 (85.6) &       4821 (89.6) &   16204 (93.4) \\
                          & Yes &                    2032 (8.1) &    99 (11.1) &    222 (14.4) &        558 (10.4) &     1153 (6.6) \\
Admissions Number (\%) & 1 &                  15471 (61.5) &   503 (56.3) &    798 (51.8) &       3005 (55.9) &   11165 (64.3) \\
                          & 2 &                   4121 (16.4) &   151 (16.9) &    283 (18.4) &        926 (17.2) &    2761 (15.9) \\
                          & 3+ &                   5578 (22.2) &   240 (26.8) &    459 (29.8) &       1448 (26.9) &    3431 (19.8) \\
LOS (days), mean (SD) &      &                     7.0 (6.1) &  21.7 (11.6) &     8.5 (6.9) &         9.0 (5.8) &      5.5 (4.3) \\
\bottomrule
\end{tabular}
\end{table}

\clearpage

\begin{table}[h!]
\setlength{\tabcolsep}{3pt}
\centering
\caption{MIMIC dataset. Summary of covariates of overall sample, among censored observations, and by event type: in-hospital death (Death), discharged to another medical facility (Another Facility), and discharge to home (Home). MCH: mean cell hemoglobin. MCHC: mean cell hemoglobin concentration. MCV: mean corpuscular volume. RDW: red blood cell Distribution Width.}
\label{table:los-labtest}
\small
\begin{tabular}{llccccc}
\toprule
                          &      &  &  &\multicolumn{3}{c}{Event Type} \\
                          &      &                       Overall &     Censored &          Death & Another Facility &           Home \\
\midrule
$n$ & {} &                         25170 &          894 &          1540 &              5379 &          17357 \\
Anion Gap (\%) & Abnormal &                    2305 (9.2) &   110 (12.3) &    401 (26.0) &        543 (10.1) &     1251 (7.2) \\
                         & Normal &                  22865 (90.8) &   784 (87.7) &   1139 (74.0) &       4836 (89.9) &   16106 (92.8) \\
Bicarbonate (\%) & Abnormal &                   6135 (24.4) &   300 (33.6) &    832 (54.0) &       1494 (27.8) &    3509 (20.2) \\
                         & Normal &                  19035 (75.6) &   594 (66.4) &    708 (46.0) &       3885 (72.2) &   13848 (79.8) \\
Calcium Total (\%) & Abnormal &                   7326 (29.1) &   365 (40.8) &    756 (49.1) &       1823 (33.9) &    4382 (25.2) \\
                         & Normal &                  17844 (70.9) &   529 (59.2) &    784 (50.9) &       3556 (66.1) &   12975 (74.8) \\
Chloride (\%) & Abnormal &                   4848 (19.3) &   255 (28.5) &    555 (36.0) &       1322 (24.6) &    2716 (15.6) \\
                         & Normal &                  20322 (80.7) &   639 (71.5) &    985 (64.0) &       4057 (75.4) &   14641 (84.4) \\
Creatinine (\%) & Abnormal &                   7124 (28.3) &   323 (36.1) &    893 (58.0) &       1945 (36.2) &    3963 (22.8) \\
                         & Normal &                  18046 (71.7) &   571 (63.9) &    647 (42.0) &       3434 (63.8) &   13394 (77.2) \\
Glucose (\%) & Abnormal &                  16426 (65.3) &   635 (71.0) &   1211 (78.6) &       3674 (68.3) &   10906 (62.8) \\
                         & Normal &                   8744 (34.7) &   259 (29.0) &    329 (21.4) &       1705 (31.7) &    6451 (37.2) \\
Magnesium (\%) & Abnormal &                    2220 (8.8) &    99 (11.1) &    234 (15.2) &         517 (9.6) &     1370 (7.9) \\
                         & Normal &                  22950 (91.2) &   795 (88.9) &   1306 (84.8) &       4862 (90.4) &   15987 (92.1) \\
Phosphate (\%) & Abnormal &                   6962 (27.7) &   313 (35.0) &    663 (43.1) &       1510 (28.1) &    4476 (25.8) \\
                         & Normal &                  18208 (72.3) &   581 (65.0) &    877 (56.9) &       3869 (71.9) &   12881 (74.2) \\
Potassium (\%) & Abnormal &                    2109 (8.4) &   110 (12.3) &    260 (16.9) &         520 (9.7) &     1219 (7.0) \\
                         & Normal &                  23061 (91.6) &   784 (87.7) &   1280 (83.1) &       4859 (90.3) &   16138 (93.0) \\
Sodium (\%) & Abnormal &                   2947 (11.7) &   171 (19.1) &    415 (26.9) &        845 (15.7) &     1516 (8.7) \\
                         & Normal &                  22223 (88.3) &   723 (80.9) &   1125 (73.1) &       4534 (84.3) &   15841 (91.3) \\
Urea Nitrogen (\%) & Abnormal &                  10032 (39.9) &   413 (46.2) &   1059 (68.8) &       2849 (53.0) &    5711 (32.9) \\
                         & Normal &                  15138 (60.1) &   481 (53.8) &    481 (31.2) &       2530 (47.0) &   11646 (67.1) \\
Hematocrit (\%) & Abnormal &                  17319 (68.8) &   691 (77.3) &   1250 (81.2) &       4111 (76.4) &   11267 (64.9) \\
                         & Normal &                   7851 (31.2) &   203 (22.7) &    290 (18.8) &       1268 (23.6) &    6090 (35.1) \\
Hemoglobin (\%) & Abnormal &                  18355 (72.9) &   735 (82.2) &   1319 (85.6) &       4320 (80.3) &   11981 (69.0) \\
                         & Normal &                   6815 (27.1) &   159 (17.8) &    221 (14.4) &       1059 (19.7) &    5376 (31.0) \\
MCH (\%) & Abnormal &                   6559 (26.1) &   306 (34.2) &    454 (29.5) &       1488 (27.7) &    4311 (24.8) \\
                         & Normal &                  18611 (73.9) &   588 (65.8) &   1086 (70.5) &       3891 (72.3) &   13046 (75.2) \\
MCHC (\%) & Abnormal &                   7762 (30.8) &   313 (35.0) &    634 (41.2) &       2033 (37.8) &    4782 (27.6) \\
                         & Normal &                  17408 (69.2) &   581 (65.0) &    906 (58.8) &       3346 (62.2) &   12575 (72.4) \\
MCV (\%) & Abnormal &                   5106 (20.3) &   243 (27.2) &    418 (27.1) &       1229 (22.8) &    3216 (18.5) \\
                         & Normal &                  20064 (79.7) &   651 (72.8) &   1122 (72.9) &       4150 (77.2) &   14141 (81.5) \\
Platelet Count (\%) & Abnormal &                   7280 (28.9) &   364 (40.7) &    688 (44.7) &       1618 (30.1) &    4610 (26.6) \\
                         & Normal &                  17890 (71.1) &   530 (59.3) &    852 (55.3) &       3761 (69.9) &   12747 (73.4) \\
RDW (\%) & Abnormal &                   7280 (28.9) &   377 (42.2) &    870 (56.5) &       2016 (37.5) &    4017 (23.1) \\
                         & Normal &                  17890 (71.1) &   517 (57.8) &    670 (43.5) &       3363 (62.5) &   13340 (76.9) \\
Red Blood Cells (\%) & Abnormal &                  19170 (76.2) &   732 (81.9) &   1341 (87.1) &       4478 (83.2) &   12619 (72.7) \\
                         & Normal &                   6000 (23.8) &   162 (18.1) &    199 (12.9) &        901 (16.8) &    4738 (27.3) \\
White Blood Cells (\%) & Abnormal &                  10013 (39.8) &   466 (52.1) &   1012 (65.7) &       2320 (43.1) &    6215 (35.8) \\
                         & Normal &                  15157 (60.2) &   428 (47.9) &    528 (34.3) &       3059 (56.9) &   11142 (64.2) \\
\bottomrule
\end{tabular}
\end{table}

\clearpage

\clearpage

\clearpage

\clearpage

\clearpage

\begin{figure}[!ht]
    \centering
    \includegraphics[width=0.8\textwidth]{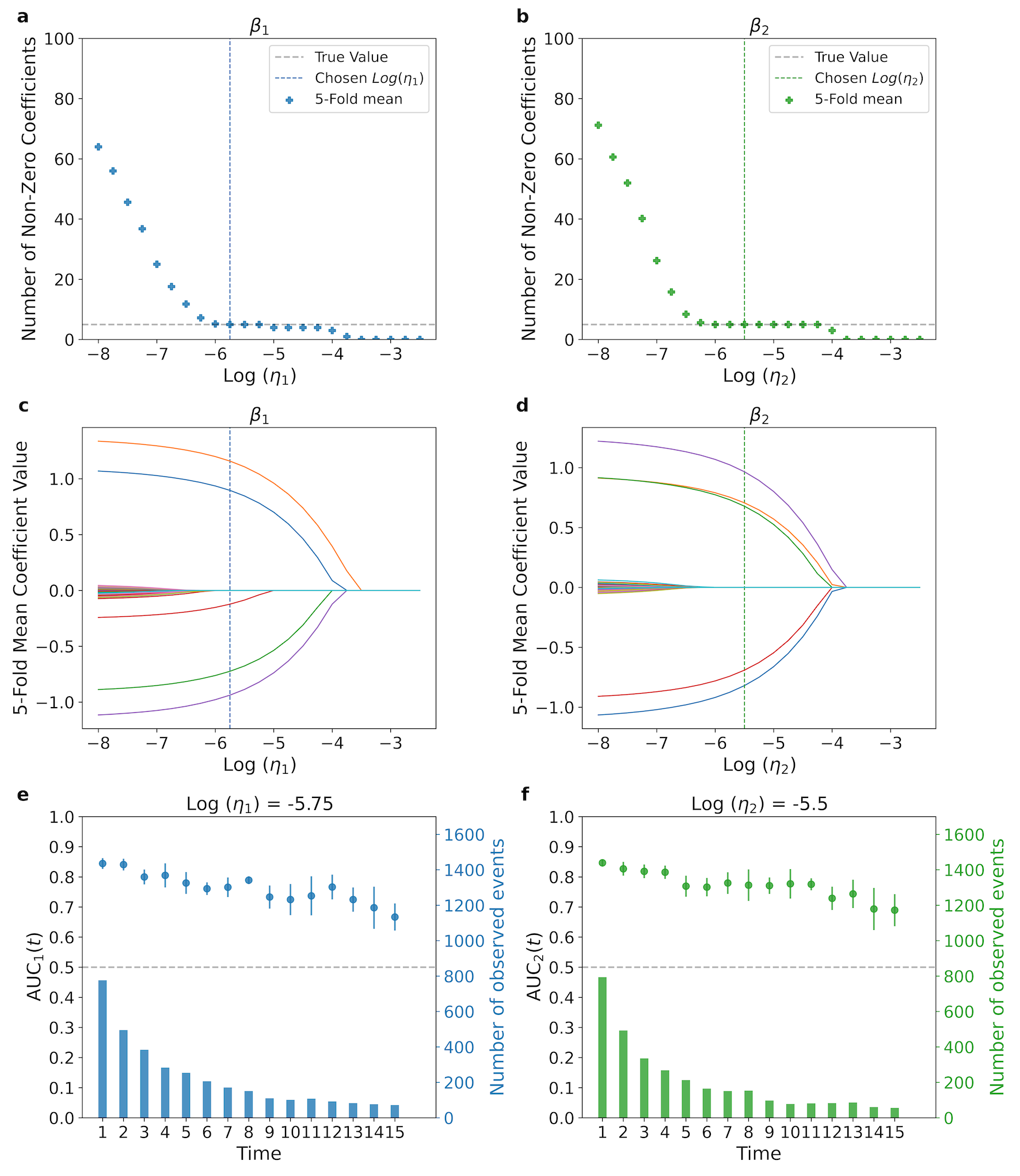}
    \caption{Lasso regularization results of one simulated random sample under setting 11 with independent covariates and five out of 100 coefficients having non-zero values.  Tuning parameters were selected through 5-fold CV, varying $\log \eta_j$ from -8 to -2.5 with a step size of 0.25.  The selected $\log \eta_j$, $j=1,2$, values, denoted by vertical dashed lines on panels \textbf{a-d}, resulted in five non-zero regression coefficients for each $j$. Panels \textbf{a-b} display the number of non-zero coefficients for events 1 and 2, respectively, with the true value (five)  shown as a horizontal dashed line. Evidently, with the chosen values of $\log \eta_j$, the analysis resulted in five non-zero regression coefficients for each $j$. The estimates of $\beta_j$ as a function of $\log \eta_j$ are depicted in panels \textbf{c} and {\bf d}. Each curve corresponds to a variable, and at the chosen $\log \eta_j$ values, each $\beta_j$ is reasonably close to its true value.  Panels \textbf{e-f} show the mean (and SD bars) of the 5-fold $\widehat{\mbox{AUC}}_1(t)$ and $\widehat{\mbox{AUC}}_2(t)$, respectively, under the selected $\log \eta_j$ values,  along with the number of observed events of event type $j$ in bars.  }  
    \label{fig:reg_sim_results}
\end{figure}

\clearpage

\begin{figure}[!ht]
    \centering
    \includegraphics[width=0.8\textwidth]{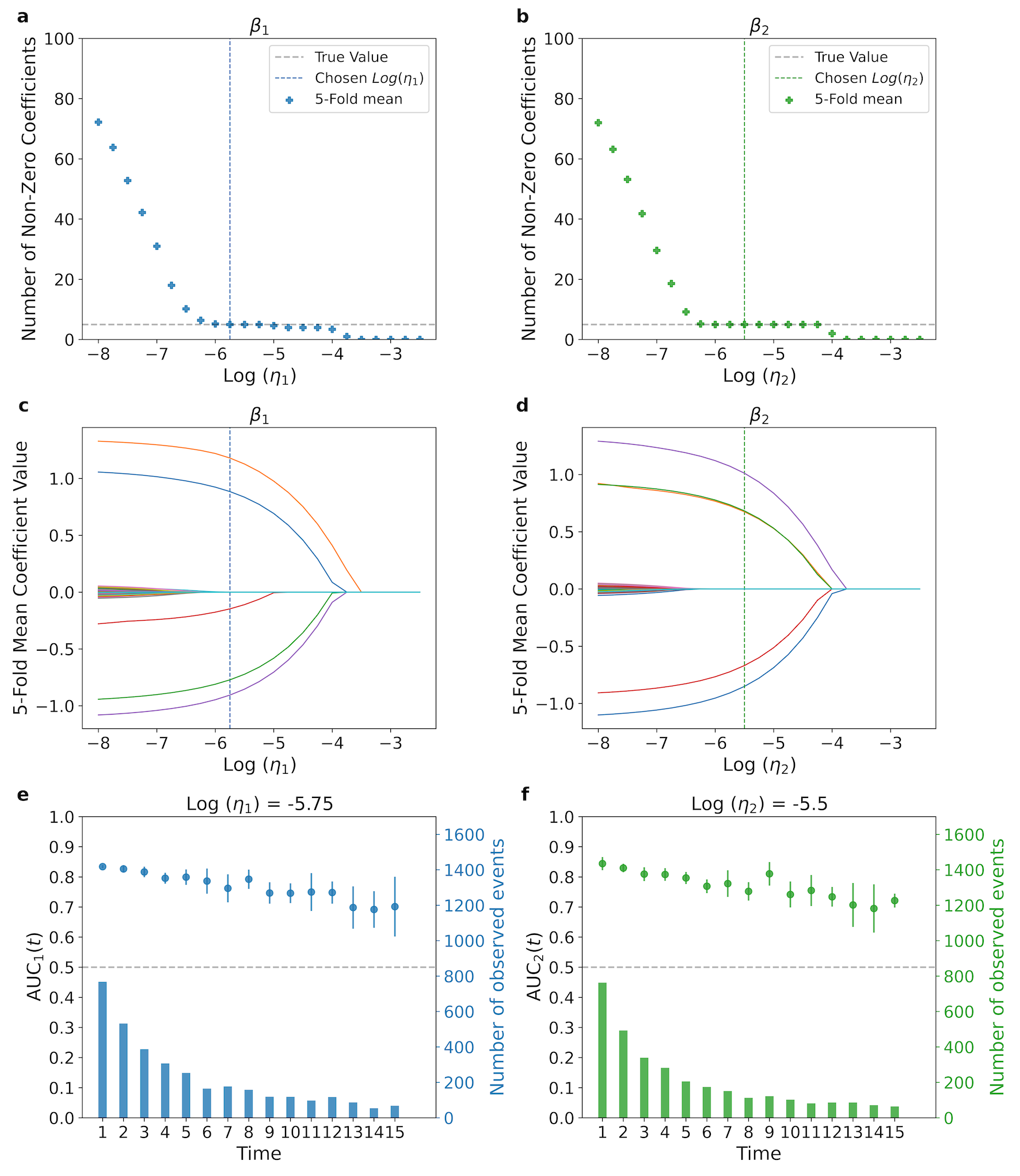}
    \caption{Lasso regularization results of one simulated random sample under setting  12 with dependent covariates, and five out of 100 coefficients having non-zero values.   Tuning parameters were selected through 5-fold cross-validation, varying $\log \eta_j$ from -8 to -2.5 with a step size of 0.25 (on a $\log$ for presentation ease).  The chosen $\log \eta_j$, $j=1,2$, values, denoted by vertical dashed lines on panels \textbf{a-d}, resulted in five non-zero regression coefficients for each $j$. Panels \textbf{a-b} display the number of non-zero coefficients for events 1 and 2, respectively, with the true value (five)  shown as a horizontal dashed line. Evidently, with the chosen values of $\log \eta_j$, the analysis resulted in five non-zero regression coefficients for each $j$. The estimates of $\beta_j$ as a function of $\log \eta_j$ are depicted in panels \textbf{c} and {\bf d}. Each curve corresponds to a variable, and at the chosen $\log \eta_j$ values, each $\beta_j$ is reasonably close to its true value.  Panels \textbf{e-f} show the mean (and SD bars) of the 5-fold $\widehat{\mbox{AUC}}_1(t)$ and $\widehat{\mbox{AUC}}_2(t)$, respectively, under the selected $\log \eta_j$ values,  along with the number of observed events of event type $j$ in bars. }
    \label{fig:reg_corr_sim_results}
\end{figure}

\clearpage

\begin{figure}[!ht]
    \centering
    \includegraphics[width=0.8\textwidth]{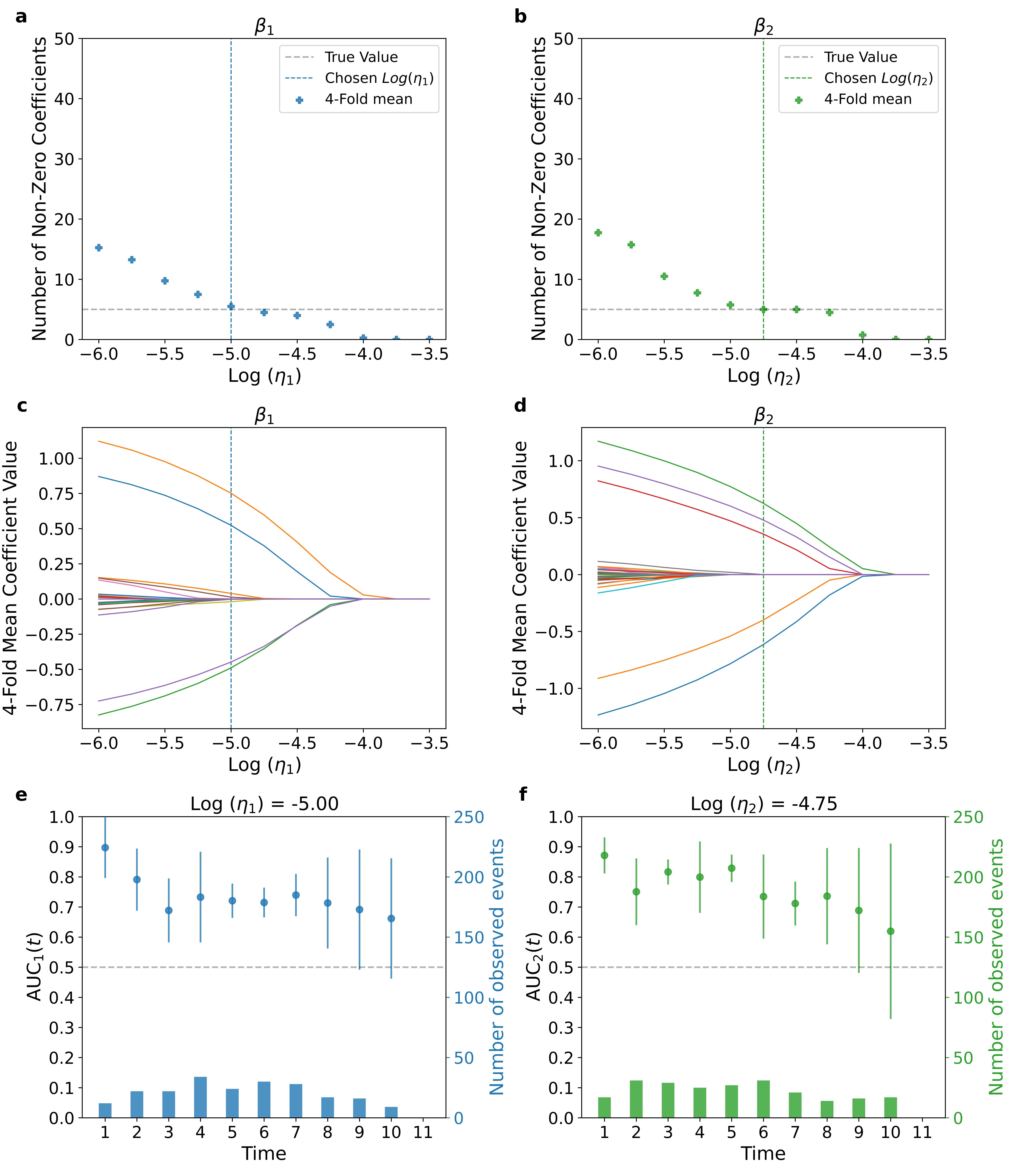}
    \caption{Lasso regularization results of one simulated random sample under setting  14 with independent covariates and five out of 35 coefficients having non-zero values.  Tuning parameters were selected through 4-fold CV, varying $\log \eta_j$ from -6 to -3.5 with a step size of 0.25.  The selected $\log \eta_j$, $j=1,2$, values are denoted by vertical dashed lines on panels \textbf{a-d}. Panels \textbf{a-b} display the mean number of non-zero coefficients for events 1 and 2, respectively, across the 4-folds, with the true value (five) shown as a horizontal dashed line. 
    Evidently, with the chosen values of $\log \eta_j$, the analysis resulted in a mean (across folds) of 5.5 and 5 non-zero regression coefficients for $j=1,2$, respectively. 
    The estimates of $\beta_j$ as a function of $\log \eta_j$ are depicted in panels \textbf{c} and {\bf d}. Each curve corresponds to a variable, and at the chosen $\log \eta_j$ values, each $\beta_j$ is reasonably close to its true value.  Panels \textbf{e-f} show the mean (and SD bars) of the 4-fold $\widehat{\mbox{AUC}}_1(t)$ and $\widehat{\mbox{AUC}}_2(t)$, respectively, under the selected $\log \eta_j$ values,  along with the number of observed events of event type $j$ in bars.  }  
    \label{fig:reg_sim_results_small_sample}
\end{figure}

\clearpage

\begin{figure}[!ht]
    \centering
    \includegraphics[width=0.8\textwidth]{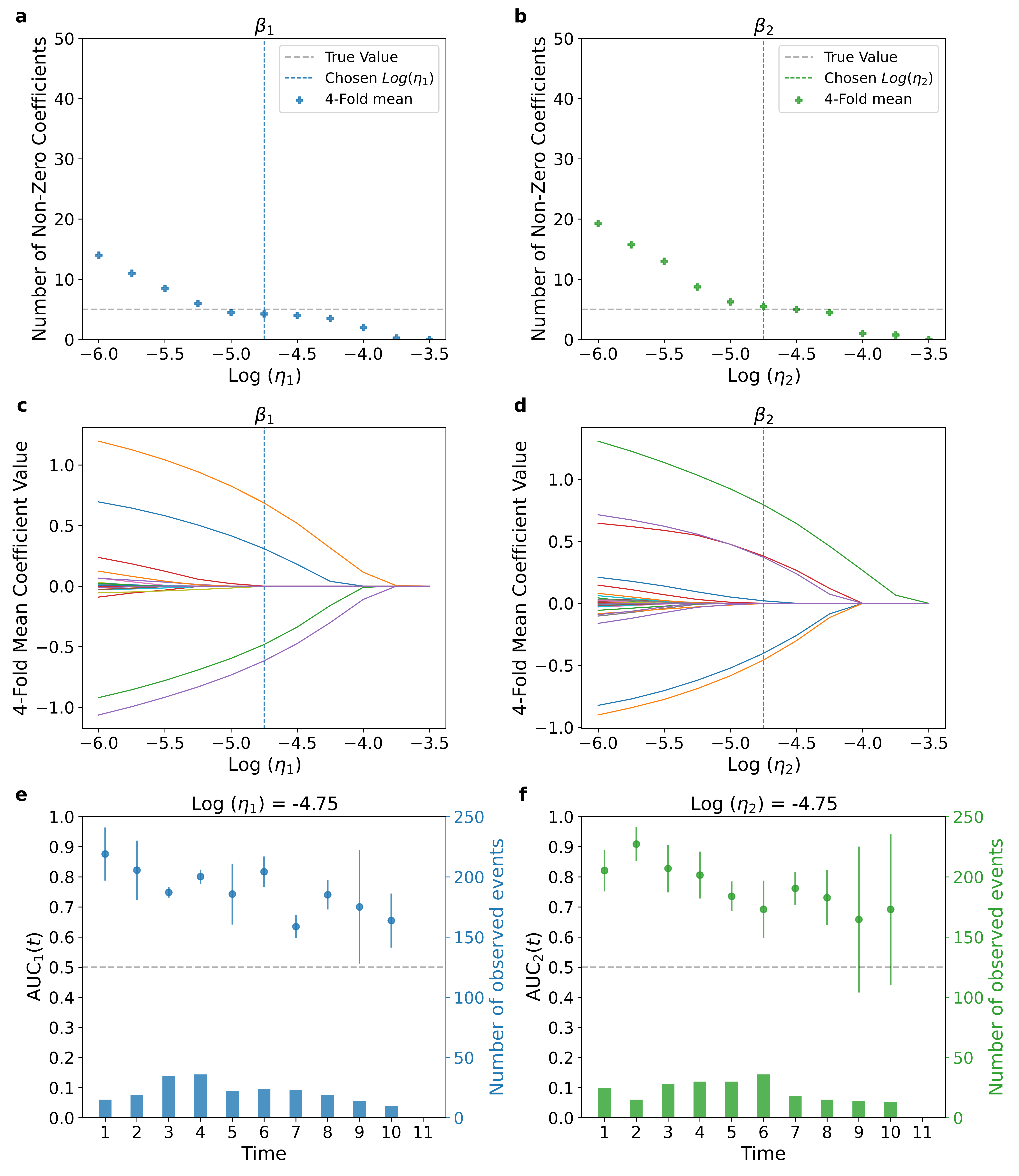}
    \caption{Lasso regularization results of one simulated random sample under setting  15 with dependent covariates and five out of 35 coefficients having non-zero values. Tuning parameters were selected through 4-fold CV, varying $\log \eta_j$ from -6 to -3.5 with a step size of 0.25.  The selected $\log \eta_j$, $j=1,2$, values are denoted by vertical dashed lines on panels \textbf{a-d}. Panels \textbf{a-b} display the number of non-zero coefficients for events 1 and 2, respectively, with the true value (five)  shown as a horizontal dashed line. 
    Evidently, with the chosen values of $\log \eta_j$, the analysis resulted in a mean (across folds) of 4.5 and 5.5 non-zero regression coefficients for $j=1,2$, respectively. 
    The estimates of $\beta_j$ as a function of $\log \eta_j$ are depicted in panels \textbf{c} and {\bf d}. Each curve corresponds to a variable, and at the chosen $\log \eta_j$ values, each $\beta_j$ is reasonably close to its true value.  Panels \textbf{e-f} show the mean (and SD bars) of the 4-fold $\widehat{\mbox{AUC}}_1(t)$ and $\widehat{\mbox{AUC}}_2(t)$, respectively, under the selected $\log \eta_j$ values,  along with the number of observed events of event type $j$ in bars.  }  
    \label{fig:reg_sim_results_small_sample_corr}
\end{figure}

\clearpage

\begin{figure}[!ht]
    \centering
    \includegraphics[width=0.8\textwidth]{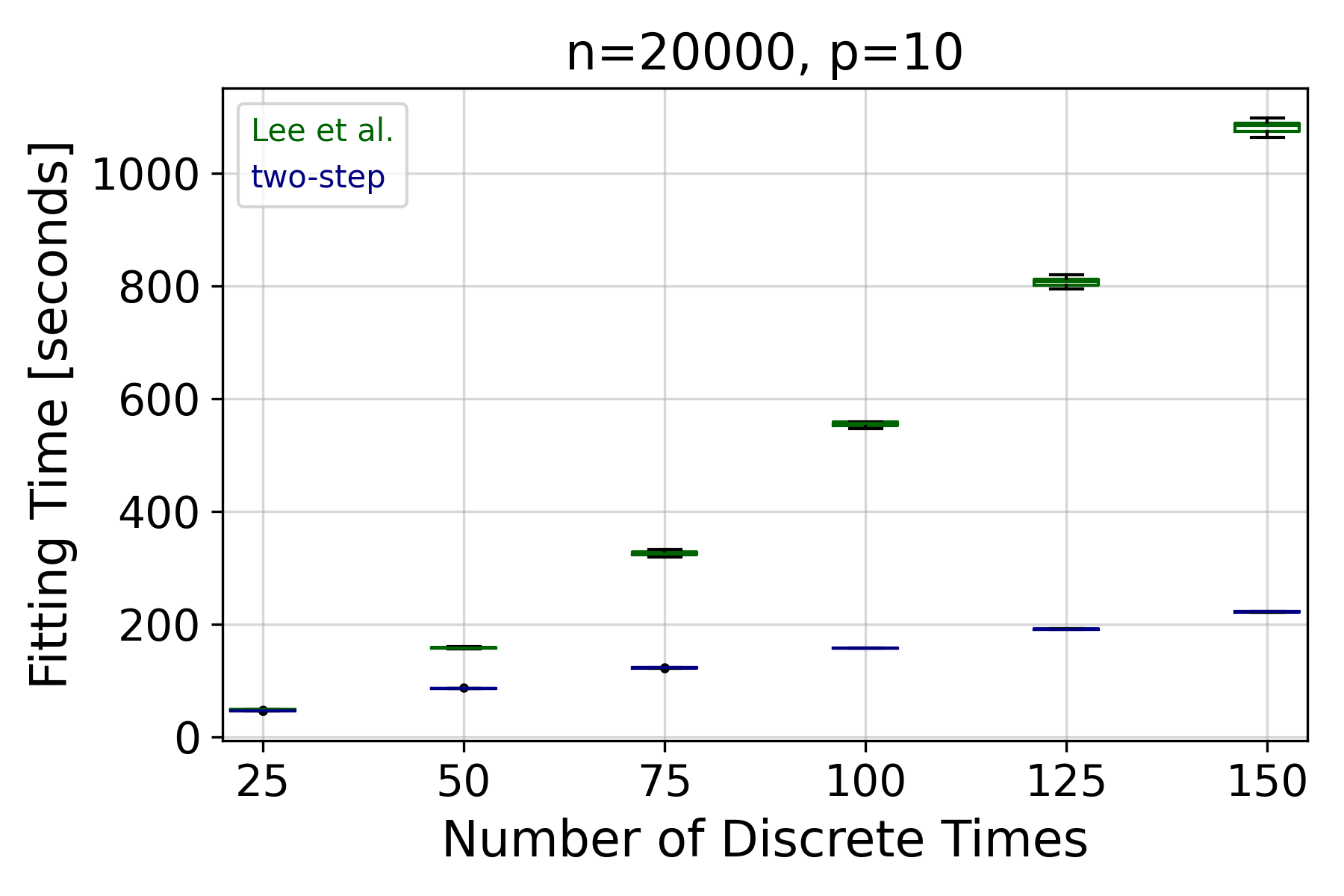}
    \caption{Simulation results: a computation time comparison between the method of Lee et al. and the proposed two-step approach.}
    \label{fig:fitting-time-comparison}
\end{figure}

\end{document}